\documentclass[rmp,aps,nofootinbib,twocolumn]{revtex4}
\usepackage{epsfig}
\usepackage{graphicx}
\usepackage{dcolumn}
\usepackage{amsfonts}
\usepackage{amssymb}

\usepackage{mathrsfs}

\usepackage{color}
\definecolor{mauve}{rgb}{0.5,0,1} 

\newcommand{\be}{\begin{equation}}
\newcommand{\ee}{\end{equation}}
\newcommand{\ba}{\begin{array}}
\newcommand{\ea}{\end{array}}
\newcommand{\bea}{\begin{eqnarray}}
\newcommand{\eea}{\end{eqnarray}}
\def\hf{\textstyle{1\over2}}
\def\half{{1\over2}}
\def\hf{\textstyle{\frac12}\displaystyle}
\def\Hb{\mathbb{H}}
\def\Rb{\mathbb{R}}

\def\afrac#1/#2{\leavevmode\kern.1em
\raise.5ex\hbox{\the\scriptfont0 #1}\kern-.1em
/\kern-.15em\lower.25ex\hbox{\the\scriptfont0 #2}}

\DeclareRobustCommand{\scrD}{\mathscr{D}}
\font\Lfrak=eufm10 at 10 pt
\newcommand{\Lfr}[1]{\mbox{\Lfrak #1}}  
\newcommand{\SD}{\Lfr D}
\newcommand{\FB}{\Lfr B}
\newcommand{\FF}{\Lfr F}
\newcommand{\Cas}{\Lfr C}

\newcommand{\Proj}{{\bf P}}

\def\pleth{\mbox{$\;\,p\hskip -8pt\bigcirc\;$}}

\newcommand{\makestrut}[2]{{\dimen12=#2
\divide\dimen12 by 4\dimen11=\dimen12\multiply\dimen11 by 3
\global\setbox#1=\hbox{\vrule height\dimen11 depth\dimen12 width0pt}}}

\newdimen\tadhdimen \newdimen\tabhdimen \newdimen\vdimen
\newdimen\smtadhdimen \newdimen\smtabhdimen
\newcount\splurt
\newbox\tadstrut \newbox\tabstrut
\newbox\smtadstrut \newbox\smtabstrut
\newbox\dtadstrut  

\newcommand{\hdashfill}{\leaders\hbox to 3.75pt{\hss
                          \vrule width1.875pt height0.4pt depth0.0pt}\hfill}
\newcommand{\vdashfill}{\leaders\vbox to 3.65pt{\vss
                          \hrule width0.4pt height1.825pt depth0.0pt}\vfil}

\newcommand{\setyoungsize}[2]{     
  \tadhdimen=#1\tabhdimen=#1\advance\tabhdimen by -0.4truept%
  \vdimen=#2%
  \makestrut\tadstrut\vdimen
  \advance\vdimen by -0.4pt%
  \makestrut\tabstrut\vdimen
  {\dimen12=\vdimen\divide\dimen12 by 4
  \global\setbox\dtadstrut=\hbox{\lower\dimen12\vbox to \vdimen{\vdashfill}}}
}

\newcommand{\youngt}[1]{%
  \vcenter{\offinterlineskip
  \halign{&\copy\tadstrut\hbox to \tadhdimen{\hss$##$\hss}\cr #1}}}
\newcommand{\youngd}[1]{%
  \vcenter{\offinterlineskip
  \halign{&\vrule##&\copy\tabstrut\hbox to \tabhdimen{\hss$##$\hss}\cr #1}}}

\newcount\mycount

\newcommand{\mymultispan}[3]{\omit\mycount #1\relax
                       \multiply\mycount by #2\relax
                       \advance\mycount by #3\relax
                       \loop\ifnum\mycount>1 \myspan\repeat}
\newcommand{\myspan}{\span\omit\advance\mycount by -1}

\newcommand{\yline}[1]{\mymultispan{#1}{2}{1}\hrulefill}

\setyoungsize{12pt}{12pt}

\begin{document}


\title{Dual pairing of symmetry groups and dynamical groups in physics}



\author{D.J.~Rowe$^1$, M.J.\ Carvalho$^2$,  
J.~Repka$^3$}
\affiliation{$^1$Department of Physics, University of Toronto, Toronto, ON M5S 1A7, Canada}
\affiliation{$^2$Department of Physics, Ryerson University, Toronto, ON M5B 2K3, Canada}
\affiliation{$^3$Department of Mathematics, University of Toronto, Toronto, ON M5S 2E4,
Canada}

\date{Accepted for publication in Reviews of Modern Physics, July 28, 2011}

\begin{abstract}

This article reviews many manifestations and applications of dual representations of pairs of groups, primarily  in atomic and nuclear physics.  Examples are given to show how such paired representations 
are  powerful aids in understanding the dynamics associated with shell-model coupling schemes and in identifying the physical situations for which a given scheme is most appropriate.
In particular, they suggest model Hamiltonians that are diagonal in the various coupling schemes.
The dual pairing of group representations has been applied profitably
in mathematics to the study of invariant theory.  We show that
parallel applications to the theory of symmetry and dynamical
groups in physics are  equally valuable.  In particular, the pairing of the
representations of a discrete group with those  of
a continuous Lie group or those of a compact Lie with those of a non-compact Lie group  makes it possible to infer many properties of difficult groups from those of simpler groups.
This review starts with the representations of the symmetric
and unitary groups, which are used extensively in the many-particle
quantum mechanics of bosonic and fermionic systems.  
It gives a summary of the many solutions and computational 
techniques for solving problems that arise in applications of symmetry methods in physics and which result from the famous Schur-Weyl  duality theorem
for the pairing of these representations.
It continues to examine many chains of symmetry groups and dual chains of dynamical groups associated with the several coupling schemes in atomic and nuclear shell models and the valuable insights and applications that result
from this examination.

\end{abstract}

\maketitle

{\bf Keywords:}   Dual pairs, complementarity, Schur-Weyl duality, Howe duality, symmetry, dynamical symmetry, nuclear models, pair-coupling models, collective models, shell-model coupling schemes, Young diagrams, Littlewood-Richardson rules, branching rules, plethysm.

\tableofcontents

\section{Introduction}
\label{sect:intro}

The value of exploiting the symmetries and algebraic structures of a
physical system in attempting to understand its properties is nowadays
much more widely appreciated than when distinguished physicists were
making disparaging remarks about {\it das Gruppenpest}.
In fact, it has become  evident that group theory and the related theory of Lie algebras underlie quantum mechanics and provide the essential language for the interpretation of physical phenomena in quantum mechanical terms.

A common strategy in seeking to understand a physical system is to start by
accumulating a large body of data  that relates to the phenomena of interest and examining it from many perspectives until it falls into recognizable patterns.
A second step is to interpret the patterns in terms of a phenomenological model; this provides some predictive capability and facilitates the search for new data that can be used to refine and develop the model.
The  challenge is then to understand the  model, and hence the data it describes,  in terms of a fundamental theory of the  system.

The problems faced in each of these steps are tailor-made for applications of group theory,  the study of symmetries and algebraic structures,  which provides a natural language for describing  the properties of physical systems and the relationships between  physical and mathematical models of such systems.  It is nowadays well recognized that solvable models invariably have simple algebraic structures; it is why they are solvable.
Moreover, the challenge of mapping a phenomenological algebraic model into a much grander algebraic scheme, such as that of many-particle quantum mechanics, is appropriately viewed as a  problem in group representation theory.

A central problem considered in this review is to take
a successful phenomenological model of some sub-dynamics of a many-particle system give it
a microscopic interpretation by identifying it with  a sub-model of many-particle quantum mechanics.  Another is to derive simple phenomenological models that exhibit the dynamics associated with particular coupling schemes for the microscopic theory.
It turns out that the algebraic methods of group theory and, in particular, the complementary concepts of \emph{symmetry groups} and 
\emph{dynamical groups} provide the basic tools  needed for these objectives.

\subsection{Algebraic models in quantum mechanics}

In quantum mechanics, the observables of a model are represented by Hermitian linear operators on a Hilbert space.
In this review, we focus  on algebraic models for which there exist a basic set of observables that span a finite-dimensional Lie algebra that we denote by 
$\Lfr{g}$. Thus, if $\hat X$ and $\hat Y$ are operators in $\Lfr{g}$ representing model observables  and  $\hat Z$ is defined by the commutation relation
\be [\hat X, \hat Y] = \hat X \hat Y - \hat Y \hat X= {\rm i} \hat Z ,\ee
then $\hat Z$ is also an element of $\Lfr{g}$.  [Note that the factor 
$ {\rm i} = \sqrt{-1}$ is needed in quantum mechanics because the commutator $[\hat X, \hat Y]$ is skew-Hermitian when $\hat X$ and $\hat Y$ are Hermitian.]
Other observables for the model are now given by Hermitian polynomials
in the elements of $\Lfr{g}$ with commutation relations inferred from the identity
\be [A,BC] = [A,B]C + B [A,C] .\ee
We then say that the full algebra of observables for such a  model is finitely generated, i.e., it is generated by the finite-dimensional subalgebra $\Lfr{g}$.

Now observe that many-particle quantum mechanics is an algebraic model
with a finitely generated Lie algebra of observables.
For example,  the position and momentum observables 
$\{x_{ni}, p_{ni}; n = 1, \dots, N, i=1,2,3\}$
of an $N$-particle system of spin-less particles in  three-space, $\Rb^3$,
are represented in quantum mechanics as operators
 $\{ \hat x_{ni}, \hat p_{ni}; n = 1, \dots, N, i=1,2,3\}$ on a Hilbert space 
$\Hb$ of normalizable wave functions according to the basic equations of quantum mechanics
\be \hat x_{nj} \Psi(x) = x_{nj}\Psi(x), \quad \hat p_{nj} \Psi(x) = 
-{\rm i}\hbar \frac{\partial}{\partial x_{nj}} \Psi(x) .\ee
These basic  observables obey commutation relations 
\be \begin{array}{c}
{[}\hat x_{nj}, \hat x_{mk} ] =  [\hat p_{nj}, \hat p_{mk} ] = 0, \\
 {[}\hat x_{nj}, \hat p_{mk} ] = {\rm i} \hbar  \delta_{n,m}\delta_{j,k} ,
 \end{array} \ee
 which close on a Heisenberg-Weyl Lie algebra.
The Hilbert space of the $N$-particle system then carries a unitary irreducible representation (irrep) of this  algebra, i.e., an irrep in which $\hat x_{ni}$ and $\hat p_{ni}$ are Hermitian operators.
For a system of many particles with intrinsic spin, it is necessary to augment this algebra by the addition of suitable spin operators.

\subsection{Symmetry groups and dynamical groups} 

\emph{A symmetry group of a system} is, by definition, a group of transformations of the system that leave its Hamiltonian invariant.  For example, a symmetry group for a system with a  Hamiltonian that is rotationally invariant is the  rotation group SO(3) (or SU(2) if particles with intrinsic spin are involved).
A given Hamiltonian may have more than one symmetry group
and a given system may have many possible choices of Hamiltonian. Thus, a system may have several possible symmetry groups.
Note that by a ``system" we mean any self-contained  model or   theory of a  physical system; thus different models of a physical system are regarded as distinct systems.

Variations on the  definition of a dynamical group for a quantum mechanical system have been given, for example, by
\textcite{DynamicalGroups}, \textcite{IBM}, \textcite{Isacker99}, and \textcite{Wulfman2011}. 
We adopt the definition that 
\emph{a dynamical group for a system} is a Lie group of  transformations of the  Hilbert space $\Hb$ of the system such that $\Hb$ carries an irreducible unitary representation of the dynamical group.
Essentially this ensures that the dynamical group is rich enough to relate all parts of the system.  This is appropriate because if a system
has parts that are not related in some way, then the parts are more usefully regarded as belonging to distinct systems.

It is also useful to define \emph{a dynamical group for a Hamiltonian} 
 $\hat H$
of a system to be a Lie group of  transformations of the  Hilbert space $\Hb$ of the system such that $\Hb$ is a sum of irreducible unitary 
 representations of this group with each  irreducible subspace being spanned by eigenstates of   $\hat H$.
The irreps of the dynamical group  for $\hat H$ then describe the  states of the system but fail to describe the relationships between states belonging to different irreps; for this one needs the full dynamical group of the system.

 Consider, for example, a  two-level BCS model of superconductivity \cite{BCS}   with a Hamiltonian
\be \hat H = \sum_{k=1}^2 \varepsilon_k \hat S_0^k 
- \sum_{i,k=1}^2 g_{ik}\hat S^i_+ \hat S^k_- ,
\label{eq:H_BCS2level}\ee
where,  for each $k=1,2$, $\hat S^k_\pm= \hat S^k_x \pm {\rm i} \hat S^k_y$, $\hat S^k_0 = \hat S^k_z$, and  
$\{ \hat S^k_x,  \hat S^k_{y}, \hat S^k_{z}\}$ 
are basis elements of an su(2)$^k$ (so-called quasi-spin) Lie algebra.
 When the model Hilbert space  is the space of a single irreducible 
 ${\rm SU(2)}^1 \times {\rm SU(2)}^2$ representation, then the group 
$G={\rm SU(2)}^1 \times {\rm SU(2)}^2$ is a dynamical group for  
both the model and the Hamiltonian.
However,  when the Hilbert space for the model carries a  sum of two or more inequivalent irreps  of $G$, then $G$ is dynamical group for the Hamiltonian 
$\hat H$ but not for the model with this Hamiltonian.
It is also seen that the subgroup  
${\rm U(1)} \subset{\rm SU(2)}^1 \times {\rm SU(2)}^2$ with a single infinitesimal generator $\hat S_0 = \hat S^1_0 + \hat S^2_0$ is a symmetry group for this Hamiltonian.

Note that a system may have many dynamical groups.  A particularly useful choice is one for which the important observables of the system have simple expressions in terms of its Lie algebra $\Lfr{g}$, e.g.,  as linear or quadratic polynomials of the algebra's operators. Different choices may suit different situations, as we  illustrate in the following.

\subsection{Coupling schemes}

Dynamical groups are  used in the construction
of basis states for the Hilbert space of a system.  
A good  basis is one for which the representation theory of a dynamical group, and especially that of its Lie algebra, facilitate
the calculation of matrix elements of physically relevant operators.
For  large (especially infinite-dimensional) Hilbert spaces, it is 
also useful if  truncations of the Hilbert space to  subspaces spanned by suitable subsets of basis states give accurate approximations 
for  states of interest.

Desirable basis states are given by a \emph{coupling scheme}  defined by a subgroup chain
\be G\supset K  \supset G_{\rm sym} ,\label{eq:DsubgChain}\ee
where $G$ is a dynamical group for the system,  $K$
is a dynamical group for some class of Hamiltonians for this system, and
$G_{\rm sym} $ is a symmetry group of transformations that leave these  Hamiltonians  invariant.
Basis states for the system, in a given irrep of $G$,
 are then given by state vectors 
$\{ |\alpha \lambda\beta\kappa\nu\rangle\}$
for a unitary irrep of $G$, where $\lambda$ labels an irrep of  $K$
 that  occurs with multiplicity indexed by $\alpha$, $\kappa$ labels an irrep of 
$G_{\rm sym}$ with multiplicity index $\beta$, and $\nu$ indexes a basis for the irrep $\kappa$.  From the definitions of $K$ and $G_{\rm sym}$,
it then follows that $\alpha$ and  $\lambda$ are good quantum numbers for the  eigenstates of any Hamiltonian  $\hat H$ for which $K$ is a dynamical group. Also $\kappa$ and $\nu$ are good quantum numbers when $G_{\rm sym}$ is a symmetry group for $\hat H$.  To determine the eigenstates of 
$\hat H$, it then  remains to diagonalize the $\alpha$ and $\nu$-independent matrices
\be M^{\lambda\kappa}_{\beta \beta'} =
\langle \alpha\lambda\beta \kappa\nu | \hat H 
| \alpha\lambda\beta' \kappa\nu\rangle .
\ee

Special cases  arise when the spectrum of a Hamiltonian doesn't depend on the multiplicity indices  $\alpha$ and $\beta$. This can happen because the  irreps of $G_{\rm sym}$ in $\Hb$ are  uniquely defined by the quantum numbers 
$\lambda$ and $\kappa$ so that  multiplicity indices are not needed, 
or because one is interested in a class of Hamiltonians for which
the basis states $\{|\alpha\lambda\beta\kappa\nu\rangle\}$ are  eigenstates for any choices of $\alpha$ and $\beta$. In either case,
\be \hat H |\alpha\lambda\beta\kappa\nu\rangle = E_{\lambda\kappa}
|\alpha\lambda\beta\kappa\nu\rangle,
\ee 
for a subset of Hamiltonians, and the  coupling scheme is said to diagonalize Hamiltonians in this subset.
A subgroup chain is often said to define a so-called  \emph{dynamical symmetry} for the class of Hamiltonians that it diagonalizes \cite{IBM}.
Thus coupling schemes defined by subgroup chains  make it possible to determine the spectra of  corresponding classes of Hamiltonians  by purely algebraic methods.
Conversely, the interpretation of a coupling scheme defined by a subgroup chain in terms of the class of Hamiltonians that it diagonalizes provides important insights into its physical significance.

In general, there will be more than one coupling scheme for a given system with a given symmetry group $G_{\rm sym}$, inasmuch as there may be more than one intermediate group $K$ between $G$ and $G_{\rm sym}$ in the subgroup chain (\ref{eq:DsubgChain}).  
However, in general, because of multiplicities, there  may not exist a coupling scheme, defined by a subgroup chain $G\supset K \supset G_{\rm sym}$, for which an arbitrary Hamiltonian for a system with symmetry group 
$G_{\rm sym}$ will automatically
be diagonal. 
The challenge is then to diagonalize  the Hamiltonian
 in a basis defined by some convenient coupling scheme.  The careful choice of  coupling scheme  is often  important,  particularly in  infinite-dimensional spaces such as those of the  atomic and nuclear shell models, for which approximations are inevitable.

A coupling scheme defined by a subgroup chain is best understood by an example.
Consider the BCS model of superconductivity with Hamiltonians of the form 
 (\ref{eq:H_BCS2level})  for which a dynamical group is given by
$G ={\rm SU(2)}^1 \times {\rm SU(2)}^2$ and a symmetry group by
$G_{\rm sym}= {\rm U}(1)$, where U(1) is the group with infinitesimal generator 
$\hat S_0 = \hat S^1_0 + \hat S^2_0$.
This model has two coupling schemes defined by the subgroup chains
\bea {\rm SU(2)}^1 \times {\rm SU(2)}^2 &\supset&
{\rm U(1)}^1 \times {\rm U(1)}^2 \supset {\rm U}(1) , \\
 {\rm SU(2)}^1 \times {\rm SU(2)}^2 &\supset&
{\rm SU(2)} \supset {\rm U}(1) ,\eea
where SU(2) is the group with Lie algebra spanned by  
$\{ \hat S_x,\hat S_y, \hat S_z\}$ with  
$\hat S_a = \hat S_a^1 + \hat S_a^2$ for $a=x,y,z$.
The first of these coupling schemes diagonalizes the subset of Hamiltonians of the form
\be \hat H_1 =  \sum_{k=1}^2 \big[ \varepsilon_k \hat S_0^k 
- g_{k}\hat S^k_+ \hat S^k_- \big], 
\label{eq:H_BCS1}
\ee
and the second diagonalizes  Hamiltonians of the form
\be
\hat H_2 =\varepsilon \sum_{k=1}^2  \hat S_0^k 
- g \sum_{i,k=1}^2 \hat S^i_+ \hat S^k_- .
\label{eq:H_BCS2}
\ee
In the two-level BCS  model, the first coupling scheme describes  the weakly-coupled limit of  two paired fermion states and  models  the normal phase of a superconductor, whereas the second coupling scheme describes the strong-coupling limit and  models the superconducting phase.

The study of coupling schemes and the Hamiltonians they diagonalize is a profitable way to expose the dynamical content of a system.
The  Interacting Boson Model \cite{IBM},  which has a U(6) dynamical group,  has been well studied in this way.
It is a model with three coupling schemes that diagonalize subsets of rotationally invariant Hamiltonians.
A remarkable result of these studies is the observation that, 
 for arbitrary rotationally invariant Hamiltonians and
 relatively large boson numbers, the low-energy eigenstates of this model 
  exist predominantly in one of the three possible phases  characteristic of its dynamical symmetries.  Similar results have also been observed in other systems \cite{RoweQDS04}.
This is remarkable because it happens even when the Hamiltonians contain significant interaction terms that mix different dynamical symmetries.  Reviews of such studies have been given in Sec.\ 7 of \textcite{RosensteelR05} 
 and by \textcite{CejnarJC10} 
  in which they are interpreted in terms of \emph{quasi-dynamical symmetries} (see Sect.\  \ref{sect:QDsymmetry}).

\subsection{Complementary symmetry groups and dynamical groups}

It is useful if a  symmetry group and a dynamical group for a Hamiltonian $\hat H$ are subgroups of a dynamical group for the whole system.  
However, although a dynamical group for $\hat H$ will usually 
contain a symmetry group of the Hamiltonian as a subgroup, as in Eq.\ 
(\ref{eq:DsubgChain}), it will not generally contain a maximal symmetry group.
As a consequence,  a given choice of dynamical group may not make it possible to take full advantage of the  symmetries of the Hamiltonian in seeking its eigenstates.
An optimal symmetry group is as close to a maximal symmetry group  as possible whereas an optimal dynamical group for a Hamiltonian is the simplest that enables the spectrum of the Hamiltonian to be computed  easily.
 Augmenting the symmetry group while decreasing the dynamical group becomes possible if one gives up the constraint that it must be a subgroup of the dynamical group.  However, if this is to be useful, the symmetry group and the dynamical group must remain compatible in such a way  that the two groups can effectively complement each other.
 Such a compatibility is  achieved if the symmetries of the Hamiltonian are   products of  subgroups of a dynamical group and  groups that commute with the dynamical group.

The rationale for seeking commuting dynamical and symmetry groups is as follows.  Suppose that $\kappa$ labels a unitary irrep of a symmetry group $G_{\rm Sym}$ of a Hamiltonian $\hat H$  and  $\nu$ labels a basis for this irrep. 
Let $\Hb^{(\kappa\nu)}$ denote the subspace of  all states in 
 $\Hb$ with the quantum numbers $\kappa$ and $\nu$.  
Then, if it should happen that each subspace $\Hb^{(\kappa\nu)}$ carries a unitary representation of a commuting
group $G_{\rm Dyn}$, this group  will be
of major assistance in determining the spectra and other properties of any Hamiltonian for which $G_{\rm Sym}$ is a symmetry group.
Moreover, the identification of a group $G_{\rm Dyn}$ whose action on a Hilbert space of interest commutes with that of a desired symmetry group provides a potentially useful way of constructing simply solvable models.

\subsection{Dual pairs of group representations}

Suppose a given Hilbert space $\Hb$ carries commuting representations of  
$G_{\rm Sym}$ and $G_{\rm Dyn}$.
A particularly valuable  situation arises  when the unitary representation  of 
$G_{\rm Dyn}$ carried by every subspace $\Hb^{(\kappa\nu)}$,
as defined above,  is irreducible and defined uniquely by $\kappa$. In this case,  a basis for the whole Hilbert space is given by a set $\{ \psi_{\mu\kappa\nu}\}$, where $\nu$ indexes a basis for an irrep $\kappa$ of $G_{\rm Sym}$ and $\mu$ indexes a basis for the irrep of  $G_{\rm Dyn}$, which can now also be labeled 
by $\kappa$.  
The pair of groups $G_{\rm Sym}$ and $G_{\rm Dyn}$   are then said to have 
\emph{dual representations} on the Hilbert space $\Hb$.

Situations of this kind might appear to be rare.  In fact,  as this review shows, 
they are common and occur for all the standard coupling schemes of the atomic and nuclear shell models.  They prove to be of profound importance. 
Moreover, it is possible to benefit from the widespead study of the dual pairing of group representations   in the mathematical field known as \emph{invariant theory} (see Sec.\ \ref{sect:bgl}). 
From such studies, several of which  were initiated in nuclear physics, some very remarkable relationships have been discovered between the properties of  very different groups that happen to have dual representations on some Hilbert spaces of relevance to physics.
Some of these properties are well-known.
For example, it is common practice to speak of a unitary representation of a many-particle system as having a given symmetry, where the symmetry referred to is a representation of the symmetric group of permutations of the particles, i.e., a group with representations that are dual to those of a unitary group.
This duality, known as Schur-Weyl duality, leads to  the  Young diagram  and many other powerful techniques that give simple solutions to  problems that arise in physics.

 Dual representations of a pair of Lie groups are defined precisely as follows:

\medskip
\emph{Two groups $G_1$ and $G_2$ are said to have dual
representations on a space $\mathbb{H}$ if the following conditions
are met: {\rm (i)} $\mathbb{H}$ is the carrier space for fully reducible
representations of both $G_1$ and $G_2$; {\rm (ii)} the actions of $G_1$
and $G_2$ on $\mathbb{H}$ commute; {\rm (iii)} the representation $\kappa$ of the direct product group $G_1\times G_2$ on $\mathbb{H}$,  defined by the
actions of $G_1$ and $G_2$ on $\mathbb{H}$, is multiplicity free; {\rm (iv)} each
irrep  of $G_1$ that occurs in the decomposition of
$\mathbb{H}$ is paired with a  single irrep of $G_2$, and vice versa.}
\medskip

Condition (i)  restricts consideration to representations of $G_1$ and $G_2$ that are expressible as direct sums of irreps. Thus, for present purposes,
we exclude representations of some non-compact groups that are direct integrals of irreps.  
For the purposes of this review, we  also restrict consideration to representations that are unitary. 

Condition (iii) states that, in the decomposition of the
representation $\kappa$ of $G_1\times G_2$ on $\mathbb{H}$ to a direct
sum of irreps, no  irrep appears more than once.
Condition (iv) guarantees that, in such a decomposition, it is
possible to identify, with a common label, paired irreps  of the two
groups (e.g., by  the angular momentum label $l$ in the SU(1,1) $\times$ SO(3) example given below).

Simple proofs of the duality theorems on which this review is based are given in another paper \cite{Thms}.

\subsection{A simple example}

Suppose we wish to determine the spectrum of a central-force Hamiltonian 
for a particle moving in ordinary three-space or for the relative motion of a diatomic molecule moving about its center of mass.  A standard practice is to seek eigenfunctions of the Hamiltonian in a basis of spherical harmonic oscillator wave functions.  Basis wave functions for the Hilbert space $\Hb$ of a spherical harmonic oscillator are given by products
\be \psi_{nlm}(r,\theta,\varphi) =
R_{nl}(r)\,Y_{lm}(\theta,\varphi),  \label{eq:1} \ee
where $\{R_{nl}\}$ are radial wave functions and $\{ Y_{lm}\}$ are spherical harmonics.
The subset of wave functions $\{ \psi_{nlm} \}$ with  $n$ fixed span a Hilbert space $\Hb^{(n)}\subset \Hb$ that carries
 an irrep of the U(3) symmetry group of a harmonic oscillator Hamiltonian. 
The subset  $\{ \psi_{nlm} \}$ with fixed values of $n$ and $l$ span a Hilbert space $\Hb^{(nl)}\subset \Hb^{(n)}\subset\Hb$   that carries an irrep of the rotation group SO(3).
Thus, the basis wave functions $\{ \psi_{nlm} \}$ are those of the coupling scheme defined by the irreps of the symmetry groups of a harmonic oscillator in the subgroup chain
\be \begin{array} {ccccc}
 {\rm U}(3) &\supset& {\rm SO}(3) & \supset &  {\rm SO}(2) \, .\\
n&& l && m 
\end{array} \label{eq:I.U3chain}\ee

As well as being a symmetry group for the spherical harmonic oscillator, the group U(3) is also a dynamical group inasmuch as its irreps are spanned by eigenstates of this Hamiltonian.  We now enquire as to whether or not there is a  group that commutes with the SO(3) symmetry group that  could serve as a dynamical group for a  general central force Hamiltonian.
We find that  SU(1,1) is such a group and that it
has a unitary irrep on each of the Hilbert spaces $\Hb^{(lm)}$ spanned by the harmonic oscillator wave functions
$\{ \psi_{nlm} \}$ with fixed values of $l$ and $m$.
Thus,  the representation theory of the group SO(3) and its SO(2) 
subgroup determines the spherical harmonics with the quantum numbers $l$ and $m$.  Moreover, the representation theory of the dynamical group SU(1,1)
and its Lie algebra can be used to determine  the radial wave functions for a general central-force Hamiltonian.

The group SU(1,1) is defined as follows.
From the vector operators  $\hat{\bf r} = (\hat x_1,\hat x_2,\hat x_3)$ and 
$\hat{\bf p}=(\hat p_1,\hat p_2,\hat p_3)$ for a particle in three-space, 
  we can form the SO(3)-invariant (i.e., scalar) operators
$\hat r^2=\hat {\bf r}\cdot \hat{\bf r}$ and $\hat p^2 = \hat{\bf p}\cdot \hat{\bf p}$.  
A linear combination of these scalars is a harmonic oscillator Hamiltonian
\be \hat H = \frac{1}{2m}\, \hat p^2 + \frac{1}{2} m\omega^2 \hat r^2, \ee
and  the commutator
\be [\hat r^2, \hat p^2] = 
2{\rm i}\hbar (\hat{\bf r} \cdot \hat{\bf p} +\hat{\bf p} \cdot \hat{\bf r} ) , \ee
is another SO(3) scalar.  Further commutators
 produce no new operators, which means that the operators
$\{ \hat r^2, \hat p^2, \hat{\bf r} \cdot \hat{\bf p} +\hat{\bf p} \cdot \hat{\bf r}\}$ are a basis for a Lie algebra.  This is the Lie algebra su(1,1) of the group SU(1,1) which is now observed to be a dynamical group for any central force Hamiltonian
\be \hat H = \frac{1}{2m}\, \hat p^2 + \hat V(r),\ee
where $V$ is a rotationally-invariant  potential energy.

The remarkable property of this SU(1,1) group is that, although its elements commute with those of SO(3), its irreps on the Hilbert space of a particle in three-space are uniquely defined by the SO(3) angular momentum quantum number 
$l$.  Moreover, any wave function in $\Hb$ that has angular momentum $l$ must belong to  irreps of both SO(3) and SU(1,1) labeled by $l$.  
This is a deep result which has its origins in the centrifugal coupling of the radial and rotational motions of the particle.

\subsection{Outline of the review}

Section II gives an historical review of the major contributions to the development and applications  of  dual group representations of which we are aware.

Section III presents the duality relationship, known as Schur-Weyl duality,  between the unitary and symmetric groups on the Hilbert spaces of many particles.  This duality  gives rise to the Young diagram methods and relationships between characters which are of enormous
 practical importance in the use of these groups in physics.

Section IV shows the power of   Schur-Weyl duality  in deriving the branching rules and tensor products that are needed in atomic and nuclear physics.

Section V presents duality relationships between pairs of unitary groups.  These duality relationships are applied to the construction of fully antisymmetric combinations of space and spin wave functions and of space, spin, and isospin wave functions.  Such methods are needed in atomic, nuclear, and elementary particle physics.

Section VI introduces methods of second quantization which provide simple ways of ensuring that wave functions of multiple identical bosons or fermions
 are automatically symmetric or  antisymmetric, respectively.

Section VII applies duality techniques to many-boson systems.  It shows the underlying algebraic structures of central force problems and the main models of nuclear collective motion, e.g., the Bohr model,  interacting boson model, and the microscopic symplectic model.

Section VIII  applies duality techniques to many-fermion systems.  It shows the relationship between pairing models and corresponding single-$j$- and multi-$j$-shell coupling schemes.  
Similar relationships are shown for  $LST$ coupling models.

Section IX  gives a brief review of other related developments in group theoretical methods.

Section X gives a  summary and some possibilities for further development and/or pursuits of  methods described in the review.


 \section{An historical perspective}
\label{sect:bgl}

Dual pairs of group representations  were first  used in the
context of invariant theory by  \textcite{Schur, Schur2}.
In constructing  the finite-dimensional irreps of
general linear groups, Schur discovered what is now called Schur-Weyl duality, which he reported in his 1901 doctoral dissertation \cite{Schur}.
 Weyl developed the theory, and applied it to diverse physical
problems \cite{WeylI}.
In his book, originally published in 1928, \textcite{WeylQM} gave wide publicity 
to Young's work  (cf.\  \textcite{YoungC}) on the symmetric groups (including coining the term Young tableau), extended the work of Schur (e.g., to include Weyl's character formula), and applied Schur's discovery in quantum mechanical contexts.

Subsequent to these  seminal works, Schur-Weyl duality has been used to
develop branching rules and tensor product decompositions for a wide
range of groups of interest to the physics community. In a series of
publications, reviewed in his book,  \textcite{Littlewood} used Schur-Weyl duality and particularly Schur function techniques 
to advance the theory of group characters. 
Work done on branching rules up to
the mid-1960s was summarized by \textcite{Whippman}. 

In more recent times, \textcite{Macdonald95, Mac1}, \textcite{King75}, 
\textcite{Wybourne93}, and colleagues
further developed the character theory of Lie groups and, in the process,
exhibited the power of Schur function techniques. The
publications \cite{King75,King70,King71,BKW},
and \cite{BW} provide entry points into their work on the character
theory of compact Lie groups. 
Schur function techniques have also been used to extend the application of
Schur-Weyl duality to the character theory of non-compact Lie groups
\cite{GW, TTW, RWB, KingW85, KW98, KW00, KW00b}, Hecke algebras
\cite{King93, Wybourne91}, and supersymmetry \cite{BR,CW1,CWa,CZ}.

\textcite{HBa, HB} have used Schur-Weyl duality to derive
relationships among coupling coefficients of symmetric and unitary groups.
Also noteworthy is the work of \textcite{DHoker} and \textcite{KT}, in which Young diagram
techniques for unitary groups are systematically extended to various classical Lie groups, and the work of \textcite{Brauer} which extended 
the theory from the symmetric group to the commutants of orthogonal and symplectic groups on tensor product spaces
with applications to quantum groups, knots and links \cite{Benk}.

In addition to the many examples already listed, a recent
application of duality to error correction in quantum computing has been
given by \textcite{JKK}. This underscores the fact that  Schur-Weyl
duality is not  only useful  in the construction of wave functions
with specified permutation symmetries; it is also  useful 
when one wishes to classify the symmetry properties of 
more general composite systems.

 The second duality relationship to be discovered was the duality between the representations of pairs of unitary groups, the so-called unitary-unitary duality. The
application of this duality relationship, which follows directly
from Schur-Weyl duality, has been widely used in the classification
of nuclear shell-model states following the introduction of Wigner's
U(4) super-multiplet group  \cite{Wigner1}. In this classification,
representations of the U(4) super-multiplet group, which contains a
U(2) isospin subgroup and a U(2) intrinsic-spin subgroup for
spin-1/2 nucleons, are combined with contragredient irreps of a
U$(n)$ group of transformations of spatial wave functions to form
the totally antisymmetric states required for a many-nucleon system.

To our knowledge, the next duality relationship to be discovered was
the compact symplectic-symplectic duality.  The utility of compact
symplectic groups in the atomic shell model was brought to the
attention of the  physics community by \textcite{Racah3}. It was
later applied to the classification of nuclear shell-model states in
$jj$ coupling by  \textcite{Flowers1,Flowers2,Flowers3}, who introduced extra group theoretical structures to account for the nucleon's isospin degrees
of freedom. In the process, Flowers recognized and exploited many
duality relationships. The symplectic-symplectic duality theorem
underlying these relationships was formulated and proved, using
character theory, by \textcite{Helm}, who described what he had
discovered as \emph{group complementarity}. Independently,
\textcite{Kerman} introduced the related and much used concept of a
quasispin group which proved to be a special case of Helmers' theorem.
The concept of USp-USp duality was reviewed in
works directed to physicists by \textcite{Parikh} and
\textcite{Lipkin}. It has been widely applied to
 fermion-pair coupling phenomena, both in nuclear systems and
superconductivity. Recent applications are described in
\cite{DH,vanI,SGD,Lorazo,Engel,PDJ,RR1}; and \cite{RR2003}.

A duality relationship between the representations of pairs of
orthogonal groups was similarly identified in the classification of
nuclear shell-model states in $LS$ coupling.  In this case, the
discovery emerged from two quite distinct methods for the
construction of shell-model basis states that diagonalize a simple
$LS$-coupling pairing Hamiltonian.  \textcite{Bayman} gave the
basis states in terms of the irreps of a symmetry group for the
Hamiltonian and, independently, \textcite{FS}
gave the same basis states in terms of the irreps of a distinct
dynamical group.
{Although the duality of these complementary
approaches has long been understood, the}
orthogonal-orthogonal duality theorem
underlying them has only recently been
formulated and proved in physics \cite{Kota,RC07,Thms}.

Independent of the discoveries in physics, the theory of dual group representations has been developed in mathematics and given a rigorous basis within the framework of invariant theory. 
The oscillator \emph{Weil representation} of the non-compact symplectic groups was particularly influential in this development.
Following the construction of the oscillator representations of the non-compact symplectic groups by \textcite{Segal}, \textcite{Shale}, and \textcite{Weil}, the well-known
 SU(1,1) $\times$ O(3) duality relationship (discussed in the introduction) was recognized as a special case of a more general symplectic-orthogonal duality relationship. 
This duality relationship was introduced into physics following its discovery by Moshinsky and Quesne \cite{MQ9, MQ0}, based on results in  Chac\'{o}n's thesis \cite{Chacon}, and derived, together with other such relationships, by 
\textcite{KV}.
In fact, apart from Schur-Weyl duality, the  duality relations described in this review are all special cases of the so-called \emph{dual reductive pairs} identified  in a paper by \textcite{Howe89} that was written in 1976 and widely circulated but not published at that time.
 Examples of dual reductive pairs were given by \textcite{Gelbart} and 
 \textcite{Howe85} and many authors have contributed to the subject.
In addition to the duality relationships considered in this review, more are known in mathematics.   Reviews of invariant theory and dual group representations in mathematics are given by \textcite{Howe92, Goodman04}, and \textcite{Li}.
For example, there are duality relationships between  U$(p,q)$ and U$(n)$ 
\cite{KV}, between compact symplectic  and non-compact SO$^*(2n)$ groups
\cite{LeungT95}, between pairs of non-compact groups (O$(p,q)$, Sp$(m,\Rb)$)
\cite{Adams83}, and also between O$(p,q)$ and an ortho-symplectic group.
The latter dual pair is shown by \textcite{LuHowe10} to be relevant to Maxwell's equations.  Howe duality has been extended to the realm of exceptional Lie groups \cite{DS} and quantum groups \cite{Green}, and has also been applied to gauge theories \cite{Schmidt} and the quantization of constrained systems \cite{Landsman}.  Its use in deriving branching rules for the harmonic series of 
Sp$(n,\Rb)$ was initiated by \textcite{RWB} and followed by a similar derivation of branching rules for U$(p,q)$ and SO$^*(2n)$ by \textcite{KingW85}.
Branching rules for many classical pairs of Lie groups have been derived by 
\textcite{HoweTW04}.

\bigskip

\section{The symmetric and unitary groups} 

\label{sect:3.SWdualitya}

The symmetric group S$_N$ is the group of permutations of the indices that label the particles of an $N$-particle system and the unitary groups are 
transformations that preserve the orthogonality relationships of quantum mechanical states.
These groups  are indispensable in quantum mechanics.
The symmetric group is a fundamental symmetry of the quantum mechanics of identical particles. 
Moreover, the Schur-Weyl duality theorem  shows that the subgroup of all unitary transformations of a many-particle system that commute with the symmetric group is the so-called group of one-body unitary transformations.  It also shows that  this subgroup and  the symmetric group have dual representations on a many-particle Hilbert space.
Thus, it identifies the group of one-body unitary transformations as the fundamental dynamical group of many-particle quantum mechanics.

Elementary particles are considered to be either boson-like or fermion-like, which means that their many-particle wave functions are totally symmetric or totally
antisymmetric, respectively, under permutation. 
However, wave functions often have  several components. For example, 
fermion wave functions may be combinations of spatial, spin, and isospin wave functions which need not be separately antisymmetric.
Thus, in calculations, it is necessary to keep track of the way the separate components transform under permutation of the particle indices so that they can be put together in antisymmetric combinations. This awe-inspiring task
is much simplified by use  of the Schur-Weyl theorem.

Suppose a space of single-particle wave functions is $n$-dimensional  and that 
U$(n)$ is the group of unitary transformations of this space.
The Schur-Weyl theorem shows that there is a duality relationship between the irreps of the unitary group U$(n)$ and those of the symmetric group S$_N$ for an $N$-particle system.  Thus, not only do the fully antisymmetric states of a many-fermion system carry a corresponding irrep of U$(n)$ but the states of any specified permutation symmetry carry a U$(n)$ irrep.  
This is a remarkable result that provides basic tools for the application
of group theory to many-particle systems and, in particular, to the development and interpretation of phenomena in terms of the various shell model  coupling schemes.
In fact, as this section will show,  Schur-Weyl duality  implies that many of the well-known properties of the discrete S$_N$ group can be used to infer corresponding properties for the continuous
U$(n)$ Lie group and vice-versa. Thus, numerous parallel techniques have been developed for the simultaneous study of the symmetric and unitary groups, e.g., within the framework of Young diagram methods and character theory.

This section makes substantial use of the review of Young diagram and related Schur function techniques given by \textcite{Wyb1}.
Sections \ref{sect:3.SWdualitya} A--C follow closely those of \textcite{RW10}.

\subsection{The Schur-Weyl  duality theorem} \label{sect:2a}

Let $\Hb_n$ denote an $n$-dimensional Hilbert space  and let
\be \Hb_n^N = \Hb_n^{\otimes N} = \underbrace{\mathbb{H}_n
\otimes \cdots \otimes \mathbb{H}_n}_{N \; {\rm copies}}  \label{eq:HnNcopies} \ee
denote the tensor product of $N$ copies of $\Hb_n$. The following Schur-Weyl theorem
is  naturally understood if $\Hb_n$ and $\Hb_n^N$ are interpreted as
spaces of single-particle and $N$-particle wave functions, respectively.

\medskip

{\bf Theorem 1 (Schur-Weyl duality):}
\emph{ The groups ${\rm S}_N$ and {\rm U}$(n)$ have dual representations on $\mathbb{H}_n^N$.%
\footnote{The theorem is somewhat more general than stated here. This is because
GL$(n,\mathbb{C})$ is the complex extension of U$(n)$ and, as a
consequence, the representations of U$(n)$ extend to (non-unitary)
representations of GL$(n,\mathbb{C})$. However, for present
purposes, we restrict consideration to the 
${\rm U}(n) \subset {\rm GL}(n,\mathbb{C})$ subgroup.
}}
\medskip

 Let  $\{\psi_1,\psi_2, \dots, \psi_n\}$ denote an orthonormal
basis of single-particle wave functions for $\Hb_n$. The group
U($n$) then has a  defining $n$-dimensional irrep $\hat U^{\{1\}}$ 
on $\mathbb{H}_n$ given by transformations of its single-particle wave functions 
\be \hat U^{\{1\}}(g)\,  
\psi_\nu = \sum_\mu \psi_\mu g_{\mu\nu} \,,\quad
g\in {\rm U}(n)\,. \label{eq:5}\ee
The corresponding  Hilbert space
$\mathbb{H}_n^N$  has an orthonormal basis given by the $N$-particle
wave functions
 \be
\Psi_{\nu_1\dots\nu_N} = \psi_{\nu_1}\otimes \psi_{\nu_2}
\otimes \dots \otimes \psi_{\nu_N} , \label{eq:7} \ee
and carries a reducible U$(n)$ tensor-product representation $\hat U ^{\{N\}}$  
given for $g\in {\rm U}(n)$ by
\be \hat U ^{(N)}(g)\, \Psi_{\nu_1\dots\nu_N} = 
 \sum_{\mu_1 \dots \mu_N}
\Psi_{\mu_1\dots\mu_N}\, g_{\mu_1\nu_1}\, g_{\mu_2\nu_2} \dots
g_{\mu_N\nu_N} . \label{eq:11a} \ee
For any positive integer $N$, this U$(n)$ representation and its
irreducible sub-representations will be referred to as \emph{tensor representations} of degree $N$.

The Hilbert space   $\mathbb{H}_n^N$ also carries a reducible
representation $\hat{P}$ of
 the symmetric group S$_N$,  defined  by the
permutations of the $N$ indices of the
$\{\Psi_{\nu_1\nu_2\dots\nu_N}\}$ basis; e.g., if $\pi_{12}\in S_N$
is the permutation that exchanges particles 1 and 2, then \be \hat
P(\pi_{12}) \Psi_{\nu_1\nu_2\dots\nu_N} =
\Psi_{\nu_2\nu_1\dots\nu_N} \,. \ee

A proof of the Schur-Weyl duality theorem can be found,
for example, in Chapter V of \textcite{WeylQM}, Chapter 5
of  \textcite{Sternberg}, and  Chapter 6 of \textcite{FH}. 
Here we highlight and explain the main points of the theorem,
some of which are immediately evident. For example, it
is readily ascertained that the actions of S$_N$ and U$(n)$ commute:
\be \hat P(\pi)\,
\hat U^{\{N\}}(g)\,  
\Psi_{\nu_1\dots\nu_N} = \hat
U ^{\{N\}}(g)\,\hat P(\pi)\, \Psi_{\nu_1\dots\nu_N} , \ee
for all $g\in {\rm U}(n)$ and all $\pi \in S_N$.  
Thus, the direct product group S$_N \times {\rm U}(n)$ has a
reducible representation $\hat T$ on $\mathbb{H}_n^N$
for $N>1$ and $n>1$, defined by 
\be \hat T(\pi,g)\, \Psi_{\nu_1\dots\nu_N} 
= \hat P(\pi)\,\hat U^{\{N\}}(g)\, \Psi_{\nu_1\dots\nu_N} ,\label{eq:14} \ee
for $\pi\in S_N$ and $g\in{\rm U}(n)$. 
Every irrep of S$_N\times {\rm U}(n)$ is then  expressed as an 
``outer product"
$\hat T = \hat{\cal P}  \times \hat{\cal U}$,  
where $\hat{\cal P}$ and $\hat{\cal U}$ 
are,  respectively,  irreps of S$_N$ and U$(n)$,  and 
\be\hat T(\pi,g) = \hat{\cal P}(\pi) \times \hat{\cal U}(g) 
\,,\quad \pi\in {\rm S}_N\,, \;\; g\in {\rm U}(n)\,. 
\ee 
(In the mathematics literature,  the irrep 
$\hat{\cal P}\times \hat{\cal U}$ 
would be denoted $ \hat{\cal P}\otimes \hat{\cal U}$.
However, we avoid this notation because of the potential confusion with the use of $\otimes$ for the standard tensor product for irreps of a single group.)

By condition (iv)  of the definition of duality, the Schur-Weyl theorem affirms
that  all irreps occurring in the decomposition of $\hat T$ are of the form 
\be\label{eq:SWdualityIrreps}
\hat T^\lambda = \hat P^{(\lambda)} \times  \hat U^{\{\lambda\}} ,  
\ee 
where $\hat P^{(\lambda)}$ and $\hat U^{\{\lambda\}}$ are, respectively, S$_N$ 
and U$(n)$ irreps that are  uniquely defined by a common label $\lambda$. 
Moreover, every irrep of S$_N$ appears in the above decomposition, provided that 
$N \le n$. Some S$_N$ irreps do not occur if $N>n$; for example, it is not possible to form a totally antisymmetric $N$-particle wave function with fewer than
$N$ linearly-independent single-particle wave functions. Thus, we
learn from the Schur-Weyl theorem
 that every S$_N$ irrep $\hat P^{(\lambda)}$, with $N\leq n$, is uniquely associated with a corresponding tensor irrep
 $\hat U^{\{\lambda\}}$ of U$(n)$.
Conversely, every tensor irrep $\hat U^{\{\lambda\}}$ of degree $N$ of U$(n)$ is uniquely associated with an S$_N$ irrep $\hat P^{(\lambda)}$. 
Although well-known and often taken for granted, we emphasize again that these are remarkable results because they mean that much of the representation theory of a family of continuous Lie groups, namely the unitary groups and their subgroups, can be inferred from the representation theory of the finite symmetric groups.

\subsection{Characterization of ${\rm U}(n)\times{\rm S}_N$  representations} 
\label{sect:IIIB}

A U$(n)$  tensor irrep is characterized in two standard ways:  
by its highest weight relative to a Cartan subalgebra  and by its S$_N$ symmetry. The relationship between these alternative characterizations exposes the duality relationship between S$_N$ and U$(n)$ representations.

The  group U$(n)$ of $n\times n$ unitary matrices has the property that a matrix 
$g\in {\rm U}(n)$ can be expressed as an exponential, 
$g= e^{{\rm i}X}$, where $X$ is an $n\times n$ Hermitian matrix. 
 Moreover, every physical observable is represented in quantum mechanics by a Hermitian operator.
Thus, it is customary in physics to define the Lie algebra, u$(n)$, of the group 
U$(n)$ as the set of Hermitian $n\times n$ matrices.

Let  $C_{\mu\nu}$ denote a matrix that has the entry 1 at the intersection of row 
$\mu$ with column $\nu$ and 0 everywhere else, i.e.,
\be \big( C_{\mu\nu} \big)_{ij} = \delta_{\mu,i} \delta_{j,\nu} .
\label{eq:4.Cmunumatrices}\ee
These matrices have commutation relations
\be [C_{\mu\nu}, C_{\kappa\lambda}] = 
\delta_{\nu,\kappa} C_{\mu\lambda}
-\delta_{\mu,\lambda} C_{\kappa\nu}  
\label{eq:4.UdCR}\ee
and span the
\emph{complex extension} of the  u$(n)$ Lie algebra.
A basis for u$(n)$ is then given, in terms of them, by the Hermitian linear combinations
\be C_{\mu\nu} + C_{\nu\mu} , \quad 
{\rm i} \big( C_{\mu\nu} - C_{\nu\mu}\big) , \quad
1\leq \mu , \nu \leq n . \label{eq:4.XY} \ee
They have $N$-particle tensor representations given by 
\be C_{\mu\nu} \to \hat C_{\mu\nu} = 
\sum_{i=1}^N \psi_\mu(i) \frac{\partial}{\partial \psi_\nu(i)} , \label{eq:3B30}    \ee
where $\partial / \partial \psi_\nu (i)$ is a functional derivative with respect to a single-particle wave function $\psi_\nu (i)$
in the $i$th factor in the tensor product space 
$\Hb_n^N$ of Eq.\ (\ref{eq:HnNcopies}).

Useful basis states for an irrep of U$(n)$ and its Lie algebra u$(n)$ are given by the simultaneous eigenstates of the subset of commuting operators
that represent the diagonal  matrices of a Cartan subalgebra of u$(n)$.
We call these operators \emph{Cartan operators}.
A basis of Cartan operators for the tensor representations and
corresponding raising and lowering operators are given by 
\bea \{ \hat{C}_{\nu\nu} , \nu = 1, \dots, n \} , && (\it Cartan\; operators),\\
 \{  \hat{C}_{\mu\nu}  ,  1\leq \mu < \nu\leq n\} , &&  (\it raising\; operators), \\
\{  \hat{C}_{\mu\nu}  ,  1\leq \nu < \mu\leq n\} , \ &&  (\it lowering\; operators) .\eea

If a state $|\Psi\rangle$ of a U$(n)$ representation is an eigenstate of the Cartan operators, i.e., 
\be \hat C_{\nu\nu} |\Psi\rangle = \lambda_\nu |\Psi\rangle , 
\quad \nu = 1, \dots, n , \ee
then the set of eigenvalues 
$\lambda = (\lambda_1, \lambda_2, \dots , \lambda_n)$ is said to be the
\emph{weight} of the state $|\Psi\rangle$.

A weight  $\lambda=(\lambda_1, \lambda_2, \dots, \lambda_n)$ is said to be  higher than a weight  $\lambda'=(\lambda'_1, \lambda'_2, \dots, \lambda'_n)$ if $\lambda_1> \lambda'_1$, or if $\lambda_1= \lambda'_1$ and
$\lambda_2> \lambda'_2$, etc.
For convenience, commas and trailing zeroes are usually omitted in specifying a weight when it would be unambiguous to do so. Thus,  a highest weight (2,1,0,0) is written as  (21),  and a highest  weight (1,0,0,0,0) is written as (1).
However, a weight (21,0,0) would be written explicitly.

Because U$(n)$ is  compact,  all its irreps are unitary and finite-dimensional. 
Thus, a U$(n)$ irrep has a highest weight.  
An important property of Lie algebra structure theory is that a finite-dimensional irrep of any semisimple or reductive Lie algebra
has a  state with a uniquely defined  highest weight;
in the case of the unitary algebra u$(n)$, the components $\lambda_i$ 
of the highest weight $\lambda$ satisfy the inequality
\be \lambda_1 \geq \lambda_2 \geq \dots \geq \lambda_n , \label{eq:3.hwtineq}\ee
where $\lambda_n \geq 0$ for a tensor irrep.
It follows that a U$(n)$ irrep is completely defined by its highest weight.

Consider, for example, the highest-weight state for the $N=3$ irrep of U$(3)$ with highest weight $(1^3) \equiv (111)$.  
It must be a linear combination of wave functions of the form
$\{ \psi_i(1) \psi_j(2) \psi_k(3) \}$, where $i$, $j$, and $k$ index three distinct single-particle states.
The particular  linear combination that is annihilated by the raising operators is given (to within a normalization factor) by the fully antisymmetric wave function
\be \Psi^{(1^3)}_{\rm h.wt.}(1,2,3) = \SD^{123}_{123}    ,\ee
where $\SD^{123}_{123}$ is the so-called \emph{Slater determinant} of a $3\times 3$ matrix
\be
\setlength{\arraycolsep}{5pt}
 \SD^{123}_{123} = 
\left|
\matrix{
  \psi_1(1) \!\!&\!\!\psi_1(2)\!\!&\!\! \psi_1(3) \cr
  \psi_2(1) \!\!&\!\!\psi_2(2)\!\!&\!\! \psi_2(3) \cr
  \psi_3(1) \!\!&\!\!\psi_3(2)\!\!&\!\! \psi_3(3) \cr
}
\right|
. \label{eq:3.111hwt}
\ee
 On the other hand, highest weight states for the $N=3$ irreps 
$\{3\}$ and $\{ 21\}$ have wave functions given, e.g., by 
\bea &\Psi_{\rm h.wt.}^{(3)}(1,2,3) = \SD^1_1 \SD^1_2 \SD^1_3 , &\\
& \Psi_{\rm h.wt.}^{(21)}(1,2,3) = \SD^{12}_{12} \SD^1_3 , & \eea
where $\SD^1_i = \psi_1(i)$ and
\be
\setlength{\arraycolsep}{5pt}
 \SD^{12}_{ij} =
\left|
\matrix{
  \psi_1(i) \!\!&\!\!\psi_1(j)\cr
  \psi_2(i) \!\!&\!\!\psi_2(j) \cr
}
\right| .
\ee
A highest-weight state for an  arbitrary U$(n)$ irrep is similarly expressed in 
terms of Slater determinants 
\be
\setlength{\arraycolsep}{5pt}
 \SD^{12 \dots k}_{i_1i_2 \cdots i_k}  = 
\left| \matrix{
  \psi_1(i_1) \!\!&\!\!\psi_1(i_2)\!\!&\!\! ...\!\!&\!\! \psi_1(i_k)\cr
  \psi_2(i_1) \!\!&\!\!\psi_2(i_2)\!\!&\!\! ...\!\!&\!\! \psi_2(i_k) \cr
  ...          \!\!&\!\!    ...       \!\!&\!\!  ...  \!\!&\!\! .... \cr
  \psi_k(i_1) \!\!&\!\!\psi_k(i_2)\!\!&\!\! ...\!\!&\!\! \psi_k(i_k) \cr
} \right| . \ee
For example, a highest-weight state for the $N=7$ irrep of U(3) with highest weight (421) is given by
\be 
\Psi^{(421)}_{\rm h.wt.}(1,2,3,4,5,6,7) =
\SD^{123}_{123}\,\SD^{12}_{45}    \,\SD^{1}_{6}\,\SD^{1}_{7} .
\label{eq:3.421hwt}  \ee

It can be seen that the wave function  $\Psi^{(421)}_{\rm h.wt.}$ is a sum of tensors products of seven single-particle wave functions corresponding to   
four particles with wave function $\psi_1$, two particles with wave function 
$\psi_2$, and one particle with wave function $\psi_3$.  Thus, it is of weight $(421)$.  Moreover, from its structure as a product of highest-weight Slater determinants, it is of highest weight.

The above construction gives a highest-weight state for any U$(n)$ tensor irrep. 
However, there are many such highest-weight states in the N-particle Hilbert space 
$\Hb_n^N$ for a given highest weight, $\lambda$. These many highest-weight  states correspond to the many possible permutations of the particle indices.
For example, the wave function 
\be
\setlength{\arraycolsep}{5pt}
\Psi^{(421)}_{\rm h.wt.}(1,2,7,6,5,4,3) = 
\SD^{123}_{127}\,\SD^{12}_{65}\,\SD^{1}_{4}\,\SD^{1}_{3} ,
\ee
obtained by permuting the particle indices in Eq.\ (\ref{eq:3.421hwt}) is of U$(n)$ highest weight (421) but is distinct from that given by Eq.\ (\ref{eq:3.421hwt}).
Clearly the set of all such wave functions of U$(n)$ highest weight (421), obtained by permutations of the particle indices, spans a representation of the symmetric group  S$_7$.
More significantly, the Schur-Weyl theorem shows that this S$_7$ representation is irreducible and dual to the U$(n)$ irrep with highest weight (421).  
In general, the Schur-Weyl theorem implies that the highest weight,
$\lambda$, for any tensor irrep of U$(n)$ on $\Hb_n^N$ defines a dual irrep of 
S$_N$ with $N= |\lambda|$ where $|\lambda|=\sum_{i=1}^n \lambda_i$.

The above example shows how an integer highest weight  $\lambda$, i.e., one with non-negative integer components that satisfy the inequality (\ref{eq:3.hwtineq}),  determines both a U$(n)$ tensor irrep and an S$_N$ irrep, where 
$N$ is the sum 
$N= |\lambda | = \sum_{i=1}^n \lambda_i$ of the components of $\lambda$.
Such a highest weight is described as an \emph{ordered partition} of $N$ 
and we write $\lambda\vdash N$. 
If such a partition has $p$ non-zero parts, with $p\leq n$, it is said to have 
\emph{length} $l(\lambda) = p$.

\subsection{Classification of S$_N$ and U$(n)$ irreps by
Young diagrams}
\label{sect:3.Ydiagrams}

The duality relationship between S$_N$ irreps 
and U$(n)$ tensor irreps  is clarified by Young diagram techniques.
As the above $N=7$, $n=3$ example illustrates, both a tensor irrep of U$(n)$ and the corresponding S$_N$ irrep are characterized by an ordered partition 
$\lambda\vdash N$.
It is conventional to label the S$_N$ and U$(n)$ irreps corresponding to an ordered  partition $\lambda$ by  $(\lambda)$ and $\{\lambda\}$, respectively.
The Schur-Weyl theorem can then be expressed as the statement that the
representation $\hat T_n^{N}$ of the direct product group S$_N\times {\rm U}(n)$ carried by the tensor product space $\Hb_n^N$ is a direct sum of
irreps given by 
\be \hat T_n^{N}= \bigoplus_{\lambda\vdash N}^{l(\lambda)\leq n}\, 
(\lambda) \times \{\lambda\} . \label{eq:4.SWeqn}\ee

The irreps of S$_N$ and U$(n)$ corresponding to an ordered partition,
$\lambda\vdash N$, are equivalently characterized by a so-called Young
diagram, $Y^{(\lambda)}$, which is an array of left adjusted boxes with $\lambda_1$ boxes in the
first row, $\lambda_2$ in the second row, \dots, $\lambda_p$ in the $p$'th row.
For example, the partition $\lambda = (421)$ is identified with the Young
diagram
\be
Y^{(421)} :=\youngd{     \yline4\cr
 &&&&&&&&\cr \yline4\cr
 &&&&\cr     \yline2\cr
 &&\cr       \yline1\cr} \; .
\ee

The boxes of a Young diagram can be regarded as containers for indices that label particles for S$_N$  and single-particle wave functions for U$(n)$.  
Such numbered diagrams are called \emph{Young tableaux}.
For example, for the  wave function $\Psi^{(421)}_{\rm h.wt.}(1,2,3,4,5,6,7)$ 
of the highest-weight state given by Eq.\ (\ref{eq:3.421hwt}), 
we can put  the particle-number indices $1, \dots, 7$ into an S$_7$ diagram 
and the single-particle state indices into a U(3) diagram as follows:
\be 
\begin{array}{cccc}
  {\rm S}_7 && {\rm U}(3)  \\
\noalign{\medskip}
\youngd{         \yline4\cr
 &1&&4&&6&&7&\cr \yline4\cr
 &2&&5&\cr       \yline2\cr
 &3&\cr          \yline1\cr}
       &\times &
\youngd{         \yline4\cr
 &1&&1&&1&&1&\cr \yline4\cr
 &2&&2&\cr       \yline2\cr
 &3&\cr          \yline1\cr} 
 ,
\end{array} 
\label{eq:4.hwttableau}
\ee
where the columns of the tableaux correspond to the determinantal factors in the wave function of Eq.\  (\ref{eq:3.421hwt}):
\be
\SD^{123}_{123} \sim
\youngd{ \yline1\cr
 &1&\cr  \yline1\cr
 &2&\cr  \yline1\cr
 &3&\cr  \yline1\cr}
\times
\youngd{ \yline1\cr
 &1&\cr  \yline1\cr
 &2&\cr  \yline1\cr
 &3&\cr  \yline1\cr} \,,
\qquad
\SD^{12}_{45} \sim 
\;\youngd{ \yline1\cr
 &4&\cr  \yline1\cr
 &5&\cr  \yline1\cr}
\times
\youngd{ \yline1\cr
 &1&\cr  \yline1\cr
 &2&\cr  \yline1\cr}\,,
\ee
\vspace{-0.3cm}
\be
\SD^1_6 \sim 
\youngd{ \yline1\cr
 &6&\cr  \yline1\cr}
\times
\youngd{ \yline1\cr
 &1&\cr  \yline1\cr} \, , 
\qquad
\SD^1_7 \sim 
\youngd{ \yline1\cr
 &7&\cr  \yline1\cr}
\times
\youngd{ \yline1\cr
 &1&\cr  \yline1\cr}\, .
\ee

The tableaux in Eq.\ (\ref{eq:4.hwttableau}) are special because 
the first represents a so-called \emph{leading state} of the S$_7$ irrep (421)
and the second  represents a  state of highest weight for the U(3) irrep $\{421\}$,
as constructed above.
 The leading state is defined, somewhat arbitrarily, to be one for which the particle indices, 1,2,\dots,7, are entered sequentially down columns, starting with the first column. 
 The Young tableau for a  U$(n)$ highest weight state is one for which the integer $i$ fills all boxes of row $i$.
Other basis states are obtained by putting the numbers into the boxes in different ways, subject to the condition that all the particle indices must be distinct. 
However, simple rules must be followed to avoid getting an over-complete set. For example, inspection of the wave function  $\Psi^{(421)}_{\rm h.wt.}$ shows that the numbers in corresponding columns of the  S$_7$ and  U(3)  tableaux, respectively,
give the particle and state  indices of the single-particle wave functions of a Slater determinant.  Interchanging their order, in either tableau, can at most change the sign of the wave function.
To obtain a linearly independent set of states, it is therefore appropriate to
impose the rule that the numbers in any column must always increase strongly
from top to bottom (increasing strongly simply means that no number is repeated whereas increasing weakly means not decreasing).
A second, less obvious rule, is that, to obtain linearly independent states,
the numbers in any row of boxes should also increase (weakly in the case of a U$(n)$ tableau and strongly in the case of S$_N$) from left to right.  Note also that, while state indices may be repeated, the particle indices must all be distinct.

It is easy to check that the above rules work out in given situations.
For example, the 4-dimensional space, $\Hb_{n=2}^{N=2}$,  is spanned by  three states of the $\lambda = (2)$ irrep of S$_2\, \times\,$U(2) and one state of the $\lambda = (11)$ irrep:
\be
\begin{array}{rclrc}
{\rm S}_2\;\;  &\times& \;  {\rm U}(2)\\
\noalign{\medskip}
 \youngd{ \yline2\cr &1&&2&\cr \yline2\cr}
&\times&
\youngd{ \yline2\cr &1&&1&\cr \yline2\cr} 
&\;\; \sim& \psi_1(1) \psi_1(2) ,
\\ 
\noalign{\medskip}
 \youngd{ \yline2\cr &1&&2&\cr \yline2\cr}
&\times&
\youngd{ \yline2\cr &1&&2&\cr \yline2\cr} 
&\;\;  \sim& \;\psi_1(1) \psi_2(2) +\psi_2(1) \psi_1(2),
\\ 
\noalign{\medskip}
\youngd{ \yline2\cr &1&&2&\cr \yline2\cr}
&\times&
\youngd{ \yline2\cr &2&&2&\cr \yline2\cr}  
&\;\;  \sim& \psi_2(1) \psi_2(2) ,
\\ 
\noalign{\medskip}
\youngd{ \yline1\cr &1&\cr \yline1\cr &2&\cr \yline1\cr}
&\times&
\youngd{ \yline1\cr &1&\cr \yline1\cr &2&\cr \yline1\cr} 
&\;\;  \sim&\quad \left|  \matrix{    \psi_1(1)&\psi_1(2) \cr
                                               \psi_2(1) &\psi_2(2) \cr  } \right|  .
\end{array} \label{eq:4.nd=2}
\ee
Note that to determine the state with Young tableau 
$\youngd{ \yline2\cr &1&&2&\cr \yline2\cr}$ for U(2), one applies the lowering operator $\hat C_{21}$, 
 defined by Eq,\ (\ref{eq:3B30}), to the state with tableau 
$\youngd{ \yline2\cr &1&&1&\cr \yline2\cr}$, i.e., 
\be \hat C_{21} \psi_1(1)\psi_1(2) = \psi_1(1) \psi_2(2) +\psi_2(1) \psi_1(2). \ee

Similarly, the 8-dimensional space  $\Hb^{N=3}_{n=2}$ is spanned by four states of the 
$\lambda = (3)$ irrep of S$_3\, \times\,$U(2) and four states of the $\lambda = (21)$ irrep:
\be
\begin{array}{rclcrcl}
 {\rm S}_3 \;\;  &\times& \;\;  {\rm U}(2) & \qquad&
 {\rm S}_3 \;\;  &\times& \;\;  {\rm U}(2) 
 \\
\noalign{\medskip}
\youngd{ \yline3\cr &1&&2&&3&\cr \yline3\cr}
&\times&\youngd{ \yline3\cr &1&&1&&1&\cr \yline3\cr}\,,
&&
\youngd{ \yline3\cr &1&&2&&3&\cr \yline3\cr}
&\times&\youngd{ \yline3\cr &1&&1&&2&\cr \yline3\cr}\, ,
\\
\noalign{\medskip}
\youngd{ \yline3\cr &1&&2&&3&\cr \yline3\cr}
&\times&\youngd{ \yline3\cr &1&&2&&2&\cr \yline3\cr} \,,
&&
\youngd{ \yline3\cr &1&&2&&3&\cr \yline3\cr}
&\times&\youngd{ \yline3\cr &2&&2&&2&\cr \yline3\cr}\,,
\\
\noalign{\medskip}
\youngd{ \yline2\cr &1&&2&\cr \yline2\cr &3&\cr \yline1\cr}
&\times& \youngd{ \yline2\cr &1&&1&\cr \yline2\cr &2&\cr \yline1\cr} \,,
&&
\youngd{ \yline2\cr &1&&3&\cr \yline2\cr &2&\cr \yline1\cr}
&\times& \youngd{ \yline2\cr &1&&1&\cr \yline2\cr &2&\cr \yline1\cr}\, , 
\\
\noalign{\medskip}
\youngd{ \yline2\cr &1&&2&\cr \yline2\cr &3&\cr \yline1\cr}
&\times& \youngd{ \yline2\cr &1&&2&\cr \yline2\cr &2&\cr \yline1\cr} \,,
&&
\youngd{ \yline2\cr &1&&3&\cr \yline2\cr &2&\cr \yline1\cr}
&\times& \youngd{ \yline2\cr &1&&2&\cr \yline2\cr &2&\cr \yline1\cr}\, . 
\end{array}
\ee

\subsection{The relationship between S$_N$ and U$(n)$ characters}
\label{sect:2.D}

A first application of Schur-Weyl duality is to obtain an algorithm for deriving characters of the unitary groups from those of the symmetric groups.
Such characters are frequently needed because
U$(n)$ plays a central role in the decomposition of tensor products of various group representations and in the reduction of an irrep of a group on restriction to a subgroup. These are frequently occurring problems in the application of symmetry to many-particle physics.

The character $\chi$ of a representation $\hat T$ of a group $G$ is a
complex-valued function $\chi :G\to \mathbb{C}$ whose value for a group element $g\in G$ is the trace
\be \chi(g) = {\rm Tr} (\hat T(g)) \,. \ee
Characters take the same values for all elements of a group  that belong
to a common conjugacy class, because, 
if two elements $g_1$ and $g_2$ of a group $G$ are conjugate, they
 are related by $g_1= gg_2g^{-1}$, for some $g\in G$. The identity
$ {\rm Tr}(\hat T(g_1))={\rm Tr}(\hat T(gg_2g^{-1}))
={\rm Tr}({ \hat T}(g) \hat{T} (g_2) \hat{T}(g)^{-1}))
= {\rm Tr}(\hat T(g_2))$
then implies that 
\be \chi(g_1)=\chi(g_2) \,.\ee

For the symmetric group S$_N$, all elements that belong to the same conjugacy class have the same {\em cycle structure\/}. This is readily seen by expressing a general permutation as a product of cyclic permutations.  An arbitrary element $g\in S_N$ can be expressed
$g=\left(\begin{array} {cccc}1&2&3&\dots\\
                           g_1& g_2& g_3 &\dots
                           \end{array}\right)$
which denotes the permutation of a set of $N$ objects in which the object in slot $i$ is moved to to slot $g_i$.
Consider, for example, the particular element
$g^\prime=\left(\begin{array} {ccccc}1&2&3&4&5\\
                           3& 4& 5 &2 &1
                           \end{array}\right)$
of S$_5$.  It corresponds to a sequence of two cyclic permutations
$g^\prime=(1\to 3\to 5\to 1) ( 2\to 4\to 2)$ 
which can be expressed efficiently with the notation
\be g^\prime \equiv (1,3,5)(2,4) .\label{eq:cycles} \ee
Such a product of a three-cycle and a two-cycle is said to have a (3,2) cycle structure.  It is then readily seen that the conjugate of $g^\prime$ by $g$ (for any $g\in \mathrm{S}_5$)is the element
\be gg^\prime g^{-1} = (g_1,g_3,g_5)(g_2,g_4)\ee
which has the same cycle structure as $g'$.   
The converse is also true:  two elements of S$_N$ with the same 
cycle structure are conjugate.
Thus, we identify a class $\rho$ of S$_N$ with a so-called {\em cycle structure\/}
$(\rho_1,\rho_2,
\dots, )$, where $\rho_i$ is the length of a cycle. For example, the class containing all
permutations $\{ g (1,2,3)(4,5) g^{-1}; g\in S_5\}$  has a cycle structure (3,2).
It may also be noted that, if the lengths of the cycles defining a class are ordered such that
\be \rho_1 \geq\rho_2 \geq\rho_3 \geq \dots
\ee
then the class label $\rho \equiv (\rho_1,\rho_2, \dots, )$ is 
an ordered partition of $N$.
It follows from this result that the number of classes of the group
S$_N$ is equal to the number of ordered partitions of $N$ which, in
turn, is equal to the number of inequivalent irreps of S$_N$. 
However,  a partition defines a class whether or not it is
ordered. Thus,  it is often useful to denote a class by the set
$\rho \equiv \{ i^{r_i}\}$, where $r_i$ denotes the
number of cycles of length $i$.

The values of
an S$_N$ character, $\{\chi_\rho\}$, for the classes
of S$_N$  are  conventionally displayed in tables  in which each
row corresponds to an irrep of S$_N$. Thus, for example, because
S$_3$ has three irreps, labeled by $\lambda = (3)$, (21), and
$(1^3)$, and three classes, $\rho =(1^3)$, $(2, 1)$, and $(3)$, its
character table is a $3\times3$ array $\{\chi^\lambda_\rho\}$ 
as shown in Table \ref{tab:III.1}.
\begin{table}[ht]  
\caption{\label{tab:III.1} The character table for S$_3$  ($n_\rho$ 
is the number of group elements in the class $\rho$).} \vspace{-0.3cm}
$$ \begin{array}{ccccccccc}
{\rm Class}&\qquad\qquad &(1^3) &\qquad\qquad&(2,1)&\qquad\qquad&(3)\\
n_\rho &&1&&3&&2
\vspace{0.2cm}\\
\chi^{(3)}&&1&&1&&1\\ 
\chi^{(21)}&&2&&0&&-1\\
\chi^{(1^3)}&&1&&-1&&1\\
\end{array}$$
\end{table}
The irreducible S$_N$ characters
satisfy the orthogonality relations
\be
\label{Eq:characterorthog} \sum_{\rho\vdash N} n_\rho
\chi^\lambda_\rho \chi^{\lambda'}_\rho = N! \delta_{\lambda,\lambda'},
  \label{eq:S_Ncharorthog}\ee
where $n_\rho$ is the number of elements of S$_N$ in the class $\rho$.

The number $n_\rho$ is determined as follows.
The number of elements of the group  S$_N$ is equal to $N!$, 
which is the number of ways of ordering the integers $1,2, \dots, N$.
Thus, the number of  elements of S$_N$ in a class is  the total number of
inequivalent ways of distributing the $N$ particle indices over the
cycle structure of the class. There are a total of $N!$ distributions. However,
different distributions that correspond simply to a permutation of
cycles of the same length correspond to the same element of S$_N$.
For example, the distributions $(5,6,7)(1,2)(3,4)$ and
$(5,6,7)(3,4)(1,2)$ over the cycle  structure $(3,2,2)$
denote identical  permutations. Thus, it is necessary to divide $N!$ by
the number of orderings of the cycles of the same length.  If
$r_i$ denotes the number of  cycles of length $i$ in the cycle
structure $\rho$, this number is $r_1! r_2! \dots r_N!$. Also, all
cyclic permutations of the numbers within a cycle correspond to the
same permutation, e.g.,\ $(1,3,5)$, $(3,5,1)$, and $(5,1,3)$ are
identical permutations as can be seen from the definition of Eq.\
(\ref{eq:cycles}).  The number of identical permutations obtained in
this way is $1^{r_1} 2^{r_2} \dots N^{r_N}$. Thus, the number of
permutations $n_\rho$ in the class $\rho \equiv \{ i^{r_i}\}$ of
S$_N$ is given by 
\be n_\rho=\frac{N!}{r_1! r_2! \dots r_N!\,
1^{r_1} 2^{r_2} \dots N^{r_N}} .  \label{eq:nrho} \ee

We now consider the characters of the group U$(n)$. First recall
that every U$(n)$ matrix can be brought to diagonal form by a
unitary  transformation. 
Such a unitary transformation does not change the trace of a matrix.
Consequently, every U$(n)$ matrix is
conjugate to a diagonal matrix of the form given by 
\be 
z = \pmatrix{z_1 &&& \cr & z_2&& \cr &&
\ddots &\cr &&& z_n\cr  }, \quad |z_i|=1. \label{eq:23} \ee
and each U$(n)$ class contains a representative
diagonal matrix with diagonal entries that we denote by a set of
complex numbers $z = \{z_i\}$.
For such a U$(n)$ matrix, it is seen from Eq.\ (\ref{eq:5}) that
\be  \hat U^{(1)}(z)\psi_{\nu} = z_\nu \psi_\nu \,,\ee 
and, from Eq.\ (\ref{eq:7}) for an $N$-particle tensor-product wave function, that
\be
 \hat U^{(N)}(z)  \Psi_{\nu_1\nu_2\dots\nu_N} =
\left(\prod_{j=1}^N z_{\nu_j}\right) \Psi_{\nu_1\nu_2\dots\nu_N} .
\ee
Thus, for an $N$-particle state $\Psi^{(\lambda)}$ of weight $\lambda$,
 of a U$(n)$ tensor representation, it is determined that
\be
 \hat U^{(N)} (z)  \Psi^{(\lambda)} =
(z_1)^{\lambda_1} (z_2)^{\lambda_2}
\dots (z_n)^{\lambda_n}   \Psi^{(\lambda)} .
\ee
The character of a U$(n)$ representation is therefore a function of the $n$ complex variables $\{ z_i\}$.
For example, the fundamental U$(n)$
irrep $\{1\}$  has character  
\be s_1  = \sum_{i=1}^n z_i \, .\ee

The character $s_{\lambda}$ of a general tensor irrep of  U$(n)$, with highest weight regarded as a partition $\lambda\vdash N$, is now obtained from the character $\chi^{\lambda}$ of the dual S$_N$ irrep as follows. First determine the character of the reducible S$_N \times {\rm U}(n)$ representation,  $\hat T$, on 
$\mathbb{H}_n^N$.
This is easy in the basis defined by Eq.~(\ref{eq:7}) because the action of a group element
$(\pi,z)\in S_N \times {\rm U}(n)$, with $z$ diagonal, is simply to permute the basis functions and multiply them by
$z$-dependent factors.  For example, the identity
\be \hat
T(\pi_{12},z) \Psi_{\nu_1\nu_2\nu_3} = z_{\nu_1}z_{\nu_2}z_{\nu_3}\,
 \Psi_{\nu_2\nu_1\nu_3} \,,
\ee
implies that (after  normalizing, as necessary,
so that  $\{\Psi_{\nu_1\nu_2\nu_3}\}$ is an
orthonormal basis)
\be \langle \Psi_{\nu_1\nu_2\nu_3},
\hat T(\pi_{12},z) \Psi_{\nu_1\nu_2\nu_3}\rangle =
 \delta_{\nu_1,\nu_2}
z^2_{\nu_1}z_{\nu_3} \ee
and hence 
\be  \sum_{\nu_1,\nu_2,\nu_3} \langle
\Psi_{\nu_1\nu_2\nu_3}, \hat T(\pi_{12},z)
\Psi_{\nu_1\nu_2\nu_3}\rangle = p_2(z) p_1(z) \,, \ee
 where $p_k(z)$, for any positive integer $k$, is
 the so-called \emph{power sum}
\be p_k(z) =\sum_{i=1}^n z_i^k \,.   \label{eq:pfn} \ee
It will be noted that the permutation $\pi_{12}\in \mathrm{S}_3$ belongs to the class with
cycle structure $(2,1)$. In general, one finds that the character $\Xi^N$ of the reducible
representation  $\hat T$ of S$_N\times {\rm U}(n)$ on $\mathbb{H}^N$ has values
\be  \Xi^N(\rho,z) = p_\rho(z) , \label{eq:XiH^N}\ee
where, for $\rho = \{ i^{r_i}\}$,
\be  p_\rho(z) =
\big(p_1(z)\big)^{r_1}\, \big(p_2(z)\big)^{r_2}\,
\big(p_3(z)\big)^{r_3}\, \dots
\label{eq:prho}\ee

Now, if $s_\lambda$ denotes the character of the U$(n)$ irrep of
highest weight $\lambda\vdash N$, it follows from the expression
 (\ref{eq:4.SWeqn}) of the Schur-Weyl theorem that
 \be
\Xi^N(\rho,z) = \sum_{\lambda\vdash N} \chi^\lambda_\rho
s_\lambda(z) \,. \label{eq:45}\ee
Thus, we obtain the simple identity 
\be
\sum_{\lambda\vdash N} \chi^\lambda_\rho s_\lambda(z) =
 p_\rho(z) . \label{eq:Frob}\ee
Use of the orthogonality relation (\ref{Eq:characterorthog}) together with
(\ref{eq:Frob}) leads
 to the well-known expression for U$(n)$ characters 
\be s_\lambda{(z)} = \frac{1}{N!} \sum_{\rho\vdash N} n_\rho
\chi^\lambda_\rho  p_\rho(z) \,, \label{eq:47}\ee
first derived by  \textcite{Schur}.

Characters of the U$(n)$ polynomial irreps, denoted by $s_{\lambda}$,
are known as {\em Schur functions\/} or {\em S functions}.
They were  studied as sets of orthogonal symmetric polynomials long before
they were determined
by Schur to be characters of the unitary groups \cite{Macdonald95}. Some of the above relationships also
appear to have been known long ago in different contexts.
For example, as pointed out by \textcite{Led} and \textcite{FH}, 
Eq.\ (\ref{eq:Frob}) can also be derived
from a formula given by \textcite{Fro} for  S$_N$ characters. What is remarkable is how simply and easily Schur's formula,   Eq.\ (\ref{eq:47}), follows from a consideration of the dual actions of the groups S$_N$ and U$(n)$ on the 
space $\Hb_n^N$.

It is worth noting that Eq.\ (\ref{eq:47}) has the remarkable and
valuable property of not only relating characters of S$_N$
and U$(n)$ irreps for all positive integer values of $N$ and $n$ but of
doing so in an $N$- and $n$-independent way.  Thus, many results
arising from character theory and Schur-Weyl duality are $N$- and
$n$-independent.

The particular significance of  Schur-Weyl duality for the nuclear and atomic shell-model is that it enables the Pauli constraints on systems of  identical particles to be taken into account by simply restricting to  appropriate combinations of unitary group representations, as the following section will show.
Further information about the many varied uses of Schur
functions in physics  is given in  the works of 
\textcite{Wyb1} and \textcite{King75}.


\section{Applications of Schur-Weyl duality} \label{sect:3.SWdualityb}

In addition to the  character formula Eq.\ (\ref{eq:47}),
the one-to-one correspondence between irreps of  the unitary and symmetric groups also implies a  linkage between  other important operations on irreps of  these groups, e.g.,\ branching rules and tensor products, which we now consider.   Such linkages, which are thematic of duality, enable results derived for the irreps of one group to be applied to the irreps of  the other. Later sections will include examples from other dual-pair situations.

Recall that a representation $\hat T$ of a group $G$ automatically defines a
representation of any subgroup $H\subseteq G$ known as the restriction of the representation $\hat T$ to the subgroup.
However, even if a representation $\hat T$ is irreducible as a
representation of $G$, its restriction to a subgroup $H\subset G$
is  generally reducible. Branching rules give expansions of the restriction of irreps of a group as direct sums of subgroup irreps.
For example, on restriction to an SO(2) subgroup, an irrep $[L]$ of the rotation group SO(3), labeled by a positive integer-valued angular momentum quantum number $L$, branches
to a direct sum of $(2L+1)$ SO(2) irreps labeled by integer-valued components of the angular momentum $M = -L, -L+1, \dots, +L$ about the axis of SO(2) rotations. This well-known result is expressed formally by the {\em branching rule}
\be {\rm SO}(3) \downarrow {\rm SO}(2)\;; \; [L]\downarrow
 \bigoplus_{M=-L}^{L} [M]\, .\ee

\subsection{Branching rules for symmetric and unitary groups}
\label{sect:4A}

This section shows that several  branching rules for symmetric and unitary groups are expressed directly in terms of the coefficients appearing in decompositions of S$_N$ and U$(n)$ tensor products.

Let  $(\kappa )$, $(\tau )$ and $(\lambda )$ denote irreps of S$_N$, each defined by an ordered partition of $N$. The reduction of the
tensor products of pairs of  S$_N$ irreps then defines a set of  
$\{\gamma^\lambda_{\kappa\tau}\}$ coefficients, with integer values, by the expansion
\be
(\kappa)\otimes (\tau) = \bigoplus_\lambda
\gamma^\lambda_{\kappa\tau} (\lambda) .\label{eq:56} \ee
Similarly,
if  $\{\kappa\}$, $\{\tau\}$ and $\{\lambda\}$ denote U$(n)$
tensor irreps, defined by ordered partitions of integers, the
tensor products of such U$(n)$ irreps determine integer-valued
$\{ \Gamma^\lambda_{\kappa\tau}\}$ coefficients
(known as Littlewood-Richardson coefficients)
 by the expansion
\be
\{\kappa\}\otimes\{\tau\} =\bigoplus_\lambda
\Gamma^\lambda_{\kappa\tau} \{\lambda\}  .
\label{eq:UNtensorprodrep}\ee
 In the latter case, if $\kappa\vdash
N_1$,  $\tau\vdash N_2$, and $\lambda\vdash N_3$, then
$\Gamma^\lambda_{\kappa\tau}$ is zero unless $N_3=N_1+N_2$, i.e., unless
$|\lambda|=|\kappa| + |\tau|$. The
above equations can be expressed equivalently in terms of
characters, i.e.,
\bea &\displaystyle\chi^\kappa_\rho \chi^\tau_\rho
= \sum_\lambda \gamma^\lambda_{\kappa\tau}
\chi^\lambda_\rho ,& \label{eq:SNtensorproduct}\\
&\displaystyle s_\kappa (z) s_\tau (z) =\sum_\lambda
\Gamma^\lambda_{\kappa\tau} s_\lambda(z).&
\label{eq:UNtensorproduct} \eea
Because the characters of S$_N$ and U$(n)$ are known, the coefficients
$\gamma^\lambda_{\kappa\tau}$ and $\Gamma^\lambda_{\kappa\tau}$  can be computed and various methods exist for their computation. For example,
the $\gamma^\lambda_{\kappa\tau}$ coefficients are readily computed by use of
the orthogonality property of S$_N$ characters and
the $\Gamma^\lambda_{\kappa\tau}$ coefficients can be determined by
manipulation of Young diagrams using the so-called  Littlewood-Richardson
rule \cite{Ham}. 

We now consider branching rules.

\medskip 
{\bf Theorem 2:}\emph{ 
The coefficients appearing in the expansions of the branching rules
\bea  
\displaystyle \mathrm{S}_{N_1+N_2} &\!\downarrow\!&
\mathrm{S}_{N_1}\times {\rm S}_{N_2}  ; \nonumber\\
(\lambda)  &\!\downarrow\!&
\bigoplus_{\kappa \vdash N_1, \tau \vdash N_2}\, 
\Gamma^\lambda_{\kappa\tau}
(\kappa) \times (\tau) , \label{eq:54a}\\
\displaystyle \mathrm{U}(mn)  &\!\downarrow\!&
 \mathrm{U}(m)\times\mathrm{U}(n)  ;  \nonumber\\
\{ \lambda\} &\!\downarrow\!&  \bigoplus_{\kappa \vdash | \lambda| , \tau
\vdash |\lambda|}\,
\gamma^\lambda_{\kappa\tau} \{\kappa\} \times \{\tau\} , 
\label{eq:55a} \\
\displaystyle \mathrm{U}(m\!+\!n)  &\!\downarrow\!& \mathrm{U}(m)\times
\mathrm{U}(n);\nonumber\\
 \{ \lambda\}  &\!\downarrow\!& \bigoplus_{N_1+N_2=|\lambda|}\;
\bigoplus_{\mu \vdash N_1,\nu\vdash N_2}
\!\!\! \Gamma^\lambda_{\mu\nu}\,
\{ \mu \} \times \{\nu\}  , \quad\label{eq:Um+n}
\eea
 for the symmetric and unitary groups are  identical to the coefficients in the ${\rm S}_N$ and ${\rm U}(n)$ tensor product reductions of Eqs.\ (\ref{eq:56}) and 
 (\ref{eq:UNtensorprodrep}). }
\medskip

{\em Proof:}
First we prove the branching rule (\ref{eq:54a}) starting with the observation that if $N=N_1+N_2$ then
\be \Hb_n^{N} = \Hb_n^{N_1}\otimes\Hb_n^{N_2}.
\label{eq:N1N2character}\ee
The space $\Hb_n^{N_1}$ carries a representation 
\be \hat T^{N_1}_n = \bigoplus_{\kappa\vdash N_1} (\kappa) \times \{\kappa\}\ee
of the group ${\rm S}_{N_1}\times {\rm U}(n)$ and the space $\Hb_n^{N_2}$  carries a representation 
\be 
\label{eq:RightFactorDecomp}
\hat T^{N_2}_n = \bigoplus_{\tau\vdash N_2} (\tau) \times \{\tau\}\ee
of the group ${\rm S}_{N_2}\times {\rm U}(n)$.
Thus, $\Hb_n^{N}$ carries a representation
\bea \label{eq:ProductGroupDecomp}
\hat T^{N_1, N_2}_n &=& \bigoplus_{\kappa\tau} (\kappa) \times (\tau)
\times \big( \{\kappa\} \otimes \{\tau\}\big)  \nonumber\\
&=&  \bigoplus_{\kappa\tau\lambda} \Gamma^\lambda_{\kappa\tau} (\kappa) \times (\tau)\times \{\lambda\}  ,
   \eea
with ${\kappa\vdash N_1}$ and ${\tau\vdash N_2}$,
of the product group ${\rm S}_{N_1}\times {\rm S}_{N_2}\times{\rm U}(n)$.
We also know that the Hilbert space $\Hb_n^{N}$ carries a representation
\be \hat T^N_n = \bigoplus_\lambda (\lambda) \times \{ \lambda\} 
\label{eq:85}\ee
of the group ${\rm S}_{N}\times {\rm U}(n)$.  
Combining 
(\ref{eq:ProductGroupDecomp}) and (\ref{eq:85}),
 we obtain the branching rule
\bea{\rm S}_N\times {\rm U}(n) 
&\downarrow& {\rm S}_{N_1}\times  {\rm S}_{N_2}\times {\rm U}(n); \nonumber \\
\hat T^N_{n} &\downarrow& \hat T^{N_1,N_2}_{n} ,\eea
and, hence, the branching rule for 
${\rm S}_{N} \downarrow {\rm S}_{N_1}\times {\rm S}_{N_2}$ given by
Eq.\ (\ref{eq:54a}).
This branching rule is equivalently expressed in terms of characters by the identity
\be \chi^\lambda_{\rho_1\rho_2} = \sum_{\kappa, \tau}
\Gamma^\lambda_{\kappa\tau}
 \chi^\kappa_{\rho_1} \chi^\tau_{\rho_2} , \label{eq:SN1+N2BR}
\ee 
where $\rho_1$ and $\rho_2$ are,
respectively, classes of S$_{N_1}$ and  S$_{N_2}$,
and $\rho_1 \rho_2$ is regarded as a class of $S_N = S_{N_1+N_2}$.

To prove the branching rule of  Eq.\ (\ref{eq:55a}),  consider the product Hilbert space
\be \mathbb{H}_{mn}^N =
(\Hb_{m}\otimes \Hb_{n})^N =
\mathbb{H}_{m}^N\otimes \mathbb{H}_{n}^N \,, \label{eq:52z}\ee where
$\Hb_{m}$ is of dimension $m$ and $\Hb_{n}$ is of dimension $n$. 

According to the Schur-Weyl theorem, the Hilbert spaces 
$\Hb_m^N$ and $\Hb_n^N$ carry representations of 
${\rm S}_{N}\times {\rm U}(m)$  and
${\rm S}_{N}\times {\rm U}(n)$ given, respectively, by
\be
\hat T^N_m= \bigoplus_{\kappa\vdash N}^{l(\kappa)\le m} 
(\kappa)\times \{\kappa\} , \quad
\hat T^N_n= \bigoplus_{\tau\vdash N}^{l(\tau)\le n} 
(\tau)\times \{\tau\} .
\ee
It follows that $\Hb_{mn}^N$ carries a representation
\bea \hat T^N_{m,n} &=&  \bigoplus_{\kappa\vdash N}^{l(\kappa)\le m} 
\bigoplus_{\tau  \vdash N}^{l(\tau)\le n}
\big( (\kappa)\otimes (\tau)\big) \times \{ \kappa\} \times \{ \tau\} 
\nonumber\\
&=& \bigoplus_\lambda
\bigoplus_{\kappa\vdash N}^{l(\kappa)\le m} 
\bigoplus_{\tau\vdash N}^{l(\tau)\le n} \gamma^\lambda_{\kappa \tau}\,
(\lambda)\times \{ \kappa\} \times \{ \tau\} 
\eea
 of the  direct product group ${\rm S}_N \times {\rm U}(m) \times {\rm U}(n)$, where $\gamma^\lambda_{\kappa \tau}$ is the coefficient 
of $\lambda$ in the reduction of the tensor product $(\kappa)\otimes (\tau)$ given  by  Eq.\ (\ref{eq:56}).
We also know from the Schur-Weyl theorem that the Hilbert space $\Hb_{mn}^N$ carries a representation of ${\rm S}_N \times {\rm U}(nm)$ given by
\be \hat T^N_{mn} = \bigoplus_{\lambda\vdash N}^{l(\lambda)\le mn}\, 
(\lambda)\times \{\lambda\}  .\ee
From these two  expressions we obtain the branching rule
\bea{\rm S}_N\times {\rm U}(mn) 
&\downarrow& {\rm S}_N\times {\rm U}(m)\times {\rm U}(n); \nonumber \\
\hat T^N_{mn} &\downarrow& \hat T^N_{m,n} ,\eea
and, hence, the branching rule for 
${\rm U}(mn) \downarrow {\rm U}(m)\times {\rm U}(n)$ given by Eq.\ (\ref{eq:55a}).

Finally, to prove the third branching rule, Eq.\ (\ref{eq:Um+n}), first observe that,
when restricted to the subgroup
${\rm U}(m)\times {\rm U}(n)$, the U$(m+n)$ character for the irrep $\{ \lambda\}$ is given by
$s_\lambda (x, y)$, where $(x, y)$ denotes the elements
$(x_1,\dots, x_m, y_1, \dots, y_n)$ of a diagonal U$(m+n)$
matrix.
From Eq.\ (\ref{eq:47}), we have the identity
\be s_\lambda(x,y) = \frac{1}{N!} \sum_\rho n_\rho
\chi^\lambda_\rho p_\rho (x,y) . \label{eq:75}\ee
Now, by making use of Eq.\ (\ref{eq:nrho}) for $n_\rho$ and Eq.\
(\ref{eq:prho}) for $p_\rho$, we have
\bea \displaystyle s_\lambda(z) &=&\frac{1}{N!} \sum_\rho \frac{N!}{r_1!\,r_2!\, r_3!
\cdots 1^{r_1} 2^{r_2} 3^{r_3} \dots } \nonumber\\
&&\times
\chi^\lambda_\rho\, \big(p_1(z)\big)^{r_1}\, \big(p_2(z)\big)^{r_2} \,
\big(p_3(z)\big)^{r_3} \dots , \label{eq:76}\eea
where $r_i$ denotes the number of cycles  of length $i$ in the cycle structure
$ (1^{r_1}, 2^{r_2}, 3^{r_3}. \dots ) $ of the class $\rho$.
Thus, with $z=(x,y)$ and the definition $ p_k(z) =\sum_{i=1}^n z_i^k$
 (cf.\ eqn.\  (\ref{eq:pfn})), it follows that
\be  p_k(z) = p_k(x,y) = p_k(x) + p_k(y) \ee
and that
\be
\big(p_k( z)\big)^{r_k} =\sum_{s_k+t_k=r_k}
\frac{r_k!}{s_k!\, t_k!}
 \big(p_k(x)\big)^{s_k}  \big(p_k(y)\big)^{t_k}.
\ee
Hence, 
\be p_\rho(x,y) = \sum_{\sigma,\tau} \frac{r_1!\, r_2!\,
r_3! \dots} {s_1!\, s_2!\, s_3! \dots \, t_1!\, t_2!\, t_3! \dots}\,
p_\sigma (x) p_\tau(y)\label{eq:79}, \ee 
where the sum is over all
classes $\sigma$ with cycle structure $(1^{s_1}, 2^{s_2}, 3^{s_3},
\dots )$ of the symmetric group S$_{N_1}$ and all classes $\tau$
with cycle structure $(1^{t_1}, 2^{t_2}, 3^{t_3}, \dots )$ of the
symmetric group S$_{N_2}$ for any $N_1, N_2$ satisfying  $N_1+N_2
=N$ and $s_k+t_k=r_k$ for all $k$. Inserting (\ref{eq:79}) into
(\ref{eq:75}),
with the identity 
\be n_\sigma=\frac{N_1!}{s_1! s_2!
\dots s_N!\, 1^{s_1} 2^{s_2} \dots N^{s_N}} , \label{eq:nsigma} \ee
 from Eq.\ (\ref{eq:nrho}),
yields 
\be s_\lambda(x,y) = \sum_{\sigma, \tau} \frac{n_\sigma
n_\tau}{ N_1!N_2!}\, \chi^\lambda_{\sigma\tau}\, p_\sigma (x)
p_\tau(y) \ee 
and, with the expansion (\ref{eq:SN1+N2BR}), we obtain
\bea
s_\lambda(x,y)\! &=&\!\!\! \!\sum_{N_1 + N_2 = N} 
\sum_{{\sigma, \tau} \atop{\mu \vdash N_1 , \nu \vdash N_2}} \!\!\!\!
\frac{n_\sigma n_\tau}{N_1!N_2!}\, \Gamma^\lambda_{\mu\nu}
 \chi^\mu_{\sigma} \chi^\nu_{\tau}\, p_\sigma (x) p_\tau(y)
 \nonumber\\
&=& \!\!\!\!\sum_{N_1+N_2=N} \sum_{\mu \vdash N_1 , \nu \vdash N_2}
\!\!\Gamma^\lambda_{\mu\nu} s_{\mu}(x) s_\nu(y) , \eea 
where the second
equality follows from Eq.\ (\ref{eq:47}). This is the expression of
the branching rule (\ref{eq:Um+n}) in terms of characters and
completes the proof of Theorem 2. \hfill $\Box$

The branching rules given by Theorem 2 are remarkable because they show that coefficients defined for the tensor products of one group determine the branching rules of a different group.  Even Eq.\ (\ref{eq:55a}), which at a superficial glance might appear to be simply the inverse of Eq.\ (\ref{eq:UNtensorprodrep}), is seen to involve outer products of distinct unitary groups in contrast to Eq.\ (\ref{eq:UNtensorprodrep})
which is concerned with tensor products of a single unitary group.

It is useful to note that the above relationships between  tensor products and branching rules for the symmetric and unitary groups hold for all values of $N$ and $n$. However, one must be mindful of the fact that results of calculations involving symmetric group characters may lead to labels for U(n) irreps that do not
exist for particular $n$ values. Consider, for example, the
following application of the Littlewood-Richardson rule: $\{ 1^2 \}
\otimes \{ 1 \} = \{ 2 1 \} \oplus \{ 1^3 \}$. This is correct for
all U$(n)$ for which $n \geq 3$. But $\{1^3\}$ does not exist as an
irrep of U(2). The irrep label $\{1^3\}$ is therefore discarded  and
the correct relation for U(2) is $\{1^2\}\otimes\{1\} = \{21\}$.

\subsection{Symmetrized tensor products -- plethysms} \label{sect:pleth}

The concept of a plethysm as a symmetrized tensor power was conceived by 
\textcite{Littlewood36}  in a natural generalization of the Schur-Weyl theorem.
Littlewood denoted the plethysm operation symbolically by 
$\{\kappa\}\otimes\{\lambda \}$.
However, in the following, we use the symbol $\pleth$  instead of $\otimes$, which we reserve for a tensor product.
Plethysm was introduced into the mainstream   of physics by 
\textcite{SmithW67,SmithW68} and \textcite{Wyb1}
and subsequently employed by many for branching rule calculations  and other applications, notably in atomic and nuclear spectroscopy.

Suppose, for example, that  $\Hb$ is a space of single-particle wave functions that carries an irrep of a group such as U(3) and one wants to know what  representation of  ${\rm S}_N\times{\rm U}(3)$ is carried by the corresponding tensor product space 
$\Hb^N=\Hb^{\otimes N}$ of $N$-particle wave functions.  When $\Hb$ is three-dimensional and the irrep of U(3) carried by $\Hb$ is the defining representation, the answer is already given by the Schur-Weyl theorem.  
In general it is given by an expansion of irreps determined by the plethysm operation.
However, as in the decomposition of a tensor product of SU(2) representations as sums of irreps obtained by the use of Clebsch-Gordan coefficients, it is not generally important to know how to derive the coefficients in the expansion of a plethysm because computer programs are available for that purpose.  It is more important to understand what the coefficients are and how to use them.  Thus, the value of plethysms lies in the availability of computer programs to evaluate them;
cf. \cite{CD1,CarvalhoD01} and references therein.
  Thus, as soon as the answer to a problem is expressed as a plethysm, it is effectively solved. 

If  $\Hb_m$ is an $m$-dimensional Hilbert space it carries the defining irrep
$\{ 1\}$ of the group U$(m)$,  
 and the tensor product  $\Hb_m^N$ of $N$  copies of 
$\Hb_m$ carries a reducible representation of ${\rm S}_N\times{\rm U}(m) $, denoted here by  $\hat T^N_1$.  According to the  Schur-Weyl theorem, this representation is a direct sum of irreps given by
\be \hat T^N_{1} = \bigoplus_{\lambda \vdash N}  (\lambda ) \times \{ \lambda\}
, \label{eq:TN1} \ee
 where the sum over partitions of $ N$ is restricted to $\lambda$ with no more than $m$ parts. It is then meaningful to regard the U$(m)$ irrep $\{ \lambda\}$ as the cofactor of the  S$_N$ irrep  $(\lambda)$ in the expansion (\ref{eq:TN1}).
Thus, we say that $\{\lambda\}$ is
a symmetrized tensor power of the irrep $\{1\}$,
and denote it by  the so-called plethysm
\be  \{1\} \pleth \{\lambda\}  = \Proj^{(\lambda)} \hat T^N_1 =\{\lambda\} ,
\label{eq:proja}\ee
where $\Proj^{(\lambda)}$ is a projection operator  that picks out the cofactor of
the S$_N$ irrep $(\lambda)$ in a representation of ${\rm S}_N\times {\rm U}(m)$.
This equation is equivalent to Eq.\ (\ref{eq:TN1}).
However, its value is that it leads to a powerful generalization of the 
Schur-Weyl theorem.
For, if $\Hb_m$ is also the carrier space for an $m$-dimensional irrep $\{\kappa\}$ of a group U$(n)$ for some $n\leq m$ 
and $\hat T^N_\kappa$ is the corresponding reducible 
representation of ${\rm S}_N\times {\rm U}(n)$ carried by $\Hb_m^N$ then, by definition, the cofactor of the S$_N$ irrep $(\lambda)$ in the expansion of this irrep defines a general plethysm for any 
$\lambda\vdash N$ by  
\be \{\kappa\} \pleth \{\lambda\} = \Proj^{(\lambda)}
\hat T^N_\kappa ,
\label{eq:projb} \ee

A plethysm defined in this way can be evaluated using S$_N$ characters.
Let $\Xi^N_\kappa$ denote the character of the 
${\rm S}_N\times {\rm U}(n)$ representation $\hat T^N_\kappa$ carried by 
$\Hb_m^N$.
Using the orthogonality relationship (\ref{eq:S_Ncharorthog}) for S$_N$ characters,
\be \chi^\lambda\cdot \chi^\kappa = \frac{1}{N!} \sum_\rho n_\rho \chi^\lambda_\rho \chi^\kappa_\rho
= \delta_{\lambda,\kappa} ,\ee
Eq.\ (\ref{eq:projb}) is  expressed in terms of characters by
\be s_\kappa\pleth s_\lambda  = \chi^\lambda \cdot \Xi_\kappa^N .
\label{eq:4.plethdef}\ee
For example, with the character $\Xi^N_1$ given by Eq.\ (\ref{eq:XiH^N}), one regains the previously derived identity
\be
s_\lambda = s_1\pleth s_\lambda =
\chi^\lambda \cdot \Xi_1^N =
\frac{1}{N!} \sum_\rho n_\rho \chi^\lambda_\rho p_\rho  .\ee

We also obtain a useful and insightful expression for the plethysm 
$s_\kappa \pleth s_\lambda$ from the observation that the unitary irrep 
$\{ \kappa\}$ of U$(n)$ on $\Hb_m$ can be regarded as a map from U$(n)$ to U$(m)$,  i.e.,  to the fundamental $m$-dimensional irrep $\{ 1\}$ of U$(m)$. 
Recall that the character of a U$(m)$ matrix is given by its trace.
Thus, the character $s_1$ of the  fundamental U$(m)$ irrep $\{1\}$ is the
 function
\be s_1(z) = \sum_{i=1}^m z_i ,\ee
of a set of variables $\{ z_1,  \dots,z_m\}$ corresponding to the diagonal entries of U$(m)$ matrices.
Because each class of a unitary group contains
a diagonal matrix, we can restrict consideration to subsets of diagonal matrices.
Thus, we consider a diagonal U$(n)$  matrix 
\be M^{(n)}(x) = \mathrm{diag} [x_1,x_2, \dots , x_n] .\ee
In  a suitable basis for an $m$-dimensional  U$(n)$ irrep $\{\kappa\}$, this matrix maps to the $m\times m$ matrix
\be  M^{(m)}(z^{(\kappa)}(x)) = \mathrm{diag} [z^{(\kappa)}_1(x),z^{(\kappa)}_2(x), \dots , z^{(\kappa)}_m(x)] . \ee
It follows that the character of the irrep $\{ \kappa\}$ is given by
\be s_\kappa(x) = \sum_{i=1}^m z^{(\kappa)}_i(x) = s_1(z^{(\kappa)}(x)) .\ee
It also follows that 
$\Xi^N_\kappa (\rho ,  x) = \Xi^N_1(\rho , z^{(\kappa)}(x))$ 
and hence, because
$\chi^\lambda \cdot \Xi_1^N = s_\lambda$, that
\be [s_\kappa\pleth s_\lambda](x)  = s_\lambda(z^{(\kappa)}(x)). \label{eq:91}\ee

As an example, let $n=3$ and $\{\kappa\}=\{ 2\}$.  The U(3) irrep $\{2\}$
has dimension $m=6$ and character given by
\be s_{2} (x) = \sum_{i\leq j}^3x_ix_j =
x_1^2+x_2^2+x_3^2+x_1x_2+x_2x_3+x_1x_3 . \ee
 It is equal to the U(6) character
 $ s_{1}(z) =\sum_i^6 z^{(2)}_i(x)$ with
\bea &z^{(2)}_1(x) = x_1^2,\;\; z^{(2)}_2(x)=x_2^2,\;\;
 z^{(2)}_3(x)=x_3^2,& \nonumber\\
&z^{(2)}_4(x) =x_1x_2,\;\; z^{(2)}_5(x)=x_2x_3,\;\; 
z^{(2)}_6(x)=x_1x_3 . &\qquad \label{eq:z(x)}\eea
 Thus, the plethysm $s_2\pleth s_2$  for U(3) is  given by
\bea [s_{2} \pleth s_{2}] (x)  = s_2 (z^{(2)}(x)) =
\sum_{i\leq j}^6 z_i^{(2)}(x) z_j^{(2)}(x)&& \nonumber\\
 = x_1^4 + x_1^2x_2^2 + x_1^2 x_3^2 + x_1^3 x_2 + x_1^2 x_2x_3 + 
x_1^3 x_3 +   \dots . && \nonumber\\
&& \eea
 Then, from a knowledge of the $S$ functions and their orthogonality properties
 \cite{Macdonald95}, one obtains
 \be s_{2} \pleth s_{2}  = s_{4} +s_{2^2} .\ee

Plethysms provide powerful tools for numerous operations arising in the applications of symmetry to quantum mechanical systems.
An early use of plethysms in the nuclear shell model by \textcite{Elliott1, Elliott2} made use of laborious hand calculations by \textcite{Ibrahim51,Ibrahim52} to determine which SU(3) irreps occur in the nuclear
$(2s1d)$ shell and with what S$_N$ symmetries.
Such calculations can now be carried out quickly and easily by use of the available computer codes.
As discussed  in Section \ref{sect:V.many-body states}, a knowledge of the 
S$_N$ symmetry of an SU(3) irrep is required in order that the SU(3) wave functions, in nuclear physics, can be combined with spin and isospin wave functions of complementary S$_N$ symmetry to form totally antisymmetric states.
Because a U(3) irrep remains irreducible on restriction to its SU(3) subgroup, the first problem was to classify the U(3) irreps that occur in the $(2s1d)$ shell by their ${\rm S}_N \times {\rm U}(3)$ symmetries.
A single nucleon in the $ (2s1d )$ 
shell belongs to the 6-dimensional U(3) irrep 
$\{2\}$ and so the $N$-nucleon states with S$_N$ symmetry $(\kappa)$ with $\kappa\vdash N$ span a U(3) representation given by  $\{2\} \pleth \{\kappa\}$.
For example, using the code of \textcite{CD1}, we obtain the U(3) plethysm
\bea \{2\}\pleth \{32\} &=& \{541\} \oplus\{442\}\oplus\{532\}\oplus\{64\}\oplus 2\{622\}
\nonumber\\  &&
\oplus \{73\}\oplus\{721\}\oplus\{82\}\oplus\{631\} .\eea
Thus,  the space of $N=5$ nucleons in the $(2s1d )$ 
shell with S$_5$ symmetry $(32)$ is a sum of 
U(3) irreps $\{ \lambda_1\lambda_2\lambda_3\}$ and, hence, of
SU(3) irreps $(\lambda\mu) = (\lambda_1-\lambda_2, \lambda_2-\lambda_3)$
given by the SU(3) plethysm
\bea (2)\pleth \{32\} &=&
(13)\oplus (02)\oplus (21)\oplus (24)\oplus 2(40) \nonumber\\
&&\oplus (43)\oplus (51)\oplus  (62)\oplus (32) , \eea
in agreement with a result given by \textcite{Elliott1, Elliott2}.
[Note that that the round brackets in this equation
 denote SU(3) irreps rather than S$_N$ ireps as used elsewhere.]

An  important application of plethysms is to the evaluation of branching rules. 
The above definition of a plethysm shows that, whereas a Hilbert space 
$\Hb_m$ that carries an $m$-dimensional irrep $\{\kappa\}$ of a group U$(n)$  is also the carrier space for an irrep $\{1\}$ of the group U$(m)$, 
a subspace  $\Hb_m^{\{\lambda\}}\subset \Hb_m^N$ that carries a 
U$(m)$  irrep $\{ \lambda\}$, with $\lambda\vdash N$,
is the carrier space for a (generally reducible) representation $\{\kappa\} \pleth \{\lambda\}$ of U$(n)$.
Thus, if U$(n)$ is a subgroup of U$(m)$ with
 an $m$-dimensional irrep $\kappa$,
the restriction of the U$(m)$ irrep $\{ \lambda\}$ to U$(n)$ is given by the plethysm $\{\kappa\} \pleth \{\lambda\}$.  In other words, the branching rule
\be {\rm U}(m) \downarrow {\rm U}(n) \;\; : \;\;  \{1\} \downarrow \{\kappa\}  
\label{eq:4B.brulea}\ee
implies the general rule
\be {\rm U}(m) \downarrow {\rm U}(n) 
\;\; : \;\;  \{\lambda\} \downarrow \{\kappa\} \pleth \{\lambda\} .
\label{eq:4B.bruleb}\ee
In fact, the above example, of computing the symmetrized tensor products of SU(3) irreps was viewed by \textcite{Elliott1, Elliott2} as a calculation of the 
${\rm U}(6) \downarrow {\rm U}(3)$ branching rules required for the classification of $(2s1d)$-shell states that reduce the subgroup chain
\be {\rm U}(6) \supset {\rm U}(3) \supset {\rm SO}(3) \supset {\rm SO}(2).\ee
Thus, because U(3) is a subgroup of the U(6) group whose defining 6-dimensional irrep $\{1\}$ satisfies the branching rule 
${\rm U}(6) \downarrow {\rm U}(3): \{1\} \downarrow \{2\}$ it follows, for example, that the U(6) irrep $\{32\}$ restricts to the U(3) representation 
$\{2\}\pleth \{32\}$.

Plethysms are also used  for calculating the properties of other groups
that  are subgroups or contain subgroups of general linear or unitary groups.
For example, if a group $G$ contains some U$(n)$ group as a subgroup
one can restrict the characters of $G$ to U$(n)$ and thereby  express them as sums of S-functions.
In this way, calculations involving  characters of irreps of a given group can be evaluated by means of operations on S-functions and the results re-expressed in terms of characters of the group under study.
Conversely, if the group is  a subgroup of a unitary group, the characters of its
irreps can be regarded as linear combinations of S-functions with their arguments restricted to the subgroup.
For example, on restriction of U(3) to its SO(3) subgroup, we have the branching rules
\bea \label{eq:U3SO3branching}
{\rm U}(3) \downarrow {\rm SO(3)} &:& \{2\} \downarrow [L=2] \oplus [L=0] ,\\
&:& \{0\} \downarrow  [L=0]  .
\eea
Thus, if $\overline{\{2\}}$ is now used to denote the restriction of the U(3) irrep $\{ 2\}$ to SO(3), we can make the identification
\be [L=2] \equiv \overline{\{2\}}- \overline{\{0\}} . \label{eq:4B.SO3pleth} \ee
This device was  introduced
 by \textcite{Littletrick}.
It makes most sense when the formulae for plethysms are written in terms of S functions because while it is not clear what the negative of a group representation means, the negative of a Schur function is well defined.

In many applications there is a need to apply plethysms sequentially and to representations which may be expressed as tensor products of other representations or linear combinations of irreps.  Moreover, as Eq.\  (\ref{eq:4B.SO3pleth}) illustrates, one encounters applications in which combinations of irreps occur with negative coefficients.  
Rules  for expressing plethysms of algebraic combinations of Schur functions
 in terms of sums and products of simple plethysms were determined by
\textcite{Littlewood44} and extended by \textcite{SmithW67,SmithW68} and others. 
The following rules, expressed in terms of functions $A$, $B$, and $C$, which may be any combinations (sums and sums of products) of S functions, are taken from the book of \textcite{Wyb1}.
The first rule
\be (A\pleth B)\pleth C  =  A\pleth (B\pleth C ) \ee
shows that while plethysms are not commutative, they are associative.
It is next observed that a plethysm is  distributive on the right with respect to addition, subtraction and multiplication, i.e., it satisfies the identities
\bea &A\pleth (BC) = (A\pleth B)(A\pleth C) , &\\
&A\pleth (B\pm C) = A\pleth B \pm A\pleth C . & \eea
Plethysms are not distributive on the left but obey the combination rules
\bea \displaystyle
(A\!+\!B) \pleth\! \{ \lambda\} \!=\! \sum_{\mu\nu} \Gamma_{\mu\nu}^\lambda
(A\pleth\! \{\mu\})(B\pleth\! \{\nu\}) , \qquad&& \label{eq:4.PRule1}\\
\displaystyle
(A\! -\! B)\pleth\! \{\lambda\} \! =\! \sum_{\mu\nu} (-1)^{N_\nu} \Gamma_{\mu\nu}^\lambda  (A\pleth\! \{\mu\})(B\pleth\! \{\tilde\nu\}) ,&\quad& \label{eq:4.PRule2} \\
\displaystyle
(AB)\pleth\!\{\lambda\} \! =\!  \sum_{\mu\nu} \gamma_{\mu\nu}^\lambda
(A\pleth\! \{\mu\})(B\pleth\! \{ \nu\}) , \qquad&& \label{eq:4.PRule3}\eea
where $\gamma_{\mu\nu}^\lambda$ and $\Gamma_{\mu\nu}^\lambda$ are, respectively, the  coefficients appearing in the S$_N$ and U$(n)$ branching rules (\ref{eq:56})   and (\ref{eq:UNtensorprodrep}), $N_\nu = \sum_i \nu_i$, and
$\{\tilde\nu\}$ denotes the irrep defined by the partition of $N_\nu$ conjugate to $\nu$ as defined by Eq.\ (\ref{eq:conjugate}).
The above rules are consistent with the relationships,
which follow from Eq.\ (\ref{eq:4.plethdef}),
\bea s_\kappa \pleth (s_\mu s_\nu) &=&
\sum_\lambda \Gamma^\lambda_{\mu\nu} s_\kappa \pleth s_\lambda ,
\label{eq:120}\\
(s_\mu s_\nu) \pleth s_\lambda &=& 
\Big(\sum_\kappa \Gamma^\kappa_{\mu\nu}
s_\kappa \Big)\pleth  s_\lambda . \label{eq:121}\eea

The partition $\tilde\nu \vdash N_\nu$ 
appearing in Eqs.\ (\ref{eq:4.PRule2})  and (\ref{eq:4.PRule3}) is conjugate to the partitions $\nu \vdash N_\nu$,  defined
such that the Young diagram for $\tilde \nu$ is obtained from the diagram for 
$\nu$ by interchanging rows with columns; e.g.,
\be 
\nu \sim\youngd{         \yline4\cr
 &&&&&&&&\cr \yline4\cr
 &&&&\cr       \yline2\cr
 &&\cr          \yline1\cr}
       \quad \Rightarrow \quad \tilde\nu \sim
\youngd{         \yline3\cr
 &&&&&&\cr \yline3\cr
 &&&&\cr       \yline2\cr
 &&\cr          \yline1\cr
 &&\cr          \yline1\cr} 
\; .
\label{eq:conjugate}
\ee

The characters of the S$_{N_\nu}$ irreps $(\nu)$ and $(\tilde\nu)$ are related by the equation
 \be \chi^{\tilde\nu}_\rho = \chi^{1^N}_\rho \chi^{\nu}_\rho =
(-1)^\rho\chi^{\nu}_\rho \,, \ee
with $(-1)^\rho=\pm 1$, according
as  $\rho$ is a class of even or odd permutations.

\section{Unitary-unitary duality}
\label{sect:V.many-body states}

An important use of Schur-Weyl duality is to derive  rules for
constructing basis wave functions for systems of indistinguishable particles when the particle  wave functions belong to tensor product spaces. For
example, each particle might have space and spin wave functions; its
wave functions would then belong to a tensor product  of spatial and spin Hilbert spaces. If the particles are boson-like, with
 integer intrinsic spins, then according to the spin-statistics theorem, their total wave functions should be symmetric under permutations. On the other hand, if they  are fermion-like, and so have half-odd integer spins, they should be antisymmetric under odd permutations. The
spin wave functions and the  spatial wave functions
separately can have other symmetries so long as their combinations are symmetric (for bosons) or antisymmetric (for fermions).
The use of Schur-Weyl duality to  construct symmetric and antisymmetric many-particle wave functions in tensor product
spaces gives rise to ${\rm U}(n)\times {\rm U}(m)$ duality relationships which play a central role in the construction of coupling schemes for many-particle calculations in atomic and subatomic physics.

\subsection{The unitary-unitary duality theorem} \label{sect:V.uud} 

Let 
\be \mathbb{H}_{mn} = \mathbb{H}_m\otimes\mathbb{H}_n\ee 
denote a tensor product of Hilbert spaces, of (say) spatial and spin wave functions for a single particle, of dimension $m$ and $n$, respectively.
Then $\mathbb{H}_m$ carries the  $m$-dimensional
irrep $\{1\}$ of U$(m)$;
$\mathbb{H}_n$ carries the  $n$-dimensional irrep $\{1\}$  of
U$(n)$; and $\mathbb{H}_{mn}$ carries the  $mn$-dimensional
irrep $\{1\}$ of U$(mn)$. We distinguish  these irreps by $\hat U^{\{1\}}_{m}$,
$\hat U^{\{1\}}_{n}$, and $\hat U^{\{1\}}_{mn}$, respectively. 

The question now arises: how does one build up symmetric and
antisymmetric many-particle  basis wave functions for irreps of 
${\rm U}(m)\times{\rm U}(n)$ from single-particle wave functions in $\Hb_{mn}$? The answer is given by the unitary-unitary duality theorem.
Let $\mathbb{H}_{mn}^N$ denote the tensor product of $N$ copies of 
$\mathbb{H}_{mn}$
\be \mathbb{H}_{mn}^N =
\underbrace{\mathbb{H}_{mn} \otimes \cdots \otimes 
\mathbb{H}_{mn}}_{N \; {\rm copies}}
=\mathbb{H}_m^N\otimes \mathbb{H}_n^N \,. \label{eq:52}\ee
According to the Schur-Weyl theorem, the subspace $\mathbb{H}_{mn}^{\{N\}}$
 of fully symmetric $N$-particle wave functions in $\mathbb{H}_{mn}^N$ carries an irrep, $(N)\times\{N\}$, of the direct product group S$_N\times{\rm U}(mn)$.
However, the fully symmetric irrep $(N)$ of S$_N$ is
one-dimensional. It follows that the subspace of fully symmetric
$N$-particle wave functions  
$\mathbb{H}_{mn}^{\{N\}}\subset \mathbb{H}_{mn}^N$ is the carrier space
for the U$(mn)$ irrep  $\{N\}$. 
Similarly, the subspace $\mathbb{H}_{mn}^{\{1^N\}}$
of fully antisymmetric $N$-particle wave functions in
$\mathbb{H}_{mn}^N$ is the carrier space for the
U$(mn)$ irrep $\{1^N\}$. 
The converse of these observations is given by the following theorem.

\medskip
{\bf Theorem  3 (unitary-unitary duality):}
\emph{
The groups ${\rm U}(m)$ and ${\rm U}(n)$ have dual representations
on the fully symmetric and fully antisymmetric subspaces of $\Hb_{mn}^N$
in accordance with the branching rules:
\bea {\rm U}(mn) \downarrow {\rm U}(m)\!\times\! {\rm U}(n) \!&:&
\{N\} \downarrow
\displaystyle\bigoplus_{\kappa\vdash N}\, \{\kappa\}
\!\times\! \{\kappa\} ,\qquad \label{eq:70}\\
\!&:& \{1^N\} \downarrow \displaystyle\bigoplus_{\kappa\vdash N}
\,\{\kappa\} \!\times\! \{\tilde\kappa\} , \qquad\label{eq:71} \eea
where the
sum in Eq.\ (\ref{eq:70}) extends over all partitions $\kappa\vdash N$
of length $l(\kappa)\leq \min (m,n)$
 and the sum in Eq.\ (\ref{eq:71}) extends over all partitions 
 $\kappa\vdash N$ for which $l(\kappa)\leq m$ and $l(\tilde\kappa)\leq n$,
where $\tilde\kappa$ is defined by Eq.\ (\ref{eq:conjugate}). }
\medskip

A direct proof of Theorem 3 is given by \textcite{Thms}.  
 The following proof shows that it is implied by the Schur-Weyl theorem.

\medskip \noindent \emph{Proof:}
According to the Schur-Weyl theorem, the Hilbert space $\Hb^N_{mn}$ carries the
${\rm S}_N\times {\rm U}(mn)$  representation
$\hat T^N_{mn} = \bigoplus_{\tau\vdash N}^{l(\tau) \le mn} (\tau) \times \{\tau\}$.
Thus, according to the branching rule of Eq.\ (\ref{eq:55a}), this representation restricts to the representation 
 \be \hat T^N_{m,n} = \bigoplus_\tau
\bigoplus_{\kappa\vdash N}^{l(\kappa)\le m} 
\bigoplus_{\lambda\vdash N}^{l(\lambda)\le n} \gamma^\tau_{\kappa\lambda}\,
(\tau)\times \{ \kappa\} \times \{ \lambda\} 
\ee
of ${\rm S}_N\times {\rm U}(m) \times {\rm U}(n)$.
The component of this representation with S$_N$ symmetry $(\tau)$ is then
\be \hat T^{(\tau)}_{m,n} = \bigoplus_{\kappa\vdash N}^{l(\kappa)\le m} 
\bigoplus_{\lambda\vdash N}^{l(\lambda)\le n} \gamma^\tau_{\kappa\lambda}\,
(\tau)\times \{ \kappa\} \times \{ \lambda\} . 
\ee
The tensor product of two S$_N$ irreps $(\kappa) \otimes (\lambda)$ 
contains a copy of the identity irrep $(N)$ if and only if
$\lambda = \kappa $  \cite{Ham}.  Similarly, it contains a copy of the antisymmetric irrep $(1^N)$ if and only if $\lambda = \tilde \kappa $ (the conjugate of $\kappa$). Moreover, no  tensor product of two S$_N$ irreps contains more than one copy of either $(N)$ or $(1^N)$.  Thus, 
\be \gamma_{\kappa\lambda}^{N}
=\delta_{\kappa,  \lambda} , \quad
\gamma_{\kappa\lambda}^{1^N}=\delta_{\kappa,\tilde{\lambda}},\ee
and we obtain
\bea \hat T^{(N)}_{m,n} &=& 
\!\bigoplus_{\kappa\vdash N}^{l(\kappa)\le m, l(\lambda)\le n} \!
(N)\times \{ \kappa\} \times \{ \kappa\}  , \\ 
\hat T^{(1^N)}_{m,n} &=&
\!\bigoplus_{\kappa\vdash N}^{l(\kappa)\le m, \kappa_1\le n} \!
(1^N)\times \{ \kappa\} \times \{ \tilde\kappa\}  .
\eea
Comparing these results with those of the Schur-Weyl expression for the representation of ${\rm S}_N \times {\rm U}(mn)$ carried by the symmetric and antisymmetric components of $\Hb^N_{mn}$, 
for which
\be \hat T^{(N)}_{mn} = (N) \times \{ N\} , \quad
\hat T^{(1^N)}_{mn} = (1^N) \times \{ 1^N\}, \ee
leads to the results of the theorem.  \hfill $\Box$
\medskip

\medskip
{\bf Corollary 3:}\emph{
Let $\Hb^{\rm S}_{mn}= \bigoplus_{N=0}^\infty \Hb_{mn}^{\{N\}}$ denote the sum of the Hilbert spaces that carry the fully symmetric irreps $\{N\}$ of 
${\rm U}(mn)$ and let
$\Hb^{\rm AS}_{mn} = \bigoplus_{N=0}^\infty \Hb_{mn}^{\{1^N\}}$ denote the sum of the Hilbert spaces that carry the fully antisymmetric irreps $\{1^N\}$ of 
${\rm U}(mn)$.
The  groups ${\rm U}(m)$ and ${\rm U}(n)$ then have
dual representations on  $\Hb^{\rm S}_{mn} $ given by
$\bigoplus_{\kappa} \{ \kappa\} \times \{\kappa\}$
and on $\Hb^{\rm AS}_{mn}$ given by
$\bigoplus_{\kappa} \{ \kappa\} \times \{\tilde \kappa\}$.}
\medskip

\emph{Proof:}  The corollary follows from the observation that each
$\{ \kappa\}$  is a partition of a non-negative integer $N$, i.e., 
$\kappa\vdash N$, and occurs once and only once in the  sum.  
\hfill $\Box$.
 \medskip

As the following examples illustrate, unitary-unitary duality  can be employed directly at an operational level to construct  appropriately symmetrized
wave functions for an $N$-particle system.

\subsection{Symmetric and anti-symmetric space-spin wave functions}
\label{sect:3c}

The above duality relationships show that if $\Hb_m$ and $\Hb_n$ are,
respectively, Hilbert spaces of spatial  and spin wave functions for a particle,
 then the  combinations of these wave functions,
appropriate for a system of $N$ bosons (with integer spins), belong to the 
fully symmetric subspace of the tensor product space 
$\Hb^N_{mn} =\Hb_m^N\otimes\Hb_n^N$.
This subspace  contains only  wave functions that are invariant under any permutation $P \in {\rm S}_N$.

We first consider a simple case in which
$\{\psi^\kappa_i\}$ and $\{\varphi^\lambda_j\}$ denote orthonormal bases for 
 S$_N$ irreps  $(\kappa)$ and $(\lambda)$, respectively.
 There is known to be
precisely one S$_N$-invariant bilinear combination of these basis functions for each $ \lambda=\kappa$. If the bases correspond, it is given by 
$\Phi^\kappa =\sum_i \psi^\kappa_i\otimes \varphi^{\kappa*}_i$,
where  $*$ denotes complex conjugation.%
\footnote{ Note that S$_N$ irreps are self-contragredient and, as a consequence, their Hilbert spaces are invariant under complex conjugation.}
This follows from the observation that, under a permutation
$P\in {\rm S}_N$, 
\be \Phi^\kappa \to \hat P \Phi^\kappa= \sum_{ijk}
\psi^\kappa_{j}\otimes\varphi^{\kappa *}_{k}\, M^{(\kappa)}_{ji}(P)
M^{(\kappa) *}_{ki}(P) ,
\ee
where $M^{(\kappa)}(P)$ is the matrix 
representing the permutation $P \in {\rm S}_N$, relative to the basis 
$\{\psi^\kappa_i\}$  or $\{ \varphi^\kappa_i\}$.
Thus, because the irrep $(\kappa)$ is unitary, we determine that
$\hat P\Phi^\kappa =\Phi^\kappa$.

It is important to note that
it is always possible, and indeed  natural, to chose bases for 
S$_N$ irreps such that the matrices $M^{(\kappa)}(P)$
are real.  
The S$_N$ invariants are then given by
$\Phi^\kappa = \sum_i \psi^\kappa_i \otimes \varphi^\kappa_i$.

Now, for arbitrary values of $m$ and $n$,
 let $\{\psi^\kappa_{i\alpha}\}$ 
denote an orthonormal basis for
$\Hb_m^N$, where  $\kappa\vdash N$  labels  
an S$_N\times {\rm U}(m)$ irrep, $i$ indexes a basis for the S$_N$ irrep $(\kappa)$, and $\alpha$ indexes a basis for the U$(m)$ irrep $\{\kappa\}$.
Let $\{\varphi^\lambda_{j\beta}\}$ denote a similarly-defined basis for
$\Hb_n^N$ which reduces the group S$_N\times {\rm U}(n)$.
The product  functions
$\{ \psi^\kappa_{i\alpha}\otimes\varphi^\lambda_{j\beta}\}$
 are then a basis for the tensor product space $\Hb_m^N\otimes\Hb_n^N$.
Thus, from the above results, an orthonormal basis for the totally symmetric subspace of $\Hb_m^N\otimes\Hb_n^N$, appropriate for a system of bosons, is given by the linear combinations
\be \Psi^\kappa_{\alpha\beta} =  \frac{1}{\sqrt{d^\kappa}} \sum_i
\psi^\kappa_{i\alpha}\otimes \varphi^{\kappa}_{i\beta} ,\ee
with $\kappa$ running over the ordered partitions of $N$ for which 
$l(\kappa)\le m$ and $l(\kappa)\le n$,
and $d^\kappa$ is the dimension of the S$_N$ irrep $(\kappa)$.
These $\Psi^\kappa_{\alpha\beta}$ wave functions  have the useful property that they reduce the subgroup chain
\be \begin{array}{ccc} {\rm U}(mn) &\supset &{\rm U}(m)\times{\rm U}(n) \\
  N && \;\kappa \quad\qquad  \kappa   
\end{array} .
\ee

The bilinear combinations of $N$-particle spatial
wave functions in $\Hb_m^N$ with $N$-particle spin wave functions in
$\Hb_n^N$, appropriate for fermions, span
the subspace of the tensor product space $\Hb_m^N\otimes\Hb_n^N$ that is antisymmetric under the action of the symmetric group S$_N$.

As noted following Theorem 3, for every S$_N$ irrep corresponding to a partition
$\kappa\vdash N$ there is a so-called conjugate irrep, corresponding to the partition $\tilde\kappa \vdash N$ defined by Eq.\ (\ref{eq:conjugate}).
Bases for such conjugate irreps,
$\{\psi^\kappa_i\}$ and $\{\varphi^{\tilde\kappa}_i\}$, are naturally put into one-to-one correspondence,
such that the matrices for these irreps are related by the identity
\be  M^{(\tilde\kappa)}_{ki}(P) = (-1)^P M^{(\kappa)}_{ki}(P) ,\ee
with $(-1)^P = \pm 1$ according as $P$ is an even or odd permutation.
It is also observed that, if $\{\psi^\kappa_i\}$ and $\{\varphi^\lambda_j\}$ are orthonormal bases for S$_N$ irreps, it is only possible to form antisymmetric  bilinear combinations of these basis functions,
$\{ \psi^\kappa_{i\alpha}\otimes\varphi^\lambda_{j\beta}\}$,
 if $\lambda= \tilde\kappa$.
The transformation of the combination
$\Phi^\kappa =\sum_i \psi^\kappa_i\otimes \varphi^{ \tilde\kappa}_i$ under a permutation $P\in {\rm S}_N$ is  then given by 
\be \Phi^\kappa \to \hat P \Phi^\kappa = \sum_{ijk}
\psi^\kappa_{j}\otimes\varphi^{ \tilde\kappa}_{k}\, M^{(\kappa)}_{ji}(P)
M^{( \tilde\kappa)}_{ki}(P) .
\ee
Thus, by choosing phases such that the matrices $M^{(\kappa)}(P)$ are both  real and unitary, it follows that  $\hat P \Phi^\kappa = (-1)^P\Phi^\kappa$.
Similarly, for arbitrary values of $m$ and $n$,
 an orthonormal basis for the antisymmetric subspace of $\Hb_m^N\otimes\Hb_n^N$
is given by the linear combinations
\be \Psi^\kappa_{\alpha\beta} =  \frac{1}{\sqrt{d^\kappa}} \sum_i
\psi^\kappa_{i\alpha}\otimes \varphi^{\tilde\kappa}_{i\beta}
\ee
that reduce the subgroup chain
\be \begin{array}{ccc}
{\rm U}(mn) &\supset &{\rm U}(m)\times{\rm U}(n) \\
1^N && \kappa \quad\qquad  \tilde\kappa 
\end{array} ,
\ee
with $\kappa$ running over the ordered partitions of $N$ for which 
$l(\kappa)\le m$ and $l(\tilde\kappa)\le n$.

 \subsection{Anti-symmetric space-spin-isospin wave functions}
 \label{sect:LSTcoupling}

The U$(m)\times{\rm U}(n)$ duality of representations can be applied
to situations where products of more than two wave functions occur.
Such situations arise, for example, in the nuclear shell model when
nucleon wave functions are products of 3 components, spatial, spin, and isospin,
or in elementary particle physics when many-quark
systems, for example, have flavor, spin, and color degrees of freedom.

To illustrate the role of duality in such cases, consider
a single-nucleon Hilbert space
\be \mathbb{H} = \mathbb{H}_L\otimes\mathbb{H}_S \otimes \mathbb{H}_T \,,
\ee
that is a tensor product of spaces $\Hb_L$ (for spatial wave functions),
$\Hb_S$ (for spin wave functions), and $\Hb_T$ (for isospin wave functions).
Because a nucleon has spin $1/2$ and isospin $1/2$, the spaces $\mathbb{H}_S$ and $\mathbb{H}_T$ have dimension two, and each carries 
a defining irrep of U(2).
Denoting  the dimension of $\Hb_L$ by $n$, it follows that  $\Hb$ is of dimension $4n$, that it carries the  defining irrep, $\{1\}$, of U$(4n)$, and that it
remains irreducible on restriction to the ${\rm U}(n) \times {\rm U}(4)$  subgroup.

A typical shell-model problem is to
define a  basis for the  nuclear subspace $\Hb^{\{1^N\}}$ of fully antisymmetric
wave functions in $\Hb^N$ and to classify such a basis by the quantum numbers associated
with irreps of U$(n)$, U(2)$_S$, and U(2)$_T$ and other useful groups.
There are many possible  coupling schemes.

The so-called {\em Wigner super-multiplet scheme\/} \cite{Wigner1,Wsup} of $LST$ coupling starts with the four-dimensional spin-isospin space
\be \Hb_{ST} = \Hb_S\otimes \Hb_T ,\ee
which carries the standard irrep of  U(4).
The construction of a basis for the totally antisymmetric subspace of
\be \Hb^N = \Hb_L^N\otimes \Hb_{ST}^N \label{lstc}\ee
is thereby reduced to the standard problem, discussed in the first application, for which ${\rm U}(n)\times {\rm U}(4)$ duality applies.
Thus, the branching rule of the duality theorem, Eq.\ (\ref{eq:71}),
\be
{\rm U}(4n)\downarrow  {\rm U}(n)\times {\rm U}(4) \; :\;
\{ 1^N\} \downarrow \bigoplus_{\lambda\vdash N}\, \{\lambda\}\times \{\tilde\lambda\} ,
\ee
implies that the fully antisymmetric irreps of ${\rm U}(n) \times {\rm U}(4)$
are given by the tensor product $\{\lambda\}\times \{\tilde\lambda\}$ irreps   carried by subspaces of the fully antisymmetric subspace 
$\Hb^{\{1^N\}}\subset \Hb^N$.

A natural basis for U(4) is one that reduces the subgroup chain
\be
\begin{array}{ccccc} {\rm U(4)} &\supset& {\rm SU(2)}_S\times{\rm
SU(2)}_T &\supset& {\rm
U(1)}_S\times{\rm U(1)}_T \\
\tilde\lambda & & S \qquad\quad T && M_S \quad\quad M_T
\end{array} .
\ee
A desirable choice for U$(n)$ is one that reduces the subgroup chain
\be
\begin{array}{ccccccc} {\rm U}(n) &\supset& {\rm SO}(3)_L &\supset& {\rm SO}(2)_L \\
\{\lambda\} && L&& M_L
\end{array} ,
\ee
where SO(3)$_L$ is the standard rotation group.
These choices then give basis states for  $\Hb^{\{1^N\}}$ that reduce the chain
\be
\begin{array}{ccccccc} {\rm U}(4n) &\!\supset\! &{\rm U}(n)\times{\rm U}(4)
&\!\supset\!& \mathrm{SO}(3)_L\times {\rm SU(2)}_S\times{\rm SU(2)}_T \\
 1^N && \lambda \qquad\; \tilde\lambda
&\!\beta \!& L, M_L \;\quad\; S,M_S \quad\;\; T, M_T 
\end{array}  .
\label{u4u2u2}\ee

Note that the label $\beta$ is included to denote the additional labels  needed to
provide a complete classification of basis states.
Additional labels are provided, for example, by including an intermediate subgroup between U$(n)$ and  SO$(3)_L$.
One possibility is to include the group O$(n)$ in the chain
\be
 {\rm U}(n) \supset {\rm O}(n)
\supset{\rm SO}(3)_L \supset {\rm SO}(2)_L  .
\ee
An alternative, for a suitable choice of $\Hb_L$, is to include in the chain the group U(3), as in Elliott's shell model of nuclear rotational states
\cite{Elliott1, Elliott2},
\be {\rm U}(n)\supset {\rm U(3)} \supset {\rm SO(3)}_L \supset {\rm SO(2)}_L ,\ee
where U(3) is the symmetry group of the spherical harmonic oscillator.

\subsection{Unitary-unitary duality in boson systems}

The above examples of the use of unitary-unitary duality for fermions have parallels in bosonic systems with multiple degrees of freedom.  For example, in the Interacting Boson Model with two kinds of boson, corresponding to neutron pairs and proton pairs  of which each carries an irrep $\{1\}$ of U(6), the states of $N$ such bosons carry an irrep $\{N\}$ of  U(12).  The $N$ boson states can then be classified by the irrep labels in the subgroup chain  
${\rm U}(12) \supset {\rm U(6)}\times {\rm U}_F(2)$ which, in accordance with Theorem 3, are given by the branching rule of Eq.\ (\ref{eq:70}) for which
$\kappa = \{ N/2+F, N/2-F\}$, where $F$ is known as $F$-spin (see, for example, \cite{VanIsacker86} for more details).


\section{Methods of second quantization}
\label{sect:6.2ndQ}

The techniques of second quantization were invented for the quantization of fields. However, they prove to be equally powerful
nd insightful in the many-body quantum mechanics  of
indistinguishable particles and in the theory of Lie algebras.
At the time the terminology was introduced, it was common to regard standard quantum mechanics, in which the dynamics of particles was replaced by wave mechanics, as {\em first quantization\/}.
On the other hand, field theory, which considers particles as the quanta
of fields, was regarded as {\em second quantization\/}.

Quantization of the electromagnetic field was achieved by \textcite{BHJ}.  Their theory can be understood, at an elementary level, as extending
the Hamiltonian for a system of harmonic oscillators
\be \hat H = \sum_\nu \hbar\omega_\nu \left( c^\dag_\nu c^\nu + \hf \right) ,
\label{eq:Vi.Hho}\ee
with raising and lowering operators that satisfy commutation relations
\be  [c_\mu^\dag, c^\dag_\nu] = [c^\mu, c^\nu] = 0 , \quad [c^\nu, c_\mu^\dag] =
\delta^\nu_{\mu} , \label{eq:BosonCR}\ee
 to an infinite number of oscillators characterized by continuously variable frequencies. The raising operators for states of the electromagnetic field are then interpreted as creation operators for photons. This provides the fundamental link between the wave and particle theories of light.
Note that we use lower indices for creation
operators, $\{c^\dag_\nu\}$, and upper indices 
for annihilation operators $\{c^\nu\}$. This is to emphasize the fact that, whereas
the creation operators transform as a basis for the standard irrep
of a unitary group U$(n)$, where $n$ is the number of indices, the annihilation operators transform as a basis for the contravariant irrep; i.e., if a creation operator $c^\dag_\nu$ transforms under an element $g\in {\rm U}(n)$ according
to the equation $c^\dag_\nu \to \sum_\mu c^\dag_\mu g_{\mu\nu}$,
then the corresponding annihilation operator transforms 
$c^\nu \to \sum_\mu c^\mu g_{\mu\nu}^*$.

At first sight, it would appear that the quantum mechanical interpretation of the particle-wave duality for fermions (e.g.,\ electrons) of non-zero rest mass is different. Unlike quantum  electrodynamics,  non-relativistic quantum mechanics begins with particles and by means of the Schr\"{o}dinger equation assigns wave functions to  them to describe their states. Thus, it  was not obvious what could be achieved by a second quantization of the fields (i.e., wave functions) to regain the particles. In retrospect, it is now realized that
a field theory of electrons is provided by the Dirac equation
\cite{Dirac1,Dirac2}. 
Moreover, the derivation of the Dirac equation by factorization of the
relativistic Hamiltonian for an electron has similarities with the factorization of the
harmonic oscillator Hamiltonian which gives the quantization of the electromagnetic field.
However, because of the somewhat misleading particle-hole interpretation of the Dirac
equation, the parallel was not recognized at the time.  The vital step towards a fermionic field theory was taken a few years later by \textcite{JW}, who introduced operators $\{ a^\dag_\nu\}$ and $\{a^\nu\}$  that, respectively, create and annihilate fermions. Their success was based on a recognition that the Pauli principle, which states that two identical fermions cannot occupy the same state, is automatically  accommodated if the fermion operators are required to satisfy
anti-commutation relations
\be  \{a_\mu^\dag, a^\dag_\nu\} = \{a^\mu, a^\nu\} = 0 , \quad \{a_\mu^\dag, a^\nu\} = \delta^\nu_{\mu} ,\ee
rather than commutation relations. This follows simply from the observation that
\be \{a_\nu^\dag, a^\dag_\nu\} = 2a^\dag_\nu  a^\dag_\nu = 0 .\ee

One of the many advantages of using the methods of second quantization in non-relativistic quantum mechanics is that it ensures the exchange symmetries of identical particles are respected without the need for labeling indistinguishable particles and symmetrizing (or anti-symmetrizing) their wave functions. Thus, when expressed in terms of creation operators that commute with one another, the many-boson wave functions for identical bosons are automatically symmetric under exchange.  Similarly, when expressed in terms of creation  operators that anti-commute with one another, the many-fermion wave functions for  identical fermions are automatically antisymmetric under exchange. For example, the sets of three-boson and three-fermion states
\bea  |\mu\nu\kappa\rangle_B = c^\dag_\mu  c^\dag_\nu  c^\dag_\kappa |0\rangle ,\\
|\mu\nu\kappa\rangle_F = a^\dag_\mu  a^\dag_\nu  a^\dag_\kappa |0\rangle ,
\eea
where $|0\rangle$ is the zero-particle vacuum state, are automatically symmetric and antisymmetric, respectively, e.g.,
$|\mu\nu\kappa\rangle_B=|\nu\mu\kappa\rangle_B$ and
$|\mu\nu\kappa\rangle_F= -|\nu\mu\kappa\rangle_F$.

Another huge advantage of the second-quantization formalism is that it provides a powerful framework for the manipulation of Lie  algebras and their representations from which the duality relationships discussed in this review emerge naturally.
Consider  a basis $\{ X_{\mu,\nu}\}$   for the complex extension of the
Lie algebra u$(n)$ of the group U$(n)$  with commutation relations
\be [X_{\mu,\nu},X_{\mu',\nu'}] = \delta_{\nu,\mu'} X_{\mu,\nu'} -
\delta_{\nu',\mu} X_{\mu',\nu} .  \label{eq:UnCR}\ee
It follows, from the boson commutation relations,  Eq.\ (\ref{eq:BosonCR}), that this Lie algebra has a \emph{boson realization} 
\be X_{\mu,\nu} \to \hat
X_{\mu,\nu} = c^\dag_\mu c^\nu .\ee
This realization automatically extends the  defining irrep $\{1\}$ of U$(n)$
on the Hilbert space $\Hb$ to the symmetric irreps, $\{N\}$, on
symmetric subspaces, $\Hb^{\{N\}}\subset \Hb^N$, for positive integer
values of $N$.
Thus, whereas the one-boson states
$\{c^\dag_\nu |0\rangle, \nu =1,\dots,n\}$ are a basis for the U$(n)$ irrep $\{1\}$,
the set of $N$-boson states,
$\{c^\dag_{\nu_1} c^\dag_{\nu_2} \dots c^\dag_{\nu_N} |0\rangle, \nu_1,\nu_2,\dots ,\nu_N =1,\dots,n\}$,
 are a basis for the U$(n)$ irrep $\{ N\}$.

It follows from these results that the unitary-unitary duality theorem has a natural expression in the language of second quantization.
This is seen for the boson operators
by regarding $\{ c^\dag_{i\nu}\}$ and $\{ c^{i\nu}\}$, with commutation relations 
\be [ c^{i\mu},c^\dag_{j\nu}] = \delta^i_{j}\delta^\mu_{\nu} , \quad
[c^\dag_{i\mu}, c^\dag_{j\nu}]=[ c^{i\mu}, c^{j\nu}] =0, \ee 
as creation and annihilation operators for the single-particle states
of the tensor product space $\Hb =\Hb_m\otimes\Hb_n$,
with $i=1,\dots, m$ indexing a basis for $\Hb_m$ and $\nu=1,\dots, n$
indexing a basis for $\Hb_n$.  
The Lie algebras u$(mn)$, u$(m)$ and u$(n)$ are then realized by the operators
\bea \hat X^{(mn)}_{i\mu,j\nu} &=& c^\dag_{i\mu} c^{j\nu} ,\nonumber\\
\hat X^{(m)}_{i,j} =\sum_\nu c^\dag_{i\nu} c^{j\nu} , \!&& \!\hat
X^{(n)}_{\mu,\nu} =\sum_i c^\dag_{i\mu} c^{i\nu} . \label{eq:3,bosonUmn}\eea
Highest-weight states for the $N$-boson ${\rm U}(m)\times {\rm U}(n)$ irreps 
given by the unitary-unitary branching rule
\be{\rm U}(mn) \downarrow {\rm U}(m)\times {\rm U}(n) \,;\, 
\{N\} \downarrow
\displaystyle\bigoplus_{\kappa\vdash N}\, \{\kappa\} \times
\{\kappa\} \ee
are constructed as follows.
Let $i=1,2,\dots, m$ and $\nu = 1,2, \dots ,n$, respectively, index basis states for the $\{1\}$ irreps of U$(m)$ and U$(n)$  in order of decreasing weight.
And let $\FB^K$ denote the determinant of boson operators
\be
\setlength{\arraycolsep}{5pt}
 \FB^{K} = 
\left|
\matrix{
  c^\dag_{11}  \!\!&\!\!c^\dag_{12}\!\!&\!\!  \cdots \!\!&\!\!c^\dag_{1K} \cr
 c^\dag_{21}  \!\!&\!\!c^\dag_{22}\!\!&\!\!  \cdots \!\!&\!\!c^\dag_{2K} \cr
 \vdots \!\!&\!\! \vdots\!\!&\!\!  \cdots \!\!&\!\! \vdots \cr
 c^\dag_{K1}  \!\!&\!\!c^\dag_{K2}\!\!&\!\!  \cdots \!\!&\!\!c^\dag_{KK} 
}
\right|
 \label{eq:3.bosonhwt}
\ee
defined in parallel with the Slater determinants of Eq.\ (\ref{eq:3.111hwt}).
If $|0\rangle$ is the  boson vacuum state and $\tilde\kappa$ is the partition conjugate to $\kappa$, then the  state 
\be |\{\kappa\} ; {\rm h.wt.}\rangle= 
\FB^{\tilde\kappa_1} \FB^{\tilde\kappa_2}  \FB^{\tilde\kappa_3} \dots  |0\rangle
\ee
is  observed to be of highest weight  $\kappa$ relative to both U$(m)$ and 
U$(n)$.

Similar results for the representations of U$(mn)$ are obtained for the fermion realizations starting from the observation that the U$(n)$
commutation relations (\ref{eq:UnCR}) are also satisfied by the fermion realization
 \be X_{\mu,\nu} \to \hat
X_{\mu,\nu} = a^\dag_\mu a^\nu .\ee
However, while the one-fermion states 
$\{a^\dag_\nu |0\rangle, \, \nu =1,\dots,n\}$ are a basis for the U$(n)$ irrep 
$\{1\}$, the $N$-fermion states,
$\{a^\dag_{\nu_1} a^\dag_{\nu_2} \dots a^\dag_{\nu_N} |0\rangle, \nu_1,\nu_2,\dots ,\nu_N =1,\dots,n\}$,
now span the U$(n)$ irrep $\{ 1^N\}$.

In parallel with Eq.\ (\ref{eq:3,bosonUmn}),
the Lie algebras  u$(mn)$, u$(m)$ and u$(n)$ also have fermion realizations
\bea \hat X^{(mn)}_{i\mu,j\nu} &=& a^\dag_{i\mu} a^{j\nu} , \nonumber\\
\hat X^{(m)}_{i,j} =\sum_\nu a^\dag_{i\nu} a^{j\nu} ,&\quad& \hat
X^{(n)}_{\mu,\nu} =\sum_i a^\dag_{i\mu} a^{i\nu} . \eea
However, the analogous extension of the irrep $\{ 1\}$ 
of U$(mn)$ with fermions gives  an antisymmetric irrep $\{1^N\}$ of U$(mn)$ 
on $\Hb^{\{ 1^N\}}$ for $N>1$, which according to the Schur-Weyl theorem satisfies the branching rule
\be{\rm U}(mn) \downarrow {\rm
U}(m)\times {\rm U}(n) \,;\, \{1^N\} \downarrow
\displaystyle\bigoplus_{\kappa\vdash N}\, \{\kappa\} \times \{\tilde
\kappa\} ,\ee
where $\tilde \kappa$ is the partition conjugate to $\kappa$.

Highest-weight states for these ${\rm U}(m)\times {\rm U}(n)$ irreps are constructed in terms of fermion operators as follows.
As for the boson basis, let $i=1,2,\dots, m$ and 
$\nu = 1,2, \dots ,n$ index basis states for the U$(m)$ and U$(n)$ $\{1\}$ irreps,  respectively, in order of decreasing weight.
Then, let $\FF^K$ denote the product of fermion operators
\be
\setlength{\arraycolsep}{5pt}
 \FF^{K}_i =  a^\dag_{i1} a^\dag_{i2} \dots a^\dag_{iK} .
 \label{eq:3.fermion}
\ee
Unlike its boson counterpart, this simple product already satisfes the antisymmetry requirement.
Thus, the state 
\be |\{\kappa\};  {\rm h.wt.}\rangle= 
\FF_1^{\tilde\kappa_1} \FF_2^{\tilde\kappa_2}  \FF_3^{\tilde\kappa_3} 
\cdots  |0\rangle
\ee
is observed to be of highest weight relative relative to both U$(m)$ and U$(n)$.

 In addition to the unitary--unitary duality (stated in Theorem 3, Sec.\ 
 \ref{sect:V.uud}), we will see in the following sections how the powerful formalism of second quantization plays a relevant role in identifying other dualities  of
importance in physical applications.


\section{Dual representations on harmonic-oscillator boson spaces}
\label{sect:VI.bosonduals}

In many-body theory, one is primarily interested in many-fermion systems.  
However, the position and momentum coordinates of fermions obey the boson commutation relations of a Heisenberg-Weyl Lie algebra.
For example, nucleons in a nucleus are described in a zero-order approximation as independent particles in a harmonic-oscillator potential.  Thus, in non-relativistic quantum mechanics, in which nucleons are neither created nor destroyed in their interactions with one another, the excitations of a system of nucleons can be described in terms of harmonic oscillator quanta.  Similarly, the vibrational excitations of a condensed matter system are often appropriately described in terms of phonons which, like harmonic-oscillator quanta, are bosonic.
Moreover, composite systems of tightly-bound fermions, such as alpha particles, behave as bosons at low densities (see Section \ref{sect:bmaps}).
The essential quality of a boson is that its creation and annihilation operators obey the same boson commutation relations as those of harmonic oscillator quanta. 

  Several pairs of groups can be found with dual representations on a given multi-dimensional harmonic oscillator space. Paramount to these dualities is the following  theorem.

\medskip
\noindent {\bf Theorem 4 (symplectic-orthogonal duality):}   \emph{The groups 
${\rm O}(N)$ and ${\rm Sp}(m,\Rb)$ have dual
representations on the Hilbert space of the $Nm$-dimensional oscillator. }
\medskip\

\noindent Note that the group denoted here by Sp$(m,\Rb)$ is the real non-compact symplectic group of rank $m$.  Many authors denote this same group by Sp$(2m,\Rb)$.  Note also that when $N$ is odd the ${\rm Sp}(m,\Rb)$ representation is \emph{projective}, i.e., a double-valued (spinor) representation.  It is a genuine representation of the two-fold cover of ${\rm Sp}(m,\Rb)$ known as a metaplectic group. Nonetheless, in the rest of this paper, both the genuine and projective oscillator representations of Sp(m,R) will be referred to without qualification  as ``representations".

Proofs of this and other duality theorems were given by \textcite{KV} and 
\textcite{Howe89}.   Less mathematically sophisticated proofs of this and the other duality theorems featured in this review were also given recently 
\cite{Thms}  in which complete sets of extremal (highest or lowest weight) states were identified for dual pairs of representations.
Special cases  are proved below
to illustrate the significance of this theorem.

From Theorem 4,
(together with the unitary-unitary duality theorem), it follows that the Hilbert space of an $Nm$-dimensional harmonic oscillator carries dual representations of the pairs of groups shown as direct products in the following chains:
\be
\begin{array}{ccc}
{\rm Sp}(Nm,\Rb) &\quad\times\quad & {\rm O}(1) \\
\cup &&\cap \\
 {\rm Sp}(N,\Rb) &\quad\times\quad & {\rm O}(m) \\
\cup &&\cap \\
 {\rm U}(N) &\quad\times\quad & {\rm U}(m) \\
\cup &&\cap \\
{\rm O}(N) &\quad\times\quad & {\rm Sp}(m,\Rb)  \\
\cup &&\cap \\
{\rm O}(1) &\quad\times\quad & {\rm Sp}(Nm,\Rb) \end{array}
\label{eq:95} \ee

All the above direct product groups have realizations on the space of an $Nm$-dimensional harmonic oscillator.  This space can be viewed as that of $N$ particles in an $m$-dimensional harmonic-oscillator space or as that of 
$m$ particles in an $N$-dimensional space.   Sometimes it is useful to think of  this space as that of $Nm$ simple harmonic oscillators or of a single $Nm$-dimensional harmonic oscillator.
The many possibilities for the interpretation of $N$ and $m$ mean that the above chains of duality relationships have many applications, examples of which are explored in this section.
  
The subspace of one-quantum states of the $Nm$-dimensional harmonic oscillator is spanned by the states $\{ c^\dag_{\alpha k}|0\rangle; 
\alpha=1,\dots,N, k= 1, \dots, m\} $.
This space carries an irrep $\{1\}$ of the group 
U$(Nm)$ as studied in Sec.\ \ref{sect:V.uud}.  
Because the boson operators 
$\{ c^\dag_{ \alpha k}\}$ 
are symmetric under exchange,
the space of $W$ harmonic oscillator quanta carries an irrep 
$\{ W\}$ of U$(Nm)$.
Thus, by the Corollary to Theorem 3, the Hilbert space of the $Nm$-dimensional harmonic oscillator carries a dual representation of the 
 paired subgroups in the
direct  product ${\rm U}(N) \times {\rm U}(m)$, given by 
$ \bigoplus_\kappa \{ \kappa\} \times \{\kappa \}$.

A remarkable characteristic of the duality relationships  between the pairs of groups listed in (\ref{eq:95}) is the see-saw relationship between the paired subgroup chains.
For example, whereas  the representation of U$(N)$ is dual to that of U$(m)$ on the $Nm$-dimensional harmonic-oscillator space, the representation of the orthogonal subgroup O$(N)$ of U$(N)$ is dual to that of an Sp$(m ,\Rb)$ 
group of which U$(m)$ is a subgroup.   We shall find that such chains of dual pairs of group representations provide powerful tools for a useful classification of states of a Hilbert space.


\subsection{${\rm O}(1)\times {\rm Sp}(m,\Rb) $ duality} \label{sect: 7a}

Starting with the lowest pair in the chain and setting $N=1$, we have a dual representation of the groups O(1) and ${\rm Sp}(m,\Rb) $ on the Hilbert space of the $m$-dimensional harmonic oscillator.  
The group O(1) is a discrete group with only two elements; the identity element and an element represented as a parity inversion operator.  Thus, O(1) has two distinct one-dimensional irreps: one spanned by a state of positive parity and the other by a state of negative parity.  However, each of these O(1) irreps occurs in the $m$-dimensional harmonic oscillator space an infinite number of times.  Thus, according to Theorem 4, all the positive parity states of the $m$-dimensional harmonic oscillator carry an irrep of the non-compact group Sp$(m,\Rb)$ and all the negative parity states carry another irrep.  These ${\rm Sp}(m,\Rb) $ irreps are understood as follows.

Let $\{ c^\dag_{i},c^{i};\, i=1,\dots,m\}$ denote the creation and annihilation operators of quanta for the $m$-dimensional harmonic oscillator. 
The Lie algebra of the group Sp$(m,\Rb)$  then has a complex extension spanned by  raising and lowering operators  
\be \hat {\mathcal{A}}_{ij} = c^\dag_i c^\dag_j , \quad \hat {\mathcal{B}}_{ij} = c^i c^j ,\ee
together with the commutators of these operators
\be [\hat {\mathcal{B}}_{ij},\hat {\mathcal{A}}_{kl}] = \delta_{j,k}  c^\dag_i c^l + \delta_{j,l} c^\dag_i c^k +
\delta_{i,k}  c^l c^\dag_j+ \delta_{i,l} c^k c^\dag_j .
\ee
Thus, it has a  u$(m)$ 
subalgebra spanned by Hermitian linear combinations of the operators
\be \hat {\mathcal{C}}_{ij} = \hf \big(c^\dag_i c^j + c^jc^\dag_i \big)
= c^\dag_i c^j +  \hf \delta_{i,j}.\ee
A state $|\phi\rangle$ for an irrep of this Lie algebra that satisfies the equations
\bea \hat {\mathcal{B}}_{ij} |\phi\rangle &=& 0 , \quad  1\leq  i,j\leq m , \\
\hat {\mathcal{C}}_{ij} |\phi\rangle &=& 0 , \quad   1\leq  i <j\leq m , \\
\hat {\mathcal{C}}_{ii} |\phi\rangle &=& \lambda_i |\phi\rangle , \quad 1\leq  i \leq m ,
\eea
is then a lowest-weight state for an Sp$(m,\Rb)$ irrep with lowest weight 
$\lambda = (\lambda_1, \dots , \lambda_m)$.
Such an Sp$(m,\Rb)$ irrep is denoted by the symbol $\langle \lambda\rangle$.

It is now seen that the two Sp$(m,\Rb)$ irreps on the Hilbert space of the $m$-dimensional harmonic oscillator have lowest weight states given by the harmonic oscillator ground state $|0\rangle$ and the one-quantum state $c^\dag_1 |0\rangle$. The corresponding irreps are then denoted by $\langle (\frac12)^m\rangle$ and 
$\langle \frac32 ,(\frac12)^{m-1}\rangle$.
To simplify the notation, Sp$(m,\Rb)$ irreps are sometimes denoted  more simply by the U(1) and SU$(m)$ quantum numbers
$\langle \lambda_m (\lambda_1- \lambda_m , \lambda_2-\lambda_m, \dots )\rangle$.  Then, the above two irreps are denoted by
\be \textstyle \langle \frac12 (0)\rangle \equiv \langle (\frac12)^m\rangle , \quad
 \langle \frac12 (1)\rangle \equiv \langle \frac32, (\frac12)^{m-1}\rangle .\ee


\subsection{${\rm O}(N)\times {\rm Sp}(1,\Rb)$ duality} \label{sect: 6A}

In this section we consider the ${\rm U}(N)\times {\rm U}(m)$  and ${\rm O}(N)\times{\rm Sp}(m,\Rb)$ dualities with $m=1$.
These direct product groups then simplify,
 respectively,  to ${\rm U}(N)\times {\rm U}(1)$ 
and ${\rm O}(N)\times{\rm SU}(1,1)$ (with Sp$(1,\Rb)$ isomorphic to SU$(1,1)$).
Thus, we consider  the paired subgroups of the chains
\be \begin{array}{ccccc}
{\rm U}(N) & \times & {\rm U}(1) \\
   \cup            &&               \cap \\
{\rm O}(N)  & \times  & {\rm SU}(1,1)
\end{array}
\ee
and show that they relate to the orbital and radial dynamics of a particle in an 
$N$-dimensional space (discussed briefly for $N=3$ in Section I).

Let $\Hb$ denote the Hilbert space of the $N$-dimensional harmonic-oscillator
and let $\{ c^\dag_{\alpha},c^{\alpha};\, {\alpha}=1,\dots,N\}$ denote the
boson creation and annihilation operators of harmonic oscillator quanta. The Hilbert space $\Hb$ is then a direct sum
\be \Hb = \bigoplus_{n=0}^\infty \Hb^{(n)} \,,
\ee
where $\Hb^{(n)}\subset\Hb$ is the subspace of states of $n$  quanta.
Each $\Hb^{(n)}$ is invariant under the U(1) group whose infinitesimal
generator is the boson number operator
\be \hat n =c^\dag\cdot c = \sum_{\alpha} c^\dag_{\alpha} c^{\alpha}. 
\label{eq:hat n}\ee
Each $\Hb^{(n)}$ is also invariant and irreducible under the group U$(N)$ and hence under 
its Lie algebra, whose complex extension is spanned by  
$\{\hat C_{\alpha\beta} = c^\dag_{\alpha} c^{\beta}\}$.
Moreover, $\hat n$ commutes with all elements of  U$(N)$ and any
state of $\Hb$ having $n$ quanta belongs  to an irrep $\{n\} \times\{n\}$  of  ${\rm U}(1)
\times {\rm U}(N)$. This is a simple application of the bosonic unitary-unitary duality relationship
of the previous section.
However,  the O$(N)$ subgroup of U$(N)$ has a
more interesting  dual partner, namely SU(1,1).
Note, however, that when $N$ is odd, the representation of SU(1,1) is projective.

The relevant O$(N)$ group is  the subgroup of U$(N)$
transformations that leave the scalar product
$c^\dag \cdot c^\dag = \sum_{\alpha} c^\dag_\alpha c^\dag_{\alpha}$,
invariant.
A basis for the  Lie algebra of this O$(N)$ group, identical to the Lie algebra so$(N)$ of SO$(N)$, is given by the
operators
\be \hat L_{{\alpha\beta}} = -{\rm i} (c^\dag_{\alpha}c^{\beta}-c^\dag_{\beta}
c^{\alpha}) ,
\quad {\alpha<\beta} .
\label{eq:ONL}\ee
which are the analogues, in $N$-dimensional space, of standard angular-momentum operators.
Thus, an O$(N)$ irrep is labeled  (not always  uniquely,
as we shall see) by its
highest weight $[v_1,v_2,\dots , v_r]$ relative to the ordered basis of
the usual Cartan subalgebra of so$(N)$,
\be \hat h_{1} = \hat L_{12} \,, \quad \hat h_2 = \hat L_{34} \,, \quad \dots \,, \quad h_r = \hat L_{2r-1,2r}  \,, \label{eq:soNCartan}\ee
where  $N=2r$ or $N=2r+1$.

 Recall that in addition to the subgroup ${\rm SO}(N) \subset {\rm O}(N)$, the group O$(N)$ also contains the discrete inversion subgroup.  Thus,  an extra label is often required in addition to the so$(N)$ highest weight  to characterize the inversion properties (i.e.,  parity) of an O$(N)$ irrep.
However, when $m=1$ and $N>2$,
this extra label is not needed because the inversion properties of an O$(N)$ irrep contained within the space of the $N$-dimensional harmonic oscillator are then
uniquely defined by the  highest weight of the irrep.
For example, the parity of a single particle state 
of the 3-dimensional harmonic oscillator with SO(3) angular momentum $l$ 
is equal to $(-1)^l$.

The one-dimensional space $\Hb^{(0)}$, spanned by the vacuum  state
$|0\rangle$ (the harmonic-oscillator ground state),  carries the identity irrep [0] of
O$(N)$ as well as the identity irrep $\{0\}$ of  U$(N)$.
 The space $\Hb^{(1)}$,  which carries the defining $N$-dimensional irrep $\{1\}$ of U$(N)$, likewise carries the  $N$-dimensional irrep [1] of O$(N)$.
 The highest-weight state for the latter O$(N)$ irrep is determined from the
observation that
\be [\hat h_i,c_1^{\dag}+{\rm i}c^\dag_2] = \delta_{i,1} (c_1^{\dag}+{\rm i}c^\dag_2 ),\ee
where $\hat h_1, \hat h_2, \ldots \hat h_r $ are the so($N$) Cartan  operators of Eq.\ (\ref{eq:soNCartan}).
Thus, the state
\be |1\rangle = \frac{1}{\sqrt{2}} (c_1^{\dag}+{\rm i}c^\dag_2) |0\rangle \ee
has O$(N)$ weight $[1, 0, 0, \dots , 0]\equiv [1]$, given by the eigenvalues of the Cartan operators, and is the highest-weight state  for the O$(N)$ irrep [1].

The two-quantum space $\Hb^{(2)}$, which carries an irrep
$\{ 2\}$ of the group U$(N)$, is reducible as the carrier space for an O$(N)$ representation.
It contains an O$(N)$ highest-weight state
\be |2\rangle = \frac{1}{\sqrt{8}} (c_1^{\dag}+{\rm i}c^\dag_2)^2 |0\rangle \ee
for an irrep [2].
However, it also contains the  state $c^\dagger \cdot c^\dagger |0\rangle$ which, because
$c^\dagger\cdot c^\dagger$ is O$(N)$ invariant, spans a one-dimensional irrep [0] isomorphic to the
 irrep spanned by the state $|0\rangle$.
Continuing the pattern, the space $\Hb^{(n)}$,
 which carries the U$(N)$ irrep $\{n\}$,
contains a unique (normalized) state of maximal  O$(N)$ highest weight
given by
\be |n\rangle = \frac{1}{\sqrt{2^n n!}} (c_1^{\dag}+{\rm i}c^\dag_2)^n
|0\rangle . \label{eq:ONhwtstates}\ee
 This state is an eigenstate of the so$(N)$ Cartan operators, 
\bea \hat h_i|n\rangle &=& \frac{1}{\sqrt{2^n n!}} [\hat h_i,(c_1^{\dag}+{\rm i}c^\dag_2)^n ]|0\rangle  \nonumber\\
&=& \delta_{i,1} n | n\rangle , \quad i = 1, \dots,r , \eea
and  the highest-weight state for the O$(N)$ irrep $[n]$.

However, there are other O$(N)$ irreps in $\Hb^{(n)}$ as indicated by the branching rule
\be {\rm U}(N) \downarrow {\rm O}(N) \, :
\, \{ n\} \downarrow [n] \oplus [n-2] \oplus [n-4] \oplus \cdots
\oplus [1] \; {\rm or}\; [0]  .
\label{eq:bruo}\ee
The highest-weight states of these O($N$) irreps are given, to within norm factors, by
\bea |n\rangle \,,\quad &&  (c^\dag\cdot c^\dag) |n-2\rangle \,,
\quad (c^\dag\cdot c^\dag)^2 |n-4\rangle \,, \quad  \nonumber \\
&&  \dots \,, \quad
(c^\dag\cdot c^\dag)^{k} |n-2k\rangle\,, \quad \dots \, ,  \label{eq:169}\eea
thus  illustrating how the reduction of a U$(N)$ irrep on restriction to O$(N)$ is obtained by factoring out  O$(N)$ scalars.

\begin{figure}
 \includegraphics[scale=0.45]{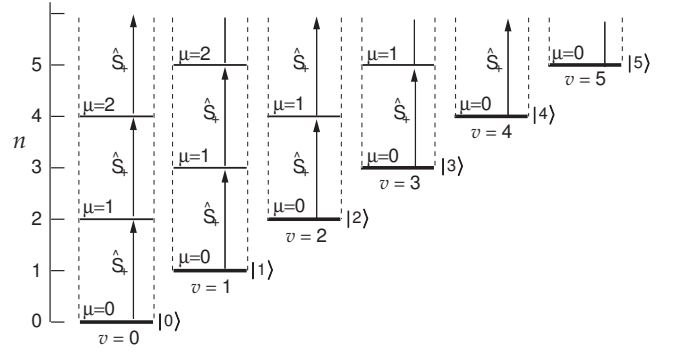}%
 \caption{\label{fig:HOirreps}  Irreps of O$(N)$ in the Hilbert space 
      of the  $N$-dimensional harmonic
      oscillator shown as horizontal lines.
      Equivalent O$(N)$ irreps are placed in a common column and
      connected by the ${\rm su(1,1)}$ raising operator
      $\hat{S}_+$.  Equivalent O$(N)$ irreps are distinguished by the
      ${\rm u(1)} \subset {\rm su(1,1)}$  quantum number $\mu$.
      The set of O$(N)$ irreps at a constant level $n$
      comprise a U$(N)$ irrep $\{n\}$ with $n=v+2\mu$.   
      The carrier space of this 
U$(N)$ irrep $\{n\}$ is $\Hb^{(n)}$,  the space spanned by the states with 
$n$ harmonic  oscillator quanta. 
}
 \end{figure}

 The decomposition of the harmonic-oscillator space $\Hb$ is illustrated in Fig.\ \ref{fig:HOirreps}.
Each horizontal level corresponds to one of the 
subspaces, $\Hb^{(n)}$,   labeled 
 by $n$ on the vertical axis.  The decomposition of 
$\Hb^{(n)}$ into irreducible O$(N)$ subspaces is shown by the 
horizontal line segments,  each of which 
represents  an O$(N)$ irrep, $[v]$, characterized by a value of $v$ which, for each value of $n$, takes either even or  odd integer values between $n$ and zero in accordance with the branching rule (\ref{eq:bruo}).
Equivalent O$(N)$ irreps, i.e., irreps sharing a common value of $v$, are
placed one above the other in a column labeled at the bottom
by the value of $v$.
It can be seen that the pattern of O$(N)$ irreps obtained in this way is independent of $N$.
For example, Fig.\ \ref{fig:HOirreps} gives the familiar  spectrum of O(3) irreps of the three-dimensional
harmonic oscillator.

Now observe  that the operator $c^\dag\cdot c^\dag$ is the raising operator of an su$(1, 1)$ Lie algebra,  whose complex extension is
 spanned by the O$(N)$-invariant operators
\bea &\hat S_+= \half c^\dag\cdot c^\dag , \quad \hat S_- = \half c\cdot c, &\nonumber\\
&\hat S_0= \frac{1}{4} (c^\dag\cdot c + c\cdot c^\dag) = \half\left(\hat n
+\frac{N}{2}\right) ,&
\label{eq:170}\eea
where $\hat n$ is the number operator for harmonic oscillator quanta, as defined by Eq.\ (\ref{eq:hat n}).
These su(1,1) operators satisfy the commutation relations
\be [\hat S_-,\hat S_+] = 2\hat S_0 \,,\quad [\hat S_0,\hat S_\pm] =
\pm\hat S_\pm \,. \label{SU11CR}\ee
 Observe also that the states 
$\{ |v\rangle; v= 0,1,2,\dots\}$, defined (with $v=n$)
by Eq.\ (\ref{eq:ONhwtstates}), are all annihilated by the su$(1,1)$ lowering operator.
 Thus, the state $|v\rangle$ is simultaneously of highest U$(N)$ weight $\{ v\}$, of O$(N)$ highest weight $[v]$, and of SU$(1,1)$ lowest weight
 $\langle \lambda(v)\rangle$, where $ \lambda(v) = v+N/2$ is an eigenvalue of 
 $2\hat S_0$,  i.e., $| v \rangle$ satisfies the equations
\be \hat S_- |v\rangle = 0 , \quad
\hat S_0 |v\rangle = \hf \lambda(v)\, |v\rangle =\hf( v+ N/2 )\,  |v\rangle \,. \ee
By construction, the  O$(N)$ irrep with highest-weight state  $|v\rangle$ 
lies lowest  on a column of equivalent O$(N)$ irreps (cf.\ Fig.\ \ref{fig:HOirreps}).
Moreover, it is seen that the highest-weight
states for the successively higher O$(N)$ irreps of a column
 are given, to within norm factors, by the states
\be |v\rangle ,\;\; (c^\dag\cdot c^\dag) |v\rangle ,\;\;
 (c^\dag\cdot c^\dag)^2 |v\rangle , \; \dots 
(c^\dag\cdot c^\dag)^\nu |v\rangle,\; \dots  \ee
Thus, the set of all O$(N)$ highest-weight states in a column span an irrep
 $ \langle \lambda(v)\rangle$ of SU(1,1) and
all the states of a column  span an irrep
$ \langle \lambda(v)\rangle \times [v]$ of the direct product group
${\rm SU}(1,1)\times {\rm O}(N)$ labeled by $v$, with $v$ taking the values $0, 1, 2, \dots$ in
successive columns.

 Finally, observe that every irrep $\langle \lambda(v)\rangle \times [v]$ of ${\rm
SU}(1,1)\times {\rm O}(N)$ is multiplicity free and every irrep $\langle \lambda(v)\rangle$ of
SU(1,1)  is uniquely paired with an irrep $[v]$ of O$(N)$. These are the properties required
to demonstrate the duality of
 ${\rm SU}(1,1)$ and  ${\rm O}(N)$ representations on
the  Hilbert space $\Hb$ of the $N$-dimensional harmonic oscillator space.


\subsection{Applications of ${\rm O}(N)\times{\rm SU}(1,1)$ duality}
 \label{sect:7B}

The duality of the representations of O$(N)$ and SU$(1,1)$ on the Hilbert space $\Hb$ of the $N$-dimensional harmonic oscillator leads to many useful relationships in physics \cite{SU11methods}. This is because  $\Hb$, the
space $\mathcal{L}^2(\Rb^N)$ of square integrable functions on the real $N$-dimensional Euclidean space $\Rb^N$, is also the Hilbert space for numerous other systems of interest.

\subsubsection{Central-force problems}

Suppose, for example, that $\hat H$ is an O$(N)$-invariant Hamiltonian on $\Hb$ that one wishes to diagonalize in a harmonic oscillator basis. Such a basis 
is defined by the irrep labels of the subgroup chain
\be \begin{array}{ccccccc}
{\rm U}(N) &\supset& {\rm O}(N) &\supset & {\rm SO(3)} &\supset&  {\rm SO(2)}
\\
 n && v & \rho& L&&M
\end{array} ,\label{eq:UNON}
\ee
where we have assumed that $N\geq 3$, so that the 
rotation group SO(3) can be defined as 
 a subgroup of O$(N)$, and $\rho$ is an
additional quantum number to distinguish any multiplicity of SO(3)
irreps that occur within a given irrep $[v]$ of O$(N)$. As a result
of the ${\rm O}(N)\times {\rm SU}(1,1)$ duality, the basis states
$\{|nv\rho LM\rangle\}$ also reduce the subgroup chain \be
\begin{array}{ccccccc} {\rm SU}(1,1) \times {\rm O}(N) &\supset&
{\rm U(1)}\times {\rm SO(3)} &
\supset& {\rm SO(2)} \\
\quad\lambda \qquad\quad v &\rho & \mu \;\;\qquad L & & M
\end{array} \label{eq:SU11ONchain}
\ee
where ${\rm U}(1)$ is the subgroup of SU$(1,1)$ with infinitesimal generator $\hat S_0$.  
The above results show that the quantum numbers  of the two chains are related by
\be \lambda = v+\hf N , \quad  n = v+2\mu ,\ee
where $v$ and $\mu$ run over all non-negative integer values.

 The classification of states by the irrep labels of the 
 subgroup chain (\ref{eq:SU11ONchain}) is particularly  useful  because the direct product structure of the two commuting groups  SU$(1,1)$ and  O$(N)$ makes it  natural to
 exploit the factorization of the wave functions for basis states into products of radial and orbital wave functions.
 Moreover, when the Hamiltonian is O$(N)$ invariant, as it is for a generalized central-force problem, the orbital wave functions are simply SO$(N)$ spherical harmonics \cite{RTR},  and the radial wave functions are eigenfunctions of a one-dimensional Schr\"odinger equation.
 In fact, as a consequence of the  ${\rm O}(N)\times {\rm SU}(1,1)$ duality, the spectral
properties of many central-force Hamiltonians can be derived by
algebraic methods using an su$(1,1)$ Lie algebra as a spectrum generating algebra \cite{SU11methods}.
This is seen by expressing the harmonic oscillator raising and lowering
operators in terms of Cartesian coordinates $\{ x_i; i=1,\dots , N\}$  for 
$\Rb^N$:
\be c^i= \frac{1}{\sqrt{2}} \left( ax_i +
\frac{1}{a}\frac{\partial}{\partial x_i} \right) , \quad
c^\dag_i = \frac{1}{\sqrt{2}} \left( ax_i - \frac{1}{a}\frac{\partial}{\partial
x_i}
\right),\ee
where $a$ is an inverse unit of length.
The su$(1,1)$ operators are then obtained in the form
\bea \hat S_\pm  &=&  {\textstyle\frac{1}{4}} \left[ a^{-2} \,\nabla^2 + a^2r^2 \mp\big(
{\bf r}\cdot\nabla +
\nabla \cdot {\bf r}\big) \right] , \label{eq:S+-}\\
\hat S_0  &=& {\textstyle\frac{1}{4}} \left[ -a^{-2} \,\nabla^2 + a^2r^2 \right] ,
\label{eq:S0}
\eea
where $r^2  ={\bf r} \cdot {\bf r} =
 \sum_i x_i^2$ and $\nabla^2 = \sum_i \partial^2/\partial x_i^2$ is the Laplacian on ${\cal L}^2(\Rb^N)$.

The representations of SU$(1,1)$ have been well studied,
cf., for example, \cite{CP,Wyb,RB98,SU11methods}.
The harmonic series of su$(1,1)$ 
irreps  are given by the equations
\be \begin{array}{ccc}
\hat S_0 |\lambda \mu\rangle = \half (\lambda+2\mu)|\lambda \mu\rangle\,,\\
\phantom{\Big|} \hat S_+ |\lambda \mu\rangle = \sqrt{(\lambda +\mu)(\mu+1)}\,  |\lambda
,\mu+1\rangle\,, \\
 \hat S_- |\lambda \mu\rangle = \sqrt{(\lambda +\mu-1)\mu}\, |\lambda
,\mu-1\rangle\,,
\end{array}  \label{eq:S0+-}
\ee
with $\lambda = v+ N/2$ and $v=0, 1, 2, \dots$.
 The operator $\hat S_0$ defined by Eq.\ (\ref{eq:S0}) is 
proportional to a harmonic oscillator Hamiltonian (\ref{eq:Vi.Hho}). 
Moreover,  $\nabla^2$ and $r^2$ are elements of the su$(1,1)$ Lie algebra and expressible as linear combinations of  $\hat S_\pm$ and $\hat S_0$ with known matrix elements in an su$(1,1)\supset {\rm u}(1)$ coupled basis.
Matrix elements of potential-energy  functions of the form
\be
V(r) = \sum_i P_i(r) e^{-\alpha_i r^2}  , \label{eq:V(r)}\ee
where each $P_i$ is an even polynomial, can also be computed with relative ease in such a basis.
 Thus, the dynamical group SU$(1,1)$ very
much simplifies the computation of matrix elements of a Hamiltonian
\be \hat H = -\frac{\hbar^2}{2M}\nabla^2 + V(r), \ee
when the potential $V(r)$ is of the form (\ref{eq:V(r)}).

Applications of SU$(1,1)$ as a spectrum generating algebra for central-force problems have been considered by many authors; e.g., \textcite{Wyb}, \textcite{CP} or, for a pedagogical review, \textcite{CW}.
The fact that these applications are valid for any positive integer $N$ and many discrete series irreps of SU$(1,1)$ makes them particularly useful.
The development of factorization methods and algebraic  methods for the evaluation of  matrix elements of SO$(N)$ tensor operators makes it possible to extend the algebraic method to a much wider class of Hamiltonian 
\cite{SU11methods}.
By such means, SU(1,1) $\times$ SO(5) has been used successfully as a dynamical group in the development of an algebraic version of the Bohr model for collective quadrupole vibrations and rotations in nuclear
physics \cite{RTR,ACM1,RT05,RoweWC09,WelshR12}.

\subsubsection{States for a six-dimensional harmonic
oscillator} \label{sect:6dimHO}

The Hilbert space of the six-dimensional harmonic oscillator, $\mathcal{L}^2(\Rb^6)$, is a tensor product,
$\mathcal{L}^2(\Rb^5)  \otimes \mathcal{L}^2(\Rb)$, of Hilbert spaces
for five-dimensional and one-dimensional harmonic oscillators. In the spirit of the
interacting boson model \cite{IBM}, to which this example applies, we may consider
the boson creation operators of these harmonic oscillators as comprising five
$\{ d^\dag_m; m=0,\pm 1,\pm 2\}$ operators, which create
harmonic-oscillator quanta
of angular momentum $L=2$ in $\mathcal{L}^2(\Rb^5)$, and
one $s^\dag$ operator, which creates  $L=0$ (angular-momentum-zero)
quanta
 in $\mathcal{L}^2(\Rb)$.
We then have the following dual pairs of group representations:
\begin{widetext}
\be \begin{array}{cccccccccccccc}
\mathcal{L}^2(\Rb^6) &= & \mathcal{L}^2(\Rb^5) &\otimes& \mathcal{L}^2(\Rb) \\
\\
\begin{array}{ccc}
{[}{\rm U}(6) &\times& {\rm U}(1)_6] \\
\cup && \cap\\ {[}{\rm O}(6) &\times& {\rm SU}(1,1)_6]
\end{array}
&&
\begin{array}{ccc}
{[}{\rm U}(5) &\times& {\rm U}(1)_5\,] \\
\cup && \cap\\ {[}{\rm O}(5) &\times& {\rm SU}(1,1)_5\,]
\end{array}
&&\begin{array}{ccc}
{[}{\rm U}(1)_1 &\times& {\rm U}(1)_1\,] \\
\cup && \cap\\ {[}{ {\rm O}(1)}_1 &\times& { {\rm SU}(1,1)_1 }\,]
\end{array}
\end{array}
\ee
\end{widetext}
Note that, for logical consistency, we show a direct product of two copies of
${\rm U}(1)_1$ although, in fact, the representations of these two groups
are dual to each other in a trivial way, i.e., they are identical.
Note also that, because the irreps of one member of a dual pair
are uniquely partnered with corresponding irreps of the other, we need only to specify the representations for one member of each pair.
Thus, we label the paired irreps by common indices as follows:
\be \begin{array}{cccccc}
{\rm U}(6) \!\times\! {\rm U}(1)_{6} & {\rm U}(5) \!\times\! {\rm U}(1)_5 & {\rm U}(1)_1
\times {\rm U}(1)_1\\ N& n& N-n\\ \\
{\rm O}(6) \!\times\! {\rm SU}(1,1)_6 & {\rm O}(5) \!\times\! {\rm SU}(1,1)_5 &
{\rm O}(1)_1 \!\times\!  {\rm SU}(1,1)_1 \\
\sigma & v& \pi
\end{array}
\ee
Note that there are only two irreps of SU$(1,1)_1$ on the one-dimensional harmonic-oscillator Hilbert space: one  carried by  states of even numbers of oscillator quanta and one by states of odd numbers of quanta.
 The discrete group O(1) also has only two irreps of parity $\pi = \pm 1$.
In $\mathcal{L}^2(\Rb)$, the even and odd boson number SU$(1,1)_1$
 irreps are paired with the $\pi =1$ and $ \pi=-1$ irreps of  O(1)$_1$, 
respectively.

The \emph{interacting boson model} has exactly solvable limits in which particular classes of Hamiltonians are diagonalized in bases that reduce corresponding subgroup chains.
In particular, there are two exactly solvable classes of O(5)-invariant Hamiltonians: one that diagonalizes the chain
\be \begin{array}{ccccccccc}
{\rm U}(6) &\supset& {\rm U}(5)\times {\rm O}(1)_1 &\supset & {\rm O}(5) &\supset &{\rm SO(3)}
&\supset& {\rm SO(2)} \\
N && n \qquad\quad \pi && v&\rho& L&&M
\end{array} \label{eq:U5}
\ee
and another that diagonalizes the chain
\be \begin{array}{ccccccccccccccc}
{\rm U}(6) &\supset& {\rm O}(6) &\supset & {\rm O}(5)\times {\rm O}(1)_1 &\supset &{\rm SO(3)}
&\supset& {\rm SO(2)} \\
N && \sigma && v\qquad\quad \pi&\rho& L&&M
\end{array}\label{eq:O6}
\ee
By using the duality relationships indicated above, we find that
the basis states $\{ |Nn\pi v\rho LM\rangle\}$ that reduce the subgroup
chain (\ref{eq:U5}) simultaneously reduce the chain
\begin{widetext}
\be \begin{array}{cccccccccccccccc}
{\rm SO}(3)&\times&{\rm SU}(1,1)_1 &\times& {\rm SU}(1,1)_5  &\supset &
{\rm SO}(2)&\times&{\rm U}(1)_1 &\times& {\rm U}(1)_5 &\supset & {\rm U}(1)_6 \\
L &&\pi\;\; && \lambda=v+\frac{5}{2} &&\, M && N-n &&n&&N
\end{array}\label{eq:189}
\ee
and the basis states $\{ |N\sigma v\pi\rho LM\rangle\}$ that reduce the
subgroup chain (\ref{eq:O6}) simultaneously reduce the chain
\be \begin{array}{cccccccccccc}
{\rm SO}(3)&\times&{\rm SU}(1,1)_1 &\times& {\rm SU}(1,1)_5  &\supset &
{\rm SO}(2) &\times &{\rm SU}(1,1)_6 &\supset & {\rm U}(1)_6  \\
L&&\pi\;\; && \lambda=v+\frac{5}{2}
&& M &&  \kappa =  \sigma +3\quad && N
\end{array}
\label{eq:190} \ee
\end{widetext}
where $ {\rm SU}(1,1)_6$ is the subgroup of 
${\rm SU}(1,1)_1 \times {\rm SU}(1,1)_5$ whose Lie algebra is spanned
by the sums $\hat S^6_k = \hat S^1_k + \hat S^5_k$,
where $\{\hat S^1_k\}$ and $\{\hat S^5_k\}$ are bases for
SU(1,1)$_1$ and SU(1,1)$_5$, respectively.

The equivalence of alternative subgroup chains for classifying basis states  means that one can choose whichever is the simplest for the purposes of diagonalizing a corresponding Hamiltonian.
Thus, although  the subgroup chains (\ref{eq:U5}) and (\ref{eq:O6})
are natural dynamical subgroup chains for diagonalizing an O(5)-invariant Hamiltonian on a U(6)-invariant subspace of the six-dimensional harmonic oscillator, it is generally much easier to use their SU$(1,1)$ counterparts in (\ref{eq:189}) and (\ref{eq:190}) \cite{Rowe04}.

It is also worth noting that one would sometimes like to know the unitary transformation between the  U(5) and O(6) bases,
i.e., the bases that, respectively, diagonalize the
subgroups chains of Eqs.\ (\ref{eq:U5}) and (\ref{eq:O6}). 
The transformation coefficients are  the $\rho LM$-independent overlaps
\be \langle Nnv\pi\rho LM|N\sigma v\pi\rho LM\rangle = \langle Nnv|N\sigma v\rangle.
\ee
A knowledge of these coefficients immediately gives an algebraic expression 
for matrix elements of any mixture of Hamiltonians that are known in either the U(5) or O(6) bases.
These coefficients can be computed with some effort by diagonalizing the O(6) Casimir operator in the U(5) basis as shown by \textcite{Castanos}. 
However, as the equivalent SU$(1,1)$ chains reveal \cite{U5toO6, RoweTh05}, they are simply equal to already-known  
${\rm SU(1,1)}$ Clebsch-Gordan coefficients:
in the $(k_1,n_1; k_2,n_2 |k,n)$ notation of \textcite{JVJ}
\be \langle Nnv|N\sigma v\rangle 
= \textstyle
\Big(\frac{v+\frac{5}{2}}{2},\frac{n-v}{2}; \frac{1}{4},\frac{N-n}{2}\, \Big|
\frac{\sigma+3}{2}, \frac{N-\sigma}{2} \Big) , 
\ee 
 for $\; (N-v) \;$ even, and 
\be \langle Nnv|N\sigma v\rangle = \textstyle
\Big(\frac{v+\frac{5}{2}}{2}, \frac{n-v}{2}; \frac{3}{4},\frac{N-n-1}{2}\, \Big|
\frac{\sigma+3}{2}, \frac{N-\sigma}{2}\Big) , \qquad\ee
for $\; (N-v)\;$ odd.

As considered further in Section X, 
this example is a prototype of many possible uses of dual subgroup chains to relate the basis states of one coupling scheme for a many-particle system to those of another.  Thus, it can provide solutions to challenging problems, e.g.,  in the calculation of matrix elements of a Hamiltonian that contains mixtures of interactions that are diagonal in different coupling schemes.

An extension of the O(5)-invariant interacting boson model, that benefits even more from these duality relationships, is a model  with mixed U(6) irreps, proposed by \textcite{Lehmann95}.
In this model, spherical neutron states mix with deformed states generated by the excitation of a proton pair into the active shell-model space from an otherwise inert closed subshell.
The spherical states of this model  are  classified by the subgroup chain 
\bea 
&\begin{array}{ccccccccc}
{\rm U(6)} & \supset & {\rm U(5)} & \supset & {\rm O}(5) \\
N && n_5 && v 
\end{array} ,&
\eea
and the deformed states, with the addition of two excited protons, are classified by the chain
\bea &\begin{array}{ccccccccc}
{\rm U(6)} & \supset & {\rm O(6)} & \supset & {\rm O}(5) \\
N+2 && \sigma && v 
\end{array} ,&
\eea
with the inclusion of multiplicity labels as needed.
These states are equivalently classified by the dual subgroup chains
\bea 
\!\!\!\!\!\! &\begin{array}{ccccccccc}
\!\!\!\!\!{\rm SU}(1,1)_1 &\!\!\!\!\times\!\!\!\!\!& {\rm SU}(1,1)_5 & \!\! \supset \!\!\!& 
{\rm U}(1)_1 &\!\!\!\!\!\!\!\!\times\!\!\!\!\!\!\!\!& {\rm U}(1)_5 &\!\!\!\supset\!\!\! \!& {\rm U}(1)_6\\
\pi &&\!\lambda= v+\frac{5}{2}\!\! && \!\!N-n_5 && n_5 &&N 
\end{array} \!\!,\quad&\\
&\begin{array}{ccccccccc}
\!\!\!\!\!\!\!{\rm SU}(1,1)_1 &\!\!\!\times\!\!\!& {\rm SU}(1,1)_5 &\!\! \!\supset\! \!\!& 
{\rm SU}(1,1)_6  &\! \!\!\supset\! \!\!&  {\rm U}(1)_6 \\
\pi && \lambda= v+\frac{5}{2}  &&\sigma &&  N+2 
\end{array} .&
\eea
The mixing of  these states is then simply described by an interaction of the form
\be \hat V = \alpha (\hat S^{(1)}_+ + \hat S^{(1)}_-)
+ \beta (\hat S^{(5)}_+ + \hat S^{(5)}_-), \ee
 where $\hat S^{(1)}_\pm$ and $\hat S^{(5)}_\pm$ are, respectively, raising and lowering operators for SU(1,1)$_1$ and SU(1,1)$_5$.

\subsubsection{The vibron model  and its q-deformed extension} 
\label{sect:4dimHO}

A parallel application by \textcite{AlvarezBS94} of the 
${\rm O}(N)\times {\rm SU}(1,1)$ duality is to the classification of states of a four-dimensional harmonic oscillator used in the vibron model of the vibrations and rotations of diatomic molecules \cite{Iachello81}.
In this application, the Hilbert space of the four-dimensional harmonic oscillator
is  regarded as a tensor product
$\mathcal{L}^2(\Rb^4)  \simeq \mathcal{L}^2(\Rb^3) \otimes \mathcal{L}^2(\Rb)$, of Hilbert spaces for three-dimensional and one-dimensional harmonic oscillators.
Thus, it is determined that basis states for this model that reduce the subgroup chain
\be 
{\rm U}(4) \supset {\rm U}(3)\times {\rm U}(1) \supset  
{\rm O}(3) \times {\rm O}(1) \supset {\rm SO(2)} \label{eq:U3}
\ee
simultaneously reduce the dual  chain
\be 
{\rm SO}(2)\times{\rm SU}(1,1)_1 \times {\rm SU}(1,1)_3  \supset 
{\rm U}(1)_1 \times {\rm U}(1)_3 \supset  {\rm U}(1)_4 . \label{eq:DualU3}
\ee
Similarly, basis states that reduce the subgroup chain
\be 
{\rm U}(4) \supset {\rm O}(4) \supset  {\rm O}(3)\times {\rm O}(1) 
\supset {\rm SO(2)} \label{eq:O4}
\ee
 simultaneously reduce the dual  chain
\be 
{\rm SO}(2)\times{\rm SU}(1,1)_1 \times {\rm SU}(1,1)_3  \supset 
 {\rm SU}(1,1)_4 \supset  {\rm U}(1)_4 . \label{eq:DualO4}
\ee

This  formulation of the vibron  model was used by \textcite{AlvarezBS94} to show that,  because the su(1,1) Lie algebra has a known q deformation 
\cite{KulishR83} to a so-called \emph{quantum algebra},  the vibron model also has a q-deformed extension.  A similar observation  applies to the interacting boson model in its U(5) and O(6) dynamical symmetry limits.


\subsection{${\rm O}(N)\times {\rm Sp}(m,\Rb)$ duality} \label{sect: ONSpm}

We now consider applications of the general ${\rm O}(N)\times{\rm Sp}(m,\Rb)$
duality relationship to the dynamics of a system of $N$ particles in an $m$-dimensional configuration space.

Let $\{ c^\dag_{\alpha i}; \alpha = 1,\dots,N; i = 1,\dots,m\}$
and $\{c^{\alpha i}; \alpha = 1,\dots,N; i = 1,\dots,m\}$
denote boson creation and annihilation
operators for an $Nm$-dimensional harmonic oscillator.
The unitary groups U$(N)$ and U$(m)$  then have dual representations
 on the Hilbert space of this harmonic oscillator with infinitesimal generators 
defined, respectively, by
\bea
\hat{C}^{(N)}_{\alpha\beta}  = \sum_{i=1}^m c^\dagger_{\alpha i}c^{\beta i},
\quad
\hat{C}^{(m)}_{ij}  = \sum_{\alpha=1}^N c^\dagger_{\alpha i}c^{\alpha j} ,
\label{eq:UNm}
\eea
cf.\ Eq.\ (\ref{eq:3,bosonUmn}).
The group O$(N)$ is the subgroup of all real orthogonal transformations in 
U$(N)$ that leave the scalar products 
\be \hat{\cal A}_{ij} = c^\dag_i \cdot c^\dag_j = \sum_\alpha c^\dagger_{\alpha i}
c^\dagger_{\alpha j}, \quad i,j = 1,\dots,m ,  \ee
invariant.
This group contains all SO$(N)$ rotations, for which infinitesimal generators are given by the generalized angular-momentum operators
\be \hat L_{\alpha\beta} = -{\rm i}
(\hat{C}^{(N)}_{\alpha\beta}-\hat{C}^{(N)}_{\beta\alpha}) ,\quad \alpha < \beta ,
\ee
and an inversion operator, that maps all the boson creation and annihilation operators to their negatives.
The  group Sp$(m,\Rb)$, dual to O$(N)$, is a simple Lie group for which infinitesimal generators are given by Hermitian linear combinations of
the O$(N)$ scalar operators
\be \begin{array} {c} \displaystyle
\hat{\cal A}_{ij} =
\sum_\alpha c^\dagger_{\alpha i} c^\dagger_{\alpha j} , \quad 
 \hat{\cal B}_{ij} = \sum_\alpha
c^{\alpha i} c^{\alpha j} ,\\
\hat{\cal C}_{ij}
= \hat{C}^{(m)}_{ij} + \hf N  \delta_{i,j} , 
\end{array}
 \label{eq:spmR}\ee
for $ i,j = 1,\dots,m$. 
Note that $\hat{\cal C}_{ij}$ and $\hat{C}^{(m)}_{ij} $, related as in  
(\ref{eq:spmR})§, are infinitesimal generators of isomorphic U($m$) groups. 

The Hilbert space of any $Nm$-dimensional harmonic oscillator carries a direct
sum of Sp$(m,\Rb)$ irreps known as positive harmonic series, which are irreps with lowest but not highest weights.
An Sp$(m,\Rb)$ irrep   on an $Nm$-dimensional harmonic oscillator space is therefore conveniently characterized by a lowest-weight state $|\lambda\rangle $ which satisfies the equations
\bea &\hat{\cal B}_{ij} |\lambda\rangle = 0, \quad 1 \leq i \leq j \leq m ,
 & \label{eq:splwt1}\\ 
&\hat{\cal C}_{ij} |\lambda\rangle = 0 ,\quad 1\leq i < j\leq m ,
\label{eq:splwt2}\\
&\hat{\cal C}_{ii} |\lambda\rangle = (\lambda_i + \frac12 N)|\lambda\rangle ,
\quad i = 1, \dots , m.\label{eq:splwt3}\eea
To simplify the notation, we denote such an irrep with lowest weight  
$(\lambda_1+\hf N, \lambda_2+\hf N, \dots , \lambda_m+\hf N )$
by $\langle \frac12 N (\lambda)\rangle$.
Note that the Sp$(m,\Rb)$ lowest-weight state $|\lambda\rangle$, defined in this way, is also the highest-weight state for an irrep $\{ \lambda\}$ of the U$(m)$ group defined by Eq.\ (\ref{eq:UNm}).

Several copies of the Sp$(m,\Rb)$ irrep $\langle \frac12 N (\lambda)\rangle$ appear in the $N$-particle, $m$-dimensional harmonic oscillator space.
However, because of the unitary-unitary duality relationship, Theorem 3, the 
U$(m)$ highest-weight  state  $|\lambda\rangle$  can be made unique  by requiring that, in addition to satisfying Eqs.\ (\ref{eq:splwt1})--(\ref{eq:splwt3}), it is also a U$(N)$ highest-weight state, i.e., it satisfies  the equations
\bea \hat{C}^{(N)}_{\alpha\beta} |\lambda\rangle &=& 0, \quad \alpha <\beta ,\\
\hat{C}^{(N)}_{\alpha\alpha} |\lambda\rangle &=& \lambda_\alpha  |\lambda\rangle, \quad
\alpha = 1, \dots, N ,
\eea
where $\lambda_\alpha = 0$ for $\alpha > m$.

There is no duality relationship between the irreps of 
Sp$(m,\Rb)$ and those of U$(N)$ because the irrep of U$(N)$ with highest weight $\lambda$ also contains states that are not of Sp$(m,\Rb)$ lowest weight, i.e., states that are not annihilated by the $\hat{\cal B}_{ij}$ lowering operators. 
However, because the $\hat{\cal B}_{ij}$ lowering operators are O$(N)$-invariant, the subset of states of the U$(N)$ irrep of highest weight $\lambda$ that are also of Sp$(m,\Rb)$ lowest weight do carry a representation of  
O$(N)$.
Moreover,  by  Theorem 4, such an  O$(N)$ representation is irreducible. 
Thus, the state $|\lambda\rangle$ is the highest-weight state for an O$(N)$ irrep 
and, simultaneously, a lowest-weight state for a dual  Sp$(m,\Rb)$ irrep.
This can be seen, for $m=1$, in Fig.\ \ref{fig:HOirreps} which shows that only the right-most  O$(N)$ irrep belonging to a single U$(N)$ irrep lies at the bottom of a column of equivalent O$(N)$ irreps that together span an Sp$(1,\Rb)$ irrep.

\subsection{Applications of ${\rm O}(N)\times {\rm Sp}(m,\Rb)$ duality} \label{sect: impONSpm}

\subsubsection{Relationships between branching rules}

The following example shows how the see-saw relationship between the
unitary-unitary and orthogonal-symplectic dual pairs
\be \begin{array}{ccc}
 {\rm U}(N) &\quad\times\quad & {\rm U}(m) \\
\cup &&\cap \\
{\rm O}(N) &\quad\times\quad & {\rm Sp}(m,\Rb) \end{array} \label{eq:seesaw}
\ee
is used to determine  ${\rm Sp}(m,\Rb) \downarrow  {\rm U}(m)$
branching rules  from known ${\rm U}(N) \downarrow {\rm O}(N)$ branching rules.
This is important for two reasons: one is that a knowledge of the 
${\rm Sp}(m,\Rb) \downarrow  {\rm U}(m)$ branching rules is needed for nuclear shell-model calculations in an 
${\rm Sp}(3,\Rb) \supset{\rm U}(3) \supset {\rm SO}(3)$ coupling scheme, appropriate for the microscopic theory of nuclear collective states; a second is that it serves as a prototype of ways to infer branching rules for a non-compact group from those of a compact group.

Note that the ${\rm U}(m) \subset {\rm Sp}(m,\Rb) $ subgroup,  
defined by Eq.\ (\ref{eq:splwt3}), has  infinitesimal generators
$\{\hat{\mathcal{C}}_{ij}= \hat{C}^{(m)}_{ij} + \hf N  \delta_{i,j}\}$
that differ from the infinitesimal generators  $\{\hat{C}^{(m)}_{ij}\}$
of the U$(m)$ group defined by Eq.\ (\ref{eq:UNm}).  The commutation relations of the $\{\hat{\mathcal{C}}_{ij}\}$ and $\{\hat{C}^{(m)}_{ij}\}$ operators are exactly the same, but they generate different, although simply related, representations when acting on the same states.  
Thus, when acting on a ${\rm U}(m) $ highest weight state 
$|\lambda\rangle$, defined by the equations
\bea 
&\hat{C}^{(m)}_{ij} |\lambda\rangle = 0 ,\quad 1\leq i < j\leq m ,
\label{eq:Umhwt1}\\
&\hat{C}^{(m)}_{ii} |\lambda\rangle = \lambda_i|\lambda\rangle ,
\quad i = 1, \dots , m ,\label{eq:Umhwt2}\eea
the ${\rm U}(m) \subset {\rm Sp}(m,\Rb) $ operators satisfy Eqs.\ (\ref{eq:splwt2}) and 
(\ref{eq:splwt3}).  Thus, they generate a U$(m)$ irrep with a \emph{shifted} 
highest weight $\lambda^{(N)}$ having components
$\lambda^{(N)}_i = \lambda_i + \hf N$.

According to the unitary-unitary duality theorem, the Hilbert space $\Hb$ of the $Nm$-dimensional harmonic oscillator
carries a direct sum   $\bigoplus_\lambda \{\lambda\} \times \{\lambda^{(N)}\}$ 
of ${\rm U}(N)\times {\rm U}(m)$ irreps.
From orthogonal-symplectic duality,  $\Hb$ also carries a direct sum 
$\bigoplus_\lambda [ \lambda ] \times \langle \half N(\lambda)\rangle$
of ${\rm O}(N)\times{\rm Sp}(m,\Rb)$ irreps.
 To relate the  ${\rm U}(N)\downarrow{\rm O}(N)$ and
 ${\rm Sp}(m,\Rb) \downarrow  {\rm U}(m)$
branching rules, we express them in the form
\bea &{\rm U}(N) \downarrow {\rm O}(N) \, ; \, \{ \lambda\} \downarrow
\bigoplus_\kappa  P_{\lambda\kappa} [\kappa] ,& \\
&{\rm Sp}(m,\Rb) \downarrow {\rm U}(m)
 \, ; \, \langle \hf N (\kappa) \rangle \downarrow
\bigoplus_\lambda
 \mathcal{P}_{\kappa\lambda}  \{ \lambda^{(N)} \}  .&
\eea
The ${\rm O}(N)\times {\rm U}(m)$ representation carried by
$\Hb$ can now be expressed as a direct sum of irreps in two ways.
One branching rule gives
\bea {\rm U}(N)\times {\rm U}(m) &\downarrow& {\rm O}(N)\times {\rm U}(m)\, ; \nonumber\\
 \bigoplus_\lambda\, \{\lambda\} \times \{\lambda^{(N)}\} &\downarrow& 
\bigoplus_{\lambda, \kappa}  P_{\lambda\kappa}\,
{[}\kappa {]}\times\{\lambda^{(N)}\} .
\eea
The other gives
\bea {\rm O}(N)\times {\rm Sp}(m, \Rb) &\downarrow&  
{\rm O}(N)\times {\rm U}(m) \, ; \nonumber\\
\bigoplus_\kappa\, [\kappa] \times \langle \hf N (\kappa) \rangle
&\downarrow& 
\bigoplus_{\lambda, \kappa} \mathcal{P}_{\kappa\lambda }\,
[ \kappa]\times\{\lambda^{(N)}\} .
\eea
Thus, comparison of these two results reveals that
\be  \mathcal{P}_{\kappa\lambda} = P_{\lambda\kappa} .
\ee
In this way, the
${\rm Sp}( m,\Rb) \downarrow {\rm U}( m)$
branching rules were determined \cite{RWB} from the 
 ${\rm U}(N)\downarrow{\rm O}(N)$ branching rules of \textcite{King75}.

Cases for which $N\geq 2m$   turns out to be particularly simple.  For example, in  the symplectic  shell-model theory of nuclear collective motion, $m=3$ is the dimension of ordinary 3-space and $N$ is the nucleon number of the nucleus.  Thus, for medium to heavy nuclei, for which the theory is most relevant, $N$ is large compared to $2m=6$.
When $N\geq 2m$, the ${\rm U}(N)\times {\rm U}(m)$ duality relationship implies that any U$(N)$ irrep $\{ \lambda\}$ carried by a subspace of the Hilbert space of the $Nm$-dimensional harmonic oscillator  is  labeled  by a partition
$\lambda$ having at most $m \leq N/2$ parts.
For such an irrep, the ${\rm U}(N) \downarrow {\rm O}(N)$ branching rule has a particularly simple expression \cite{King75}.  The dual
${\rm Sp}(m,\Rb) \downarrow {\rm U}(m)$ branching rule is then  equally 
simple  and given  by
\be {\rm Sp}(m,\Rb) \downarrow {\rm U}(m)\; :\;
\langle \hf N(\kappa)\rangle \downarrow \{ \lambda^{(N)}\} \otimes \{D_m\} ,\ee
where $\{D_m\} = \sum_{n=0}^{\infty} \{ 2\} \pleth \{2n\}$
is the direct sum of the infinite sequence of
${\rm U}(m)$ irreps given by partitions whose parts are all even non-negative integers, i.e.,
\bea \{ D_m\} &=& \{0\} \oplus \{ 2\} \oplus \{ 4\} \oplus
\{2^2\} \oplus \{ 6\} \nonumber\\
&&\oplus \{ 42\} \oplus
\{2^3\} \oplus \{ 8\} \oplus \{ 62\} \oplus \cdots ,\quad \eea
with the understanding that the number of parts must not exceed $m$.%
\footnote{\label{ftn:2}  These Sp$(m,\Rb)$ irreps are the subset of positive harmonic series irreps that belong to the discrete series.  It also follows from these results that all of the positive holomorphic discrete series of  Sp$(m,\Rb)$ irreps (not including its double-valued projective irreps) are realized within the Hilbert space of $N=2m$ particles in an $m$-dimensional harmonic oscillator 
(cf.\ \textcite{Gelbart}).}

\subsubsection{Model spaces}

A model space for a Lie group $G$, and/or its Lie algebra, 
 is defined  \cite{BGG} as a Hilbert space that carries precisely one copy from every equivalence class of a specified set of irreps of $G$.
For example, a model space for SU(3) is obtained as a subspace of all states of
the Hilbert space $\Hb$ for the six-dimensional harmonic oscillator that
are annihilated by the raising operator of a dual U(2) group. 
This follows because  $\Hb= \bigoplus_\lambda \Hb^{\{\lambda\}}$ is the Hilbert space for  a direct sum $\bigoplus_\lambda \{ \lambda\} \times\{\lambda\}$
of all irreps of ${\rm U(2)}\times{\rm U(3)}$ that are labeled by partitions 
$\{\lambda\} = \{\lambda_1\lambda_2\}$ with no more than two integer parts. 
The subspace of $\Hb$ that is annihilated by U(2) raising operators is a
model space for SU(3) because the branching rule for the restriction
of U(3) to its SU(3) subgroup, 
\be  {\rm U(3)} \downarrow {\rm SU(3)}\; ; \; 
\{\lambda_1 \lambda_2 \lambda_3\} \downarrow \{
\lambda_1-\lambda_2, \lambda_2-\lambda_3\} 
\ee 
implies that the set of U(3) irreps with $\lambda_3=0$ restricts to a complete set of SU(3) irreps.
Such a model is useful for the calculation of the subset of Clebsch-Gordan coefficients for the U(3) couplings
\be \{ \kappa_1\kappa_2\} \otimes \{ \lambda_1\lambda_2\}
= \bigoplus_\mu \{ \mu_1\mu_2\mu_3\} \ee
for which $\mu_3=0$
(see further comments in Section X).

A similar example is given by the Hilbert space $\Hb$ of the
$2m^2$-dimensional harmonic oscillator on which 
all the holomorphic discrete series irreps of Sp$(m,\Rb)$ are realized (see Footnote \ref{ftn:2}).
 The subspace of states in $\Hb$ that are annihilated by the Sp$(m,\Rb)$ lowering operators is a model space for O$(2m)$ and, conversely, the subspace of all states of  $\Hb$ that are annihilated by the O$(2m)$ raising 
operators is a model space for  the  holomorphic discrete series irreps of   
Sp$(m,\Rb)$.  These model spaces were used in a study by  \textcite{Gelbart}.

\subsubsection{The microscopic theory of nuclear collective dynamics} 
\label{sect:Sp(3,R)}

The  ${\rm O}(A)\times{\rm Sp}(3,\Rb)$ duality is central to the  microscopic theory of nuclear collective dynamics \cite{Rowe85}
in which Sp$(3,\Rb)$ is a dynamical group for an $A$-nucleon collective model Hamiltonian and O$(A)$ is a symmetry group.

A group $G$ of canonical transformations of a classical
many-particle phase space is said to generate \emph{collective motions} 
 if it transforms the phase-space coordinates of all particles in the
same way. Thus, if an element $g\in G$ maps a set of phase space
coordinates $\xi =(x,y,z,p_x,p_y,p_z)$ for a particle to a new set,
denoted by $g\cdot \xi$,  the corresponding collective
transformation of an $A$-particle system is given by
\be (\xi_1,\xi_2, \ldots ,\xi_A)\to
(g\cdot \xi_1, g\cdot\xi_2, \ldots, g\cdot\xi_A) \, .\ee
Thus, by definition, a group of collective transformations of a many-particle system is  a representation of a group of transformations of a single-particle phase space.

For a classical dynamical system described by Hamilton equations of motion,
the possible motions are generated by groups of canonical  (i.e., symplectic)
transformations.
Such  dynamics satisfy Liouville's theorem, i.e., they preserve volumes in  phase space, and are
said to be \emph{Hamiltonian}.
Thus, the  Sp$(3,\Rb)$ symplectic group, defined as the set of all linear canonical transformations of the phase space of a single particle in 3-space, is fundamental to the theory of collective structure.

Infinitesimal generators of  Sp$(m,\Rb)$  are defined by
Eq.\ (\ref{eq:spmR}) in terms of harmonic oscillator raising and lowering operators.  However, their physical significance is more apparent when expressed in terms of particle position and momentum coordinates.
Thus, if $\{ x_{i};  i=1,2,3\}$ are Cartesian coordinates for a single particle in 
$\Rb^{3}$ and $\{ p_i; i = 1,2,3\}$ are the 
corresponding momentum coordinates,
 a basis for a unitary representation of the Lie algebra 
sp$(3,\Rb)$ is given by the operators 
\be \hat K_{ij} = \hat p_{i}\hat p_{j},
\quad \hat Q_{ij} = \hat x_{i} \hat x_{j} ,\quad \hat T_{ij} =
\hat x_{i} \hat p_j +\hat p_j \hat x_{i}  , \ee 
on the single-particle Hilbert space $\Hb_L=\mathcal{L}^2(\Rb^{3})$,
where $\hat x_i$ and $\hat p_i$ satisfy the standard  commutation relations
\be [\hat x_i , \hat x_j] = [\hat p_i , \hat p_j] = 0, \quad
[\hat x_i , \hat p_j] =  {\rm i} \hbar  \delta_{i,j}.\ee
Now, if $\{ x_{n i}; n = 1,\dots, A; i=1,2,3\}$ are  Cartesian coordinates for $A$ particles and $\{ p_{n i}; n = 1,\dots, A; i=1,2,3\}$ are the corresponding momentum coordinates,  infinitesimal generators for an $A$-particle representation of Sp$(3,\Rb)$ are given by
the O$(A)$  scalar operators on $\mathcal{L}^2(\Rb^{3A})$ 
\bea &\displaystyle \hat K_{ij} = \sum_n\hat p_{ni}\hat p_{nj},
\quad \hat Q_{ij} = \sum_n \hat x_{ni} \hat x_{nj} ,& \nonumber\\
&\displaystyle\hat T_{ij} =
\sum_n \big(\hat x_{ni} \hat p_{nj}+\hat p_{nj}\hat x_{ni}\big)  . &\eea

The group Sp$(3,\Rb)$ proves to be just what is needed for a practical microscopic theory of nuclear  collective motion. 
It has the particularly valuable property that the full many-particle kinetic energy 
$ \frac{1}{2M} \sum_{n i} \hat p_{n i}^2$ is an element of its Lie
algebra. Potential energy functions of the nuclear quadrupole
moments $\{Q_{ij}\}$ can then be added to this kinetic energy to form  
collective model Hamiltonians. The
sp$(3,\Rb)$ Lie algebra also contains the Hamiltonian of the
spherical harmonic oscillator 
\be \hat H_{\rm HO} = \frac{1}{2M}
\sum_{ni} \hat p_{ni}^2 + \half M\omega^2 
\sum_{n i} 
\hat x_{n i}^2 \, , \ee
which means that it provides a natural unification of the collective
model with the harmonic-oscillator shell model.
Combined with the fact that Sp$(3,\Rb)$ is a simple Lie group
 and that the representations and coupling coefficients of its
${\rm SO(3)} \subset {\rm U}(3) $
 subgroups are already well known, these properties mean that it
is straightforward to compute the matrix elements for irreps of the 
sp$(3,\Rb)$ Lie algebra in the harmonic-oscillator representations of 
Sp$(3,\Rb)$; they are most simply computed by so-called \emph{vector-coherent-state} methods 
\cite{Rowe84,RoweRC84} as outlined in \cite{Rowe85}
(see Sect.\ \ref{sect:CSreps} of this review).

We now show that the ${\rm O}(A)\times{\rm Sp}(3,\Rb)$ duality on the Hilbert space $\mathcal{L}^2(\Rb^{3A})$ of spatial wave functions for $A$ particles facilitates the construction of a shell-model coupling scheme with basis states that are products of center-of-mass states and antisymmetric
combinations of spin, isospin, and spatial states in an Sp$(3,\Rb)$ basis.

Separation of  center-of-mass states is accomplished by the factorization 
$\mathcal{L}^2(\Rb^{3A})
= \mathcal{L}^2(\Rb^{3})\otimes \mathcal{L}^2(\Rb^{3(A-1)})$, 
where  $\mathcal{L}^2(\Rb^{3})$ is the Hilbert space of center-of-mass states and $\mathcal{L}^2(\Rb^{3(A-1)})$ is the complementary space
 for $A$ nucleons relative to their center of mass.
It remains to characterize the Sp$(3,\Rb)$ irreps in 
$\mathcal{L}^2(\Rb^{3(A-1)})$ by their S$_A$ symmetries so that they may  be combined with spin-isospin irreps of conjugate symmetry to form totally antisymmetric states.
This is achieved by standard shell-model techniques for Sp$(3,\Rb)$ irreps for which the center of mass isf  in its harmonic oscillator ground state.
More generally, it is made possible by the ${\rm O}(A-1)\times{\rm Sp}(3,\Rb)$ duality relationship.

By duality,  $\mathcal{L}^2(\Rb^{3(A-1)})$ carries a multiplicity-free direct sum of irreps of the group ${\rm O}(A-1) \times {\rm Sp}(3,\Rb)$. 
Because the symmetric group S$_A$ is a subgroup of O$(A-1)$, 
it is then possible to construct basis states for 
$\mathcal{L}^2(\Rb^{3(A-1)})$ that reduce the subgroup chain
\be
\begin{array}{ccccccc}
\Big(&{\rm O}(A-1) &\supset&{\rm S}_A &\Big)&   \times &{\rm Sp}(3,\Rb)  \\
&{[}\kappa{]}&& (\lambda)  &&& \langle \half (A-1)(\kappa)\rangle  \end{array}
\label{eq:OA-1>SA}\ee
and carry ${\rm S}_A$ irreps
corresponding to partitions $\lambda \vdash A$.
Thus, Sp$(3,\Rb)$ irreps are determined with well-defined S$_A$ symmetry and can be coupled to spin-isospin irreps of conjugate S$_A$ symmetry.
The ${\rm O}(A-1) \downarrow S_A$ branching rules needed for this purpose have been given by \textcite{BK}, \textcite{DW}, and \textcite{MJC90}.

\section{Dual representations on fermion spaces} \label{sect.fermion_dualities}

We now consider duality relationships that are specific to fermions.
A primary difference between the dual representations expressed in terms of boson operators and those expressed in terms of fermion operators is that the latter usually involve  finite-dimensional representations.
Thus, the fermionic counterparts of the bosonic
 duality relationships on harmonic-oscillator  spaces involve
 groups with dual representations on a  finite-dimensional fermion Fock space 
 $\mathbb{F}^{(wN)}$ spanned by multi-fermion states
 \be
|0\rangle , \quad a^\dag_{\mu} |0\rangle , \quad
a^\dag_{\nu}a^\dag_{\mu} |0\rangle , \quad  \cdots ,\quad a^\dag_{1}
a^\dag_{2} \cdots a^\dag_{wN}|0\rangle   \, . \label{eq:mfs} \ee

Pairs of groups with dual representations of relevance to fermion systems are found among the chains of subgroups  
\be
\begin{array}{ccc}
{\rm SO}(2wN) &\quad\times\quad & {\rm O}(1) \\
\cup &&\cap \\
{\rm USp}(2w) &\quad\times\quad & {\rm USp}(N) \\
\cup &&\cap \\
 {\rm U}(w) &\quad\times\quad & {\rm U}(N) \\
\cup &&\cap \\
 {\rm U}(1) &\quad\times\quad & {\rm U}(wN)\, , \\
\end{array} \label{eq:109}
\ee
for $N$ even, and
 \be
\begin{array}{ccc}
{\rm SO}(2wN) &\quad\times\quad & {\rm O}(1) \\
\cup &&\cap \\
{\rm SO}(2w) &\quad\times\quad & {\rm O}(N) \\
\cup &&\cap \\
 {\rm U}(w) &\quad\times\quad & {\rm U}(N) \\
\cup &&\cap \\
 {\rm U}(1) &\quad\times\quad & {\rm U}(wN)\,, \\
\end{array} \label{eq:110}
\ee
for  $N$ even or odd.
The groups ${\rm SO}(2wN) \subset {\rm O}(2wN)$ are defined below as  groups of Bogoljubov-Valatin transformations and the group U$(wN)$ has a Lie algebra whose complex extension is spanned by the operators $\{ a^\dag_\mu a^\nu\}$.

The duality of  U$(w)$ and U$(N)$ representations on
$\mathbb{F}^{(wN)}$ follows once again from Theorem 3. The
dualities between representations of USp$(2w)$ and USp$(N)$ for $N$
even, and between representations of ${\rm SO}(2w)$ and ${\rm O}(N)$
for $N$ even or odd, are established, respectively, in the following two
theorems whose proofs are given elsewhere \cite{Helm,Thms}.

\medskip
\noindent {\bf Theorem 5 (Helmers):}  \emph{
The groups ${\rm USp}(2w)$ and ${\rm USp}(N)$
have dual representations on $\mathbb{F}^{(wN)}$ for $N$ even.}

\medskip \noindent{\bf Theorem 6: } \emph{
The groups ${\rm SO}(2w)$ and ${\rm O}(N)$ have dual representations on 
$\mathbb{F}^{(wN)}$ for $N$ even or odd.}
\medskip

Before discussing the relevance of these dual pairs in physical
applications, we first define the group  of Bogoljubov-Valatin transformations, 
O$(2wN)$.

\subsection{The group of Bogoljubov-Valatin transformations}

Let $\{ a^\dag_\nu, a^\nu ; \nu = 1, \dots, wN\}$ denote a set of
fermion creation and annihilation operators  that satisfy the
anti-commutation relations 
\be \{ a^\mu, a^\dag_\nu\} = \delta^\mu_\nu, \quad
 \{ a^\dag_\mu, a^\dag_\nu\}= \{ a^\mu, a^\nu\}=0 , \label{eq:6.fermioncomrelns}\ee
and the Hermiticity relations \be \big( a^\nu \big)^\dag =
a^\dag_\nu . \ee
The group of Bogoljubov-Valatin transformations, $G$, is then the subset of complex-linear transformations
\be \begin{array}{ccc} 
a^\dag_\nu &\to& \sum_\mu (a^\dag_\mu u_{\mu\nu} + a^\mu v_{\mu\nu} ) ,  \\ \\
a^\nu &\to& \sum_\mu (a^\dag_\mu v^*_{\mu\nu} + a^\mu u^*_{\mu\nu} ) ,
\end{array} \label{eq:6.223} 
\ee 
that preserve the fermion anti-commutation relations
(\ref{eq:6.fermioncomrelns}) and the Hermiticity relationship.

To identify this group, consider  its application to the Hermitian
operators
\be \hat{\cal Q}_\nu = \frac{1}{\sqrt{2}}
(a^\dag_\nu + a^\nu), \quad \hat{\cal P}_\nu = \frac{{\rm i}}{\sqrt{2}}
(a^\dag_\nu - a^\nu) ,\ee
which satisfy the anti-commutation relations
\be \{ \hat{\cal Q}_\mu , \hat{\cal Q}_\nu \}
 = \{ \hat{\cal P}_\mu , \hat{\cal P}_\nu \}
=\delta_{\mu, \nu} , \quad \{ \hat{\cal Q}_\mu , \hat{\cal P}_\nu \} =0 .\label{eq:6.224}\ee
 If $\Rb^{2wN}$ is the real vector space spanned by the operators 
$\{\hat Z_i \} = \{\hat{\cal Q}_1, \hat{\cal Q}_2, \dots, 
\hat{\cal P}_1, \hat{\cal P}_2, \dots \}$, then
$G$ is the set of linear transformations of  $\Rb^{2wN}$ that preserve the anti-commutation relations $\{ \hat Z_i, \hat Z_j \} = \delta_{i,j}$.
These transformations  must be real; otherwise they would not preserve the Hermiticity of the $\{ \hat{\cal Q}_\nu\}$ and $\{\hat {\cal P}_\nu\}$ operators.
It follows that $G$ is the  subgroup of real
 linear transformations, $\hat Z \mapsto \hat Zg$,
 that satisfy the condition
\be \textstyle \{  \sum_i \hat Z_i g_{ij} , \sum_k \hat Z_k g_{k l }\} =
\sum_i g_{ij} g_{il} = \delta_{j,l} .\ee
This is the real orthogonal group O$(2wN)$.

A representation of  this O$(2wN)$ group, of relevance to the quantum mechanics of many-fermion systems, is carried by the  Fock space 
$\mathbb{F}^{(wN)}$.
In this representation,  the  Lie algebra so$(2wN)$ of the group O$(2wN)$, known as the {\em fermion pair\/} algebra,
is spanned by the Hermitian linear combinations of the operators
 \be
a^\dag_\mu a^\dag_\nu, \quad a^\mu a^\nu , \quad a^\dag_\mu a^\nu
-a^\nu a^\dag_\mu . \label{eq:O2Nbasis}\ee

  The Bogoljubov-Valatin group can be extended to a  full dynamical group
 O$(2wN+1)$ for the fermion system that includes both the even and odd fermion states of the Fock space $\mathbb{F}^{(wN)}$, by the addition of the operators 
 $\{\hat{\cal Q}_\nu\}$ and $\{\hat{\cal P}_\nu\}$ to its Lie algebra.   Such an addition was proposed by  \textcite{Fukutome77} based on the observation that the commutator  $[a^\dag_\mu, a^\nu] =  a^\dag_\mu a^\nu - a^\nu a^\dag_\mu$ of a fermion creation and a fermion annihilation operator
is in the complex extension  of so$(2wN)$.

Note that, if we replace the index $\nu$, 
that labels single-fermion states, by a double index,
$\nu \to(\tau m)$ where $\tau$ and $m$ take $w$ and  $N$ values, respectively,
the space $\mathbb{F}^{(wN)}$  is seen as a tensor product. 
For example, $\tau$ might index the isospin states, $\tau = \pm 1/2$, of an isospin $T=1/2$ nucleon and $m$ might index a nucleon's angular-momentum states,  $m = -j, j+1,\dots, +j$ with $j$ a half-odd positive integer. In this case
$w=2T+1=2$, and $N=2j+1$ is an even integer. In another example,
$\tau$ might index the four spin-isospin states  of a nucleon (with
spin and isospin $S=1/2$, $T=1/2$)  and $m$ might index the orbital angular momentum states with labels $-l, \dots, +l$,
where $l$ is a non-negative integer. In this case $w=4$, and $N=2l+1$
is an odd integer.

\subsection{Pair-coupling schemes for fermions of a single species}
\label{sect:qsirrep}

The special case of a ${\rm USp}(2w)\times{\rm USp}(N)$ duality,
with $w=1$ and $N=2j+1$, where $j$ (a half-odd positive integer) is
the angular momentum of a single fermion (a neutron, proton,
or electron), is of historical significance because, as far as we
know,  it  was  the first duality relationship, independent of
Schur-Weyl duality, to be identified in physics \cite{Helm}. 
Because of its simplicity, the ${\rm USp}(2j+1) \subset {\rm U}(2j+1)$ subgroup chain defines the coupling scheme most commonly used in the atomic  shell model \cite{Racah} and in the nuclear shell model for  nuclei 
with either neutron or proton closed shells \cite{Flowers1,Flowers2,Flowers3,JBF,Talmi}.
It also plays a central role in models of pairing and superconductivity in
atomic and nuclear physics  in which  USp$(2j+1)$ is dual to a so-called
SU(2)$_{\rm qs}\simeq {\rm USp(2)}$ quasi-spin group 
\cite{Anderson,Kerman,Kerman2}.
The Lie algebra su(2)$_{\rm qs}$          
is important for understanding situations in which the coupling of fermions to
form angular-momentum-zero (Cooper) pairs is energetically favored over other couplings.

A system of $n$ identical fermions of  angular momentum $j$ carries a fully antisymmetric irrep $\{ 1^n\}$ of U$(2j+1)$  
with $n$ restricted to the range $0\leq n \leq 2j+1$,  
Thus, if the Hamiltonian is rotationally invariant, we  seek
basis states for this irrep that reduce the rotation subgroup 
${\rm SU(2)}_J \subset {\rm U}(2j+1)$
and have conserved angular-momentum quantum numbers. 
(Note that the rotation group for fermions in SU(2) rather than SO(3).) As observed by \textcite{Racah}, additional quantum numbers are supplied by inclusion of the group USp$(2j+1)$ in the subgroup chain
\bea \begin{array}{ccccccc}
\mathrm{U}(2j+1) & \!\supset\! & \mathrm{USp}(2j+1) &\!\supset\!& \mathrm{SU}(2)_{J} &
\!\supset\! &\mathrm{U}(1)_{J}\\
n && v && J&&M
\end{array} .\quad \label{eq:USpchain}
\eea

Now, from the paired subgroups of O$(2(2j+1))$ with dual representations on the Fock space $\mathbb{F}^{(2j+1)}$ shown in Eq.\ (\ref{eq:109}), it is seen that 
USp$(2j+1)$ has dual representations with an 
${\rm SU(2)}_{\rm qs}\simeq {\rm USp(2)}$ quasi-spin group
and that U$(2j+1)$ has dual representations with the subgroup 
${\rm U(1)}_{\rm qs}\subset {\rm SU(2)}_{\rm qs}$. 
 It  follows that the basis states of the coupling scheme (\ref{eq:USpchain}) are identical with basis states of the Fock space $\mathbb{F}^{(2j+1)}$ that reduce the dual subgroup chain
\bea
\begin{array}{ccccccc} \mathrm{SU}(2)_{\rm qs}&\times&
\mathrm{SU}(2)_{J} & \supset&
\mathrm{U}(1)_{\rm qs}&\times&\mathrm{U}(1)_{J} \\
s && J && s_0 &&M \end{array}. \label{eq:qspinchain} \eea
We show in the following that the quantum numbers for these two chains are related by
\be    s(v) = \hf (j+\hf-v),\quad  s_0(n) = \hf (n-j-\hf) .
\label{eq:ss0}\ee
(Note that the lowest-weight {SU}(2)$_{\rm qs}$ state for the $v=0$ irrep is the $n=0$ state for which $s_0 = -s$.)
We also show that the ${\rm USp}(2j+1)$ group in the chain (\ref{eq:USpchain}) has the physical significance of being the subgroup of U$(2j+1)$  transformations that leave the creation operator for  a Cooper pair (i.e.,
an angular-momentum-zero  fermion-pair) invariant.  Moreover, this pair creation operator and the corresponding annihilation operator are shown to generate the Lie algebra 
su(2)$_{\rm qs}$ of the group SU(2)$_{\rm qs}$ that commutes with 
${\rm USp}(2j+1)$.
It follows that USp$(2j+1)$ is a symmetry group of any Hamiltonian
 that is defined in terms of the su(2)$_{\rm qs}$ pair operators.
Thus, we have two groups, SU(2)$_{\rm qs}$ and USp$(2j+1)$, with commuting actions on the Fock space $\mathbb{F}^{(2j+1)}$,  of which SU(2)$_{\rm qs}$ is a dynamical group and  USp$(2j+1)$ is a symmetry group for a class of pairing  model Hamiltonians.

 \subsubsection{The su(2$)_{\rm qs}$ and usp$(2j+1)$  Lie algebras}

The  fermion-pair creation operator used to define the USp$(2j+1)$ group is the operator
\bea \hat{\mathcal{A}} 
&=& \sqrt{2j+1}\, \sum_m (j,-m; j,m |00)\, a^\dag_{m} a^\dag_{-m}
\nonumber\\
&=& \sum_m (-1)^{j+m} a^\dag_{m} a^\dag_{-m}  ,
\eea
 where $\{ a^\dag_{m}; m = -j, \dots,+j\}$ is a set of $2j+1$ creation operators for fermions of angular momentum $j$ and $(j,-m; j,m |00) = (-1)^{j+m}$ is an SU(2) Clebsch-Gordan coefficient.
The operator $\hat{\mathcal{A}}$ is the raising operator of the su(2)$_{\rm qs}$ quasi-spin Lie algebra spanned by the operators 
\bea &\displaystyle
\hat S_+ = \hf \hat{\mathcal{A}} = \sum_{m>0} a^\dag_m a^\dag_{\bar m} , \quad \hat S_-= \sum_{m>0} a^{\bar m} a^m , &\nonumber\\
&\displaystyle 2\hat S_0 =  \sum_{m>0} (a^\dag_m a^m - a^{\bar m}
a^\dag_{\bar m}) = \hat n -\hf (2j+1)\hat I ,& \qquad \label{eq:quasispinops}\eea
where
\be  a^\dag_{\bar m} \equiv
(-1)^{j+m}a^\dag_{-m}, \quad a^{\bar m} \equiv (-1)^{j+m}a^{-m},
\ee
$\hat n=  \sum_{m=-j}^{j}a^\dag_m a^m$
is the fermion number operator, and $\hat I$ is the identity operator.%
\footnote{The bar operation, $a^\dag_m \to a^\dag_{\bar m}$,  is equivalent to a rotation through angle $\pi$.
Therefore, because $2j$ is odd, applying it twice changes the sign of a fermion operator, i.e., $a^\dag_{\bar{\bar m}} = -a^\dag_m$ and  $a^{\bar{\bar m}} = -a^m$.} 
The quasi-spin operators satisfy the commutation relations
\be [\hat S_+ , \hat S_-] = 2\hat S_0 , \quad [\hat S_0 , \hat S_\pm ] = \pm \hat S_\pm .\ee

We now identify the operators of the usp$(2j+1)$ Lie algebra.
 First observe that, for each $m$, the fermion creation and annihilation operators
$a^\dag_m$ and $a^{\bar m}$, are $\pm \frac12$ components of a
quasi-spin-$\frac12$ tensor, 
 $\hat \xi_m= (a^\dag_m,a^{\bar m})$,
as can be seen from the commutation relations
\bea
\begin{array}{lcl}
[\hat S_+,a^{\bar m}] =  a^\dag_{m}, &\quad& {[} \hat  S_-, a^\dag_m] = a^{\bar m}, \\
  {[} \hat S_{0},a^{\bar m}] = - \hf a^{\bar m} ,
 &\quad& 
 {[} \hat S_0,a^\dag_{m}] = \hf a^\dag_{m} ,  \\
 {[}\hat S_+,a^\dag_ m] = 0, &\quad& {[} \hat  S_-, a^{\bar m}] = 0,
 \end{array} \label{eq:qshalfA}
\eea 
which apply for each $m$ in the range $-j \leq m \leq +j$.
Next observe that pairs of these
quasi-spin-$\frac12$ tensors can be 
coupled with SU(2) Clebsch-Gordan coefficients to quasi-spin-scalar operators, i.e., $[\hat \xi_m \otimes \hat \xi_p]_0
=\frac{1}{\sqrt{2}}\, (a^\dag_m a^{\bar p} - a^{\bar m} a^\dag_p)$.
Moreover, to within numerical constants given, for example, by the anti-commutators $\{a^\dag_m, a^m\} = 1$,
these are the only bilinear combinations of the fermion creation and annihilation operators that can commute with the su(2)$_{\rm qs}$ operators.
Thus, in the present context, the ${\rm usp}(2j+1)$
 Lie algebra is spanned by the Hermitian linear combinations of the 
 quasi-spin scalar operators
\bea    \hat A_{mp} =  a^\dag_m a^{\bar p} + a^\dag_p a^{\bar m}  , 
&\quad&   m\geq p>0 ,  \label{eq:USpN1a}\\
 \hat C_{mp} = a^\dag_m a^{p} -  a^\dag_{\bar p} a^{\bar m} 
   , &\quad&
m,p>0 \label{eq:USpN2a}\\
 \hat  B_{mp} =  a^\dag_{\bar m}  a^{p} + a^\dag_{\bar p}  a^{ m}     , 
&&      m\geq p>0,   \label{eq:USpN3a}
\eea 
A basis for the Cartan subalgebra for this realization of usp$(2j+1)$ is given by the operators
\be \hat C_{mm} = a^\dag_m a^m - a^\dag_{\bar m} a^{\bar m} , \quad m >0 . \label{eq:Cartanusp1}\ee

\subsubsection{Labels for SU(2$)_{\rm qs}$ and USp$(2j+1)$ irreps}

Quantum numbers for basis states defined by subgroup chains are given as usual by the labels for the irreps of the groups in the chain.  Irreps of
SU(2)$_{\rm qs}$ and USp$(2j+1)$ are conveniently  labeled 
 by their lowest and highest weights, respectively.
A complete set of states in  the Fock space $\mathbb{F}^{(2j+1)}$ that are simultaneously of lowest SU(2)$_{\rm qs}$ and  highest USp$(2j+1)$ weight are the states
\bea 
 &|0\rangle, \quad |1\rangle = a^\dag_j |0\rangle , 
\quad |2\rangle =  a^\dag_j  a^\dag_{j-1} |0\rangle , \quad\cdots , & \nonumber\\
& \quad|v\rangle = a^\dag_{j} a^\dag_{j-1} \dots a^\dag_{j-v+1} |0\rangle ,\quad\cdots , \quad |j+\hf\rangle ,\quad& 
\label{eq:vstates} 
\eea
for $v$ an integer in the range $0\leq v \leq j+\hf$, which we describe as \emph{extremal states}.

A USp$(2j+1)$ irrep with highest-weight state $|v\rangle$ has highest weight 
determined by 
\be \langle v | \hat{C}_{mm} |v\rangle = \cases{ 1 &for $j\geq m\geq j-v+1$
\cr 0 &for $\hf\leq m <j-v+1$.} \ee
Thus, it has highest weight $\langle 1^v\rangle$ and  states of this irrep are labeled  by the quantum number $v$.
On the other hand, an SU(2)$_{\rm qs}$ irrep with lowest-weight state $|v\rangle$ has highest weight  defined by the quasi spin
\be s(v) = - \langle v| \hat S_0 |v\rangle = \hf (j+\hf-v) , \label{eq:sv}
\ee 
consistent with Eq.\ (\ref{eq:ss0}).
The quantum  number of the ${\rm U(1)}_{\rm qs} \subset {\rm SU(2)}_{\rm qs}$ subgroup with infinitesimal generator $\hat S_0$ is similarly given by Eq.\ (\ref{eq:ss0}), i.e.,
$s_0(n) = \hf (n-j-\hf)$.

\begin{figure}
 \includegraphics[scale=0.51]{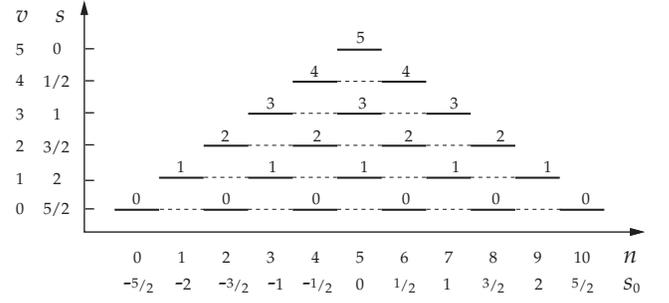}%
 \caption{\label{fig:virreps} The spectrum of USp(10) irreps in the $j=9/2$ shell labeled by
values of the seniority quantum number $v$.
The combined  USp(10) irreps having a common value of the particle
number $n$ span  the ${\rm U}(10)$ irrep $\{1^n\}$.
The combined states of all USp(10) irreps having a common value of the seniority $v$, connected by dotted lines,
span  an ${\rm SU}(2)_{\rm qs} \times {\rm USp}(10)$ irrep.    }
 \end{figure}

The integer $v$, which labels the extremal states of Eq.\ (\ref{eq:vstates}) and both the SU(2)$_{\rm qs}$ and USp$(2j+1)$ irreps, is known as a \emph{seniority} quantum number; it takes the values $v=0. \dots, j+\hf$ and has a physical interpretation as the number of unpaired particles in any state of an 
${\rm SU}(2)_{\rm qs} \times {\rm USp}(2j+1)$ irrep; it is
equal to the number of particles in an SU(2)$_{\rm qs}$ lowest-weight state.
This can be seen in Fig.\ \ref{fig:virreps}.

\subsubsection{A specific example and some general results}

Suppose, for example, that $j=9/2$. 
The  states available to a many-fermion system (either protons, neutrons or electrons)
occupying the single particle states of a $j=9/2$ nuclear or atomic shell are 
shown in Fig.\ \ref{fig:virreps}, by short lines, as subsets of states that span irreps of  USp(10).  The subsets belonging to equivalent USp(10) irreps are linked by dashed lines and together span irreps of 
${\rm USp}(10)\times {\rm SU}(2)_{\rm qs}$.
By ${\rm U}(1)_{\rm qs}\times{\rm U}(10)$ duality, all states belonging to a column of
USp(10) irreps of particle number $n$ belong to the single U(10) irrep $\{1^n\}$ that is determined by the U(1)$_{\rm qs}$ irrep $\{ n\}$.
Thus, the figure shows that a subset of 
${\rm U}(10)\downarrow {\rm USp}(10)$
branching rules are given, for $n\leq 5$, by
\be \{ 1^n\} \downarrow  \langle 1^n\rangle\oplus 
       \langle 1^{n-2}\rangle \oplus \cdots \oplus 
       \langle 1\rangle \;   {\rm or}\; \langle 0\rangle 
\ee
and, for $n >5$, by
\be \{ 1^n\} \downarrow  \langle 1^{10-n}\rangle\oplus 
      \langle 1^{8-n}\rangle \oplus \cdots \oplus 
       \langle 1\rangle \;   {\rm or}\; \langle 0\rangle .
\ee

In fact, the duality relationships between the subgroup chains 
${\rm SU}(2)_{\rm qs}\supset {\rm U}(1)_{\rm qs}$ and
${\rm U}(2j+1)\supset{\rm USp}(2j+1)$, imply that the coefficients in the two sets of branching rules
\bea
{\rm SU}(2)_{\rm qs} &\!\downarrow\! & {\rm U}(1)_{\rm qs} \,;\,
\big( s(v)\big)\downarrow \bigoplus_n k_{vn}\, 
\big( s_0(n)\big) ,  \\
{\rm U}(2j+1) &\!\downarrow\! & {\rm USp}(2j+1) \, ;\,
 \{ 1^n\}\downarrow \bigoplus_v k_{vn} \,\langle 1^v\rangle ,\quad
 \eea
are identical for any value of $j$. Thus, from the known coefficients for the 
${\rm SU}(2)_{\rm qs}\supset {\rm U}(1)_{\rm qs}$ branching rules, it is determined that a subset of ${\rm U}(N)\downarrow {\rm USp}(N)$ branching rules  is given, for any even positive integer $N$, by
\be  \{ 1^n\} \downarrow 
\cases{ \displaystyle
 \langle 1^{n}\rangle  \oplus \langle 1^{n-2}\rangle \oplus \cdots
 \oplus \langle 1\rangle \; {\rm or} \; \langle 0\rangle ,
& for $n \leq N/2$, \cr
\displaystyle
 \langle 1^{N-n}\rangle    \oplus \cdots
 \oplus \langle 1\rangle \; {\rm or} \; \langle 0\rangle ,
 & for $n > N/2$ ,}  \label{eq:U10USpBrule}
\ee

Branching rules such as these provide powerful tools for deriving many needed results in shell-model and other applications.  For example, they make
it possible to infer the angular momentum states
contained within a USp($2j+1$) irrep $\langle 1^n\rangle$ from a knowledge of the
${\rm U}(2j+1)\downarrow{\rm SU}(2)_J$ branching rules. Suppose these branching rules are expressed 
for $n\leq j+\hf$ by
\be {\rm U}(2j+1)\downarrow{\rm SU}(2)_J\; ;\; \{ 1^n\} \downarrow
\bigoplus_J F^n_J\, (J) , \ee
where we here use the symbol $(J)$ to denote an SU$(2)_J$ irrep of angular momentum $J$ (corresponding to the U(2) irrep $\{ 2J\}$).
It  follows from Eq.\ (\ref{eq:U10USpBrule}) that, for $n\leq j+\hf$,
\bea {\rm USp}(2j+1)\downarrow{\rm SU}(2)_J&\!;\!&
\langle 0\rangle\downarrow (0)  \nonumber\\
&\!;\!& \langle 1\rangle\downarrow (j) \label{eq:USptoSU}\\
&\!;\!&\langle 1^n\rangle\downarrow
\bigoplus_J (F^n_J - F^{n-2}_J)\, (J) . \quad \nonumber
\eea
This relationship is confirmed for $j=9/2$ from Fig.\ \ref{fig:virreps}.

The $ F^n_J $ coefficients can be evaluated by means of a plethysm.
First observe that the single-particle $n=1$
states all have angular momentum $J=j$. This means that they span a  
${\rm U}(2)$ irrep labeled by the partition $\{ 2j\}$ and that the restriction of
the fundamental U($2j+1$) irrep
$\{1\}$ to U(2) satisfies the branching rule
\be {\rm U}(2j+1)\downarrow{\rm U}(2)\; ;\; \{ 1\} \downarrow \{2j\}.
\ee
 It follows that $n$-particle irreps of U($2j+1$) contain
the angular momentum states given by the plethysm
\be{\rm U}(2j+1)\downarrow{\rm U}(2) \; ;\; \{ 1^n\} \downarrow \{2j\} \pleth
\{ 1^n\} .
\ee
Each U(2) irrep $\{\lambda_1\lambda_2\}$ then restricts to an SU(2)$_J$ irrep of angular momentum $J= \hf(\lambda_1-\lambda_2)$.

For example, using the plethysm code of  \textcite{CarvalhoD01}, it is determined that
\bea {\rm U}(10)\downarrow {\rm U}(2) \; ;\; \{ 1^2\} &\downarrow& \{17,1\} \oplus  \{15,3\}   \oplus \{13,5\}\nonumber\\
&& \oplus \{11,7\} \oplus \{9,9\}
\eea
and, hence, that
\be{\rm U}(10)\downarrow{\rm SU}(2)_J \; ;\; \{ 1^2\} \downarrow (8) \oplus  (6)
\oplus (4) \oplus (2) \oplus (0).
\ee
It follows from Eq.\ (\ref{eq:USptoSU}) that
\be {\rm USp}(10)\downarrow {\rm SU}(2)_J \; ;\; \langle 1^2\rangle \downarrow
(8) \oplus  (6) \oplus (4) \oplus (2).\ee
Repeating this process for other values of $n$ and $v$,
we obtain the spectrum of angular momentum states contained in the USp(10) irreps of the $j=9/2$ shell shown in Table \ref{tab:qsirrepsj9}.

\begin{table}[ht]
\caption{The spectrum of angular momentum states contained in the USp(10)
irreps of the $j=9/2$ shell \label{tab:qsirrepsj9}}
\begin{tabular}{ccccc} \hline\hline
$v$ &\quad &$s$ & \quad & J \\   \tableline
0&&$\frac{5}{2}$&&0\\
1&&2 && $\frac{9}{2}$ \\
2&&$\frac{3}{2}$ && 2, 4, 6, 8\\
3&&1 && $\frac{3}{2}$, $\frac{5}{2}$, $\frac{7}{2}$, $\frac{9}{2}$, $\frac{11}{2}$,
$\frac{13}{2}$, $\frac{15}{2}$, $\frac{17}{2}$, $\frac{21}{2}$ \\
4&& $\frac{1}{2}$ && 0, 2, 3, $4^2$, 5,
6$^2$, 7, 8, 9, 10, 12\\
5&& 0 && $\frac{1}{2}$, $\frac{5}{2}$, $\frac{7}{2}$, $\frac{9}{2}$, $\frac{11}{2}$,
$\frac{13}{2}$, $\frac{15}{2}$, $\frac{17}{2}$, $\frac{19}{2}$, $\frac{25}{2}$
\medskip\\
\hline\hline
\end{tabular}
\end{table}

\begin{figure*}
 \includegraphics[scale=0.65]{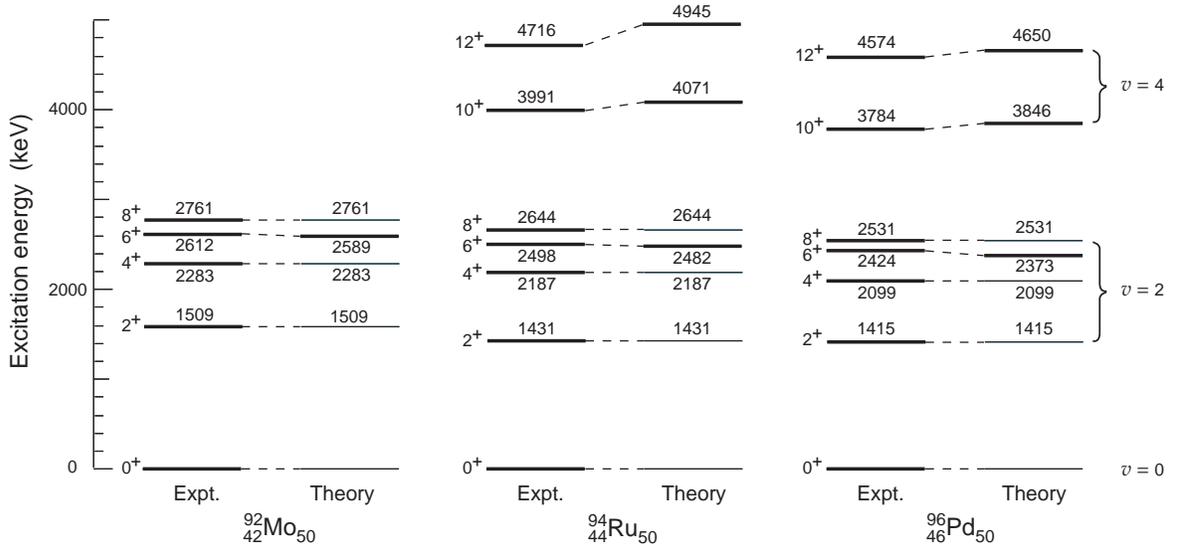}%
 \caption{\label{fig:isotones} Low-lying energy levels of some $N=50$ isotones.
The four parameters of a seniority-conserving interaction between protons in a $g_{9/2}$ shell were chosen to fit
the energies of the lowest $J=0, 2, 4$, and 8 states shown by light (Theory)
lines for each nucleus. The remaining energy levels shown by heavy (Theory)
lines were then predicted for these interactions. (Results taken from
\textcite{RR2003}.)   }
 \end{figure*}

The above results show the duality relationship between the irreps of 
USp$(2j+1)$ and SU(2)$_{\rm qs}$ to be
effective at giving a simple pairing model a microscopic expression
 within the framework of the many-nucleon shell model.
They also reveal the more general circumstances under which the USp$(2j+1)$ symmetry of a pair-coupling
model is preserved.
In particular,  a number-conserving interaction that is expressible as a polynomial in the $\mathrm{usp}(2j+1)$ and su(2)$_{\rm qs}$ Lie algebras  cannot  mix states belonging to different $\mathrm{USp}(2j+1)$ irreps. Such an interaction therefore conserves the seniority quantum number 
\cite{JBF,Talmi,RR2003}.
A remarkably large number of nuclear interactions have this property.
For example, Fig.\ \ref{fig:isotones} shows the low-lying states of
three isotones of neutron number $N=50$ with their energy levels
modeled in terms of 2, 4, and 6 protons in $j=9/2$ single-particle
states, outside of an $N=50$, $Z=40$ closed-shell core, and with
states described by  $\mathrm{U}(10) \supset \mathrm{USp}(10)
\supset \mathrm{SU}(2)_{J} \supset \mathrm{U}(1)_{J}$ quantum
numbers. The Hamiltonian used to fit the energy levels was 
chosen to comprise rotationally-invariant and seniority-conserving
two-body interactions that fit the lowest $J=0, 2, 4$, and 8
states.  Figure \ref{fig:isotones} shows the 
success of such a Hamiltonian in predicting the excitation energies
of the other observed low-lying energy levels.

 \subsection{Dual groups for a multi-shell BCS Hamiltonians} 
 \label{sect: multishellBCS}

The above coupling scheme for a single $j$-shell has an interesting extension to a multi-shell (also called multi-level) scheme.  Multi-level pair-coupling models have long been of interest both in nuclear and condensed matter physics as models of superconductivity. They are typically solved in the BCS approximation 
\cite{BCS} which is an approximation that violates particle number conservation and other symmetries of the model.  
However, it has recently been rediscovered that a method proposed by 
\textcite{RichardsonS64} and \textcite{Richardson65} and pursued from a different perspective by \textcite{Gaudin76} and \textcite{Cambiaggio97} shows that a class of BCS Hamiltonians are integrable and have formally exact solutions.%
\footnote{Solutions to the R-G equations are described as ``formally exact"
because the equations can only be solved numerically.}  
The Richardson-Gaudin method and its many applications have been reviewed by \textcite{DukelskyPS04}.

A general BCS pairing Hamiltonian is of the form
\be \hat H = \sum_{k=1}^p \varepsilon_k \hat S_0^k - \sum_{i, k=1}^p g_{ik}\hat S^i_+ \hat S^k_- ,
\label{eq:H_{BCS}}\ee
where $\hat S^k_0$ and  $\hat S^k_\pm$ span the complex extension of a quasi-spin algebra SU(2)$_k$ and $k$ indexes the levels of the  system.
Such a Hamiltonian has a dynamical group
\ \be G_2={\rm SU}(2)_1 \times {\rm SU}(2)_2 \times \dots \times {\rm SU}(2)_p .\ee
With the quasi-spin operators $\hat S^k_\pm$ interpreted as creation and annihilation operators for a pair of particles in level $k$, as defined by  Eq.\ 
 (\ref{eq:quasispinops}), $\hat H$ conserves the total particle number of the system.
 Thus, the ${\rm U(1)}\subset G_2$ subgroup, with infinitesimal generator
 $\hat S_0 = \sum_k \hat S^k_0$, is a symmetry group of $\hat H$.
However, the Hamiltonian $\hat H$  has a much larger symmetry group 
 \be G_1 ={\rm USp}(2j_1+1) \times  \dots \times{\rm USp}(2j_p+1) ,
\label{eq:sumUsP}\ee
where $j_k$ is the angular momentum of a particle in level $k$,
that is dual to the dynamical group $G_2$.
Thus, because the groups ${\rm U(1)}$ and U$\Big(\sum_k(2j_k+1)\Big)$ form a dual pair, it is determined that a BCS Hamiltonian is diagonal in a basis that simultaneously reduces the subgroup chains
$G_2\supset {\rm U}(1)$ and U$\Big(\sum_k(2j_k+1)\Big)\supset G_1$.

These observations indicate ways to extend single-$j$ sub-shell coupling schemes to multi-$j$ sub-shell schemes.
For weak pairing correlations, i.e., when the off-diagonal elements $g_{ik}$ are negligible for $i\not= k$,  the eigenstates of $\hat H$ diagonalize the dynamical subgroup chain
\be G_2
\supset  {\rm U}(1)_1 \times  \dots \times {\rm U}(1)_p \supset {\rm}U(1). 
\ee
The dual chain that defines a corresponding shell-model coupling scheme is then the subgroup chain
\be {\rm U}\Big(\sum_k (2j_k+1)\Big) \supset \prod_k {\rm U}(2j_k+1)
\supset G_1 .
\label{eq:USpsum}
\ee
This is a standard shell-model coupling scheme.  
However,  in the strong-coupling limit when $\varepsilon_k$  and $g_{ik}$ 
take $i$ and $k$-independent values, the eigenstates of $\hat H$ diagonalize the 
dynamical subgroup chain
\be G_2 \supset  {\rm SU}(2) \supset {\rm U}(1), 
\ee
where ${\rm SU(2)}\subset G_2$ is the subgroup with infinitesimal generators
 $\hat S_i = \sum_k \hat S^k_i$.
They also simultaneously diagonalize the dual subgroup chain of the shell-model coupling scheme defined by
\be {\rm U}\Big(\sum_k (2j_k+1) \Big)\supset {\rm USp}\Big(\sum_k (2j_k+1)\Big)
 \supset G_1 .
\ee

Other possible coupling schemes are available when more than two $j$-sub-shells are involved in which some sub-shells are coupled strongly and others weakly.
These several  coupling schemes  enable the selection of  basis states for shell-model calculations that can be effectively truncated to include the dominant states coupled by an interaction with strong pairing components.

\subsection{Isospin-invariant pair coupling in nuclei}
\label{sect:isqpirrep}

In this section, we extend the duality relationships relevant for a valence shell
of neutrons, protons, or electrons,  to a system of neutrons and protons.
We show that such a system, with $w=2$ and $N=(2j+1)$,
 provides  a ${\rm USp}(4)\times{\rm USp}(2j+1)$ duality on a nuclear Fock space $\mathbb{F}^{(2(2j+1))}$. The group  USp(4) $\cong$ SO(5)  is a straightforward  extension of the  USp(2) $\cong$ SU(2) quasi-spin group for pair-coupling  of a single nucleon species to a dynamical group for an isospin-invariant neutron-proton pairing model.  The duality of its representations with those of USp$(2j+1)$  provides a practical interpretation and useful relationships for the application of the standard shell model  in the $jj$- and isospin-coupled basis of \textcite{Flowers1,Flowers2,Flowers3} and \textcite{JBF}.

To track the symmetries of a system of neutrons and protons,
it is appropriate to regard neutrons and protons as different states   of  nucleons of isospin $T=1/2$, labeled by $T_0 = \pm 1/2$ respectively.
A basis for the Fock space $\mathbb{F}^{(2(2j+1))}$ is then one that reduces the subgroup chain
\begin{widetext}
\be \begin{array}{cccccccccccccccc}
\mathrm{U}(2(2j+1)) & \supset &{\rm U}(2j+1)&\times& \mathrm{U}(2)_T
&\supset& {\rm USp}(2j+1)&\times& \mathrm{SU}(2)_T &\supset&
\mathrm{SU}(2)_{J} \\
n&&\lambda&& \tilde\lambda && \kappa &&T\, T_0\;\; &&  J\,  M\;\;
\end{array} \label{eq:265}
\ee
\end{widetext}
with quantum numbers defined by the  representation labels shown.
The group U(2)$_T$ in this chain is the group of unitary transformations in two-dimensional isospin space.

From the paired subgroups of O$(4(2j+1))$ with dual representations on the Fock space $\mathbb{F}^{(2(2j+1))}$ shown in Eq.\ (\ref{eq:109}), it is now seen that USp$(2j+1)$ and the ${\rm USp(4)} \simeq  {\rm SO(5)}$ quasi-spin group 
have dual representations as do U$(2j+1)$ and the isospin subgroup 
${\rm U}(2)_T \subset {\rm SO(5)}$.
It  follows that the basis states of the coupling scheme (\ref{eq:265}) are identical to basis states of the Fock space $\mathbb{F}^{(2(2j+1))}$ that reduce the dual subgroup chain
\be
\begin{array}{ccccccccccccc} 
\!\mathrm{SO}(5)\! &\times&\mathrm{SU(2)}_J\!\!&\!\supset\!&\! \mathrm{U}(2)_T \! &\!\supset \!&\!{\rm SU}(2)_T\!&\!\times\! &\!\mathrm{U}(1) \\
(v,t)\! \!\!&&  J\, M\;\; &&\! \! \tilde\lambda && T\,T_0\;\;&&n \end{array} ,
\label{eq:chain275}\ee
 where $v$ and $t$ are related, as shown in the following, to the elements of the partition $\kappa$.

\subsubsection{The Lie algebras of SO(5) and USp$(2j+1)$ and their irreps}

The  so(5)   Lie algebra is the natural  extension of su(2)$_{\rm qs}$ to include all $J=0$ pair creation operators that become possible when there are two kinds of fermion: neutrons and protons.
Thus, its complex extension  is spanned by the  angular-momentum $J=0$ operators 
\bea \hat{\mathcal{A}}_{\sigma\tau} &=& \sum_{m>0} \big(
a^\dag_{\sigma m}a^\dag_{\tau \bar{m}} +
a^\dag_{\tau {m}} a^\dag_{\sigma\bar{m}}  \big),\\
\hat{\mathcal{B}}_{\sigma\tau} &=& \sum_{m>0} \big(
a^{\tau \bar{m}} a^{\sigma m} + a^{\sigma\bar{m}} a^{\tau{m}}  \big), \\
\hat{\mathcal{C}}_{\sigma\tau} &=& \sum_{m>0}
(a^\dag_{\sigma m} a^{\tau m}  -  a^{\tau \bar{m}} a^\dag_{\sigma \bar{m}})
\nonumber\\
&=& \sum_{m} a^\dag_{\sigma m} a^{\tau m} - \hf (2j+1)\delta_{\sigma,\tau} , \eea
where 
\bea & a^\dag_{\sigma \bar{m}} \equiv
(-1)^{j+m}a^\dag_{\sigma,-m}, \quad a^{\sigma\bar{m}} \equiv (-1)^{j+m}a^{\sigma,-m} , & \nonumber \\
&\sigma,\tau = 1,2, \; m = -j, \dots, +j .& \eea
The ${\rm u}(2)_T \subset {\rm so}(5)$ subalgebra is spanned by the Hermitian linear combinations of the
$\{ \hat{\mathcal{C}}_{\sigma\tau}\}$ operators.

The operators of the usp$(2j+1)$ Lie algebra are now defined
as the subset of u$(2j+1)$ operators that commute with these so(5) operators.
 They are simply obtained by  adding each infinitesimal generator of a neutron realization of usp$(2j+1)$  to the corresponding infinitesimal generator
 of a proton realization.
For example, a basis for the Cartan subalgebra of a combined neutron-proton realization of usp$(2j+1)$ is given by
\be \hat C_{mm} = \sum_\tau (a^\dag_{\tau m} a^{\tau m}
- a^\dag_{\tau \bar m} a^{\tau\bar m}) , \quad m >0 . \label{eq:Cartanusp2}\ee

A dual pair of  SO(5) and USp$(2j+1)$ irreps  on the Fock space
$\mathbb{F}^{(2(2j+1))}$,  labeled  by a USp$(2j+1)$ highest weight $\kappa$, is defined by a state
\be
 |\kappa\rangle =  a^\dag_{1,j}a^\dag_{1,j-1} \dots 
a^\dag_{1,j+1-\tilde\kappa_1}  |0\rangle ,
\label{eq:328a} \ee
when $\tilde \kappa_2 = 0$, and by
\bea
 |\kappa\rangle &=& \big( a^\dag_{1,j}a^\dag_{1,j-1} \dots 
a^\dag_{1,j+1-\tilde\kappa_1}\big) \nonumber\\
&& \times  \big( a^\dag_{2,j}  a^\dag_{2,j-1}
\dots a^\dag_{2,j+1-\tilde\kappa_2} \big)  |0\rangle ,  \label{eq:328b}
\eea
when $\tilde \kappa_2 \geq 1$, 
where $\tilde \kappa$ is the two-row partition conjugate to $\kappa$.
Such a state is simultaneously of lowest SO(5) weight and highest USp$(2j+1)$ weight.  Its SO(5) lowest weight is given by the non-zero expectation values of the
u(2)$_T$ operators
\bea
\langle \kappa |\hat\mathcal{C}_{\tau\tau} |\kappa\rangle &\!=\!&
\langle \kappa | \textstyle \sum_{m} a^\dag_{\tau m} a^{\tau m} -
\hf (2j+1) |\kappa\rangle \nonumber\\
 &\!=\!& \tilde \kappa_\tau - \hf(2j+1),
 \quad {\rm for}\;\tau = 1,2, \label{eq:146}
\eea
and its USp$(2j+1)$ highest weight is given by the expectation values of 
\be \kappa_{k} =\langle \kappa |\hat C_{j+1-k, j+1-k} |\kappa\rangle , \quad
k=1, \dots j + \hf .\ee
Thus, it is ascertained that
\be \kappa_{k} = \cases{ 2 & for $1 \leq k \leq\tilde \kappa_2 ,$\cr
                                        1&  for $ \tilde\kappa_2 < k \leq\tilde \kappa_1 ,$\cr
                                        0&  for $k> \tilde \kappa_1 .$}
\ee

The equivalent  \emph{seniority} and 
\emph{reduced isospin} labels $v = \tilde\kappa_1+ \tilde\kappa_2$
and $t=\hf (\tilde\kappa_1- \tilde\kappa_2)$ are understood physically as the particle number and isospin of the extremal state $|\kappa\rangle$.

\subsubsection{Tabulation of  basis states in $jj$ coupling}

The quantum numbers of basis states defined by the subgroup chain 
(\ref{eq:265}) signify irreps of the corresponding groups in the chains as follows.
The $n$-nucleon Hilbert  $\mathbb{H}^{(n)}$  carries an irrep $\{ 1^n\}$ of 
U$(2(2j+1))$.  According to Theorem 3, it also carries a direct sum of irreps of the direct product group ${\rm U}(2j+1)\times {\rm U}(2)_T$ given by the
branching rules
\be \begin{array}{rcl}
\mathrm{U}(2(2j+1)) &\downarrow& {\rm U}(2j+1)\times \mathrm{U}(2)_T \\
\;:\; \{ 1^n\} &\downarrow& \bigoplus_{\lambda\vdash n}\, \{ \lambda\} \times \{
\tilde\lambda\} ,\end{array} 
\label{eq:brule265}\ee
with $\lambda$ and $\tilde\lambda$ restricted to partitions having
no more than $2j+1$ and  $2$ parts, respectively.
The range of values of the SU(2)$_T$ isospin quantum
number $T$ is given by the branching rule
\be {\rm U}(2)_T \downarrow {\rm SU}(2)_T \; :\;
\{ \tilde\lambda\}\downarrow T = \hf \left( \right.
\tilde\lambda_1-\tilde\lambda_2 \left. \right) . \ee
The  U$(2j+1)$ irreps  and isospins for $j=3/2$ and $n\leq 4$ are shown, for example, in the first column of Table \ref{tab:isoqs}.
\begin{table}[htp]
\caption{  Irreps of the subgroups in the chains (\ref{eq:265}) and (\ref{eq:chain275}) to which the states of nucleon  number $n=1,\ldots,4$ and
$j=\afrac3/2$ belong.
 For $n = 5, \ldots , 8$, the states are mirror images of the
  $n=4,\ldots,1$ states, as can be seen in Fig.\ \protect\ref{fig:so5irreps}.
  Lowest-$n$ multiplets of 
  states of an ${\rm SO}(5)\times {\rm USp}(4)$
  irrep are denoted by an asterisk in the last column. 
   These are the lowest-$n$ multiplets of states among sets of states 
  linked by dotted lines in  Fig.\ \protect\ref{fig:so5irreps}.
 These lowest-$n$ multiplets of states contain a lowest-weight 
  state for an ${\rm SO}(5)\times {\rm USp}(4)$ irrep.  
\label{tab:isoqs}}
$\begin{array}{|c|c|c|c|c|c|c|c|}\hline
 \;{n}\phantom{x} &  \{\lambda\}  
 & \; T\phantom{x}&\;\langle\kappa\rangle\phantom{x}
&\; v\phantom{x}&\; t \phantom{x}&\; J \phantom{x}&
\; {\rm l.wt.}\phantom{\big|} \\ \hline
0& \{00\}   &      0     &  \langle 00\rangle  &  0  &    0    &     0    & *\\ \hline
1  & \{10\}   & \afrac1/2 &  \langle 10\rangle  &  1  &\afrac1/2 & \afrac3/2 & *\\ \hline
2  & \{20\}   &     0     &  \langle 20\rangle  &  2  &    0    &  1,\;3   & *\\ \hline
2 &  \{1^2\} &    1     &  \langle 00\rangle  &  0  &    0    &     0    &  \\ \hline
 \raisebox{-0.6ex}{\texttt{"}}   &\raisebox{-0.6ex}{\texttt{"}} &\raisebox{-0.6ex}{\texttt{"}} &  \langle 11\rangle  &  2  &    1    &     2    & *\\ \hline
3 &  \{21\}  & \afrac1/2 &  \langle 10\rangle  &  1  &\afrac1/2 & \afrac3/2 & \\ \hline
\raisebox{-0.6ex}{\texttt{"}} & \raisebox{-0.6ex}{\texttt{"}}  &\raisebox{-0.6ex}{\texttt{"}} &  \langle 21\rangle  &  3  &\afrac1/2 & \afrac1/2,\; \afrac5/2,\;
\afrac7/2 & * \\
\hline
3 & \{1^3\} &  \afrac3/2 &  \langle 10\rangle  &  1  &\afrac1/2 & \afrac3/2 &  \\
\hline
4 & \{2^2\} &     0     &  \langle 00\rangle  &  0  &    0    &     0    &  \\ \hline
\raisebox{-0.6ex}{\texttt{"}} &\raisebox{-0.6ex}{\texttt{"}} & \raisebox{-0.6ex}{\texttt{"}} &  \langle 11\rangle  &  2  &    1    &     2    &  \\ \hline
\raisebox{-0.6ex}{\texttt{"}}  &\raisebox{-0.6ex}{\texttt{"}} &\raisebox{-0.6ex}{\texttt{"}}  &  \langle 22\rangle  &  4  &    0    &  2,\;4   & * \\ \hline
4 & \{21^2\}&     1     &  \langle 20\rangle  &  2  &    0    &  1,\;3   &  \\ \hline
\raisebox{-0.6ex}{\texttt{"}} &\raisebox{-0.6ex}{\texttt{"}} &\raisebox{-0.6ex}{\texttt{"}} &  \langle 11\rangle  &  2  &    1    &     2    &  \\ \hline
4 &  \{1^4\} &    2     &  \langle 00\rangle  &  0  &    0    &     0    &  \\ \hline
\end{array}$
\end{table}

To determine the USp$(2j+1)$ irreps $\langle \kappa\rangle$ in a given 
U$(2j+1)$ irrep $\{\lambda\}$, one needs  the coefficients in the
\be {\rm U}(2j+1)\downarrow{\rm USp}(2j+1) \; :\;
 \{ \lambda\} \downarrow \bigoplus_{\kappa} F_{\kappa\lambda} \,
\langle\kappa\rangle 
\ee
 branching rule.    
These can be obtained from the corresponding 
 SO(5) $\simeq$ USp(4) $\downarrow$ U(2)$_T$ branching rules or, more simply,  from an algorithm  given by \textcite{King75} 
(and summarized in \textcite{RW10}) which then determines the 
SO(5) $\downarrow$ U(2)$_T$ branching rules.
The values of $v$ and $t$ for each USp$(2j+1)$ irrep equal  the values of $n$ and $T$, respectively, for the lowest value of $n$ for which the USp$(2j+1)$ irrep $\langle\kappa\rangle$ occurs. This  irrep contains an 
${\rm SO}(5) \times {\rm USp}(2j+1)$ extremal state and is noted in the last column of Table \ref{tab:isoqs} by an asterisk.

The SU(2)$_J$ irreps contained in an 
${\rm USp}(2j+1)$ irrep can be obtained recursively
from the ${\rm U}(2j+1)\downarrow{\rm USp}(2j+1)$ 
and ${\rm U}(2j+1) \downarrow {\rm U}(2)_J$
branching rules, where the latter is given by the plethysm
\bea {\rm U}(2j+1) \downarrow {\rm U}(2)_J & : & \{ 1\} \downarrow \{ 2j\} \\
&: & \{ \lambda \} \downarrow \{ 2j\} \pleth \{ \lambda\} . \label{eq:336}\eea
Suppose, for example, that we have already  determined the
${\rm USp}(2j+1) \downarrow {\rm SU(2)}_J$ branching rules for the
USp$(2j+1)$ irreps that occur for $n<4$ and we now wish to determine the rules for $n=4$.
The ${\rm U}(2j+1)\downarrow{\rm USp}(2j+1)$ branching rule for the $\{ 2^2\}$ irrep is given by
\be \{ 2^2\} \downarrow \langle 2^2\rangle \oplus \langle 1^2\rangle \oplus 
\langle 0\rangle \ee
and by the plethysm of Eq.\ (\ref{eq:336}), we derive 
\be {\rm U}(2j+1)\downarrow{\rm SU(2)}_J \; : \; 
\{ 2^2\} \downarrow (0)  \oplus 2(2) \oplus  (4) .\ee
Thus, knowing the branching rules for $n=0$ and 2
\bea {\rm USp}(2j+1)\downarrow{\rm SU(2)}_J &:& 
\langle1^2\rangle \downarrow (2) ,\\
&:& \langle 0\rangle \downarrow (0),
\eea 
it follows that
\be {\rm USp}(2j+1)\downarrow{\rm SU(2)}_J \; :\;  \langle 2^2\rangle \downarrow (2) \oplus (4).\ee 
Note that,  for any $n$ only one new USp$(2j+1)$ irrep ever occurs in any given U$(2j+1)$ irrep.

\begin{figure*}[thp]
\includegraphics[scale=0.50]{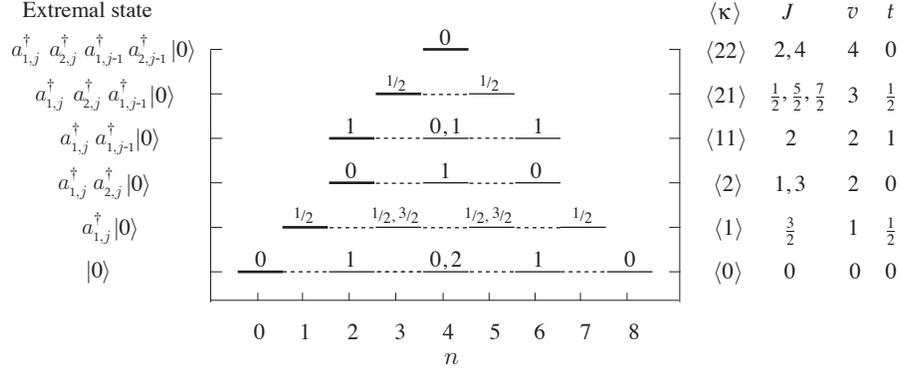}%
\caption{The spectrum of USp$(2j+1)$ irreps, shown as lines, for $j=3/2$.
Each USp$(2j+1)$ irrep, labeled by $\langle\kappa\rangle$, contains states with the angular-momentum values $J$ as shown on the right side of the figure.  These irreps are also labeled by the SO(5) quantum numbers: seniority, $v$, and reduced isospin, $t$. The basis states for each SO(5)  irrep are labeled by  isospin $T$ (shown above each level), and nucleon number $n$.
Note that all the USp(4) irreps corresponding to lines at the same horizontal level and connected by dotted lines share common values of $v$ and $t$, and common distributions of $J$ values.
They combine to span an ${\rm SO(5)}\times {\rm USp}(2j+1)$ irrep with $j=3/2$.
SO(5) lowest-weight states  lie in the left-most USp(4) irreps (shown as thick lines).
Thus, it is shown that the many-nucleon states of a $j=3/2$ shell fall into six distinct ${\rm SO(5)}\times{\rm USp}(2j+1)$ irreps.
\label{fig:so5irreps}}
\end{figure*}

\begin{figure*}[thp]
 \includegraphics[scale=0.50]{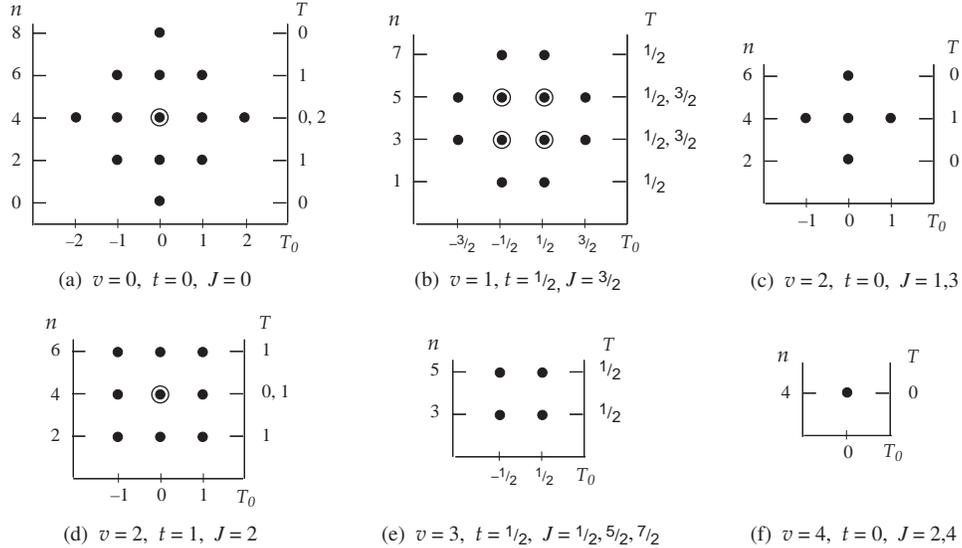}%
\caption{Weight diagrams for the states of SO(5) $\simeq$ USp(4) irreps whose lowest-weight states
have no more than four particles. States are
characterized by dots and labeled by their
${\rm U}(2)_T \supset {\rm U}(1)$ quantum numbers, i.e., the nucleon number
$n$, isospin $T$, and component of isospin $T_0$.  Weights with a multiplicity of two are denoted by dots with circles around them.  For example, the $n=4$ states of the $\langle 00\rangle$
irrep contain two states of $n=4$ and $T_0=0$; one has isospin $T=0$ and the other $T=2$.
\label{fig:so5weights}}
\end{figure*}

By the above means, we obtain a complete classification of shell-model states in any $j^n$ configuration.
For example, each USp$(2j+1)$ irrep $\langle\kappa\rangle$,
listed for $j=3/2$ in Table \ref{tab:isoqs}, 
is shown as a short line in Fig.\ \ref{fig:so5irreps} with equivalent irreps linked by dotted lines.  In combination, the states of each set of equivalent 
USp$(2j+1)$ irreps span an ${\rm SO(5)} \times {\rm USp}(2j+1)$ irrep.  The  seniority $v$,  reduced isospin $t$, and  angular momentum values are shown for these USp$(2j+1)$ irreps and the extremal state is given explicitly for each 
${\rm SO(5)} \times {\rm USp}(2j+1)$ irrep. 
The weight diagrams for the SO(5) irreps defined by the horizontal rows of Fig.\ \ref{fig:so5irreps} are shown in Fig.\ \ref{fig:so5weights}.

\subsubsection{A simple SO(5) model}

The  subgroup chain (\ref{eq:chain275}) diagonalizes a simple isospin-invariant pairing model with Hamiltonian 
\be \hat H = \varepsilon\, \hat n
 - \chi\sum_{\sigma\tau} \mathcal{A}_{\sigma\tau}  \mathcal{B}_{\sigma\tau}, \label{eq:so5pairing}\ee
 where $\chi$ is a coupling constant;  $\hat H$  has  eigenvalues \cite{RW10} given in terms of the quantum numbers of the  chain   
(\ref{eq:chain275}) by 
 \bea E_{vtnJT} &=& \textstyle  \varepsilon n - 
 \chi\big[ t(t+1)-T(T+1) + \frac32(n-v) \nonumber\\
 &&\textstyle  - \frac14 (n-v) (n + v - 4j - 2) \big] .\eea
This model may provide an acceptable description of some doubly 
open-shell nuclei in which seniority and isospin are expected to be approximately conserved.   
However, its primary value is to give  a physical interpretation of the kind of Hamiltonian that is diagonalized by the classification of nuclear shell model states in the $jj$-coupling scheme  
\cite{Flowers1,Flowers2,Flowers3,JBF} defined by the subgroup chain 
(\ref{eq:265}). 
This coupling scheme is most commonly used in the nuclear shell model because it is the simplest and is one for which fractional parentage coefficients are readily available.   
Moreover, the SO(5) model gives a direct indication of the kinds of shell-model configurations needed to describe pairing correlations in doubly open-shell nuclei, which is important to know about even when these correlations are not dominant.

As shown for the SU(2) quasi-spin pairing model in Sect.\ 
\ref{sect: multishellBCS}, the above  SO(5)  model has  a multi-shell 
extension and corresponding shell-model coupling schemes.
Interest in such  extensions  is stimulated by the recent observation that some such SO(5) models are integrable \cite{Links2002,Dukelsky2006}.

\medskip


\subsection{Pairing in $LST$-coupling}
\label{sect:OtimesOdualilty}

 In  $LST$ coupling,  a many-nucleon wave function is a combination of spatial, spin, and isospin wave functions.  There is more than one $LST$ coupling scheme, as discussed in the following section.  Here we consider the coupling scheme for nucleons occupying a single sub-shell of fixed orbital angular momentum $l$.

The primary objective is to construct wave functions that are totally antisymmetric and at the same time have good total angular momentum and isospin.  This is achieved by use of  Wigner's supermultiplet theory in concert with duality relationships as follows.

For nucleons of spin $s=\hf$ and isospin $t=\hf$, so that
$(2s+1)(2t+1)=4$, the space of single-particle wave functions of orbital angular momentum $l$ is of dimension $4(2l+1)$.
Thus, whereas in $jj$ coupling the wave functions for  $n$ nucleons of angular momentum $j$  transform according to an irrep $\{ 1^n\}$ of the unitary group U$( 2(2j+1))$, the $LST$-coupled wave functions that we now consider transform  according to an irrep $\{ 1^n\}$ of the unitary group U$(4(2l+1))$.
A desirable basis for the corresponding Fock space, 
$\mathbb{F}^{(4(2l+1))}$, is  one that reduces the subgroup chain
\begin{widetext}
\be
\begin{array}{ccccccccccccc}
{\rm U}(4(2l+1))  &\supset& {\rm U}(2l+1)& \times& {\rm U}(4)&\supset
& {\rm O}(2l+1)&\times& {\rm SU}(2)_{\rm S}
&\times&{\rm SU}(2)_{\rm T} &\supset & {\rm O(3)}_{\rm L} \\
n && \lambda &&\tilde\lambda&&\kappa&& S && T&&\pi\, L\, M
\end{array} ,
\label{eq:193z} \ee
\end{widetext}
where U(4) is Wigner's super-multiplet group \cite{Wigner1,Wsup} i.e., the group of unitary transformations of the  spin-isospin states of a nucleon.
This subgroup chain was proposed for the definition of an $LST$ coupling scheme by \textcite{Bayman}.   
The orbital $L$ and spin $S$ angular momenta can now be coupled to good total angular momentum $J$ in the knowledge that the antisymmetric requirement is looked after.

The dynamical content of this coupling scheme can be understood from a consideration of the dual subgroup chain.
From the paired subgroups of O$(8(2l+1))$ with dual representations on the Fock space $\mathbb{F}^{(4(2l+1))}$ shown for $w=4$ in Eq.\ (\ref{eq:110}), it is seen that  
O$(2l+1)$ and an SO(8) group have dual representations and that
U$(2l+1)$ and the U(4) supermultiplet group have dual representations.
It follows that the basis states of the coupling scheme  (\ref{eq:193z}) are identical to basis states for
the Fock space $\mathbb{F}^{(4(2l+1))}$ that reduce the subgroup chain
\be
\begin{array}{cccccccccccccccc}
{\rm O}(3)_L 
&\!\!\times\!\!&{\rm SO}(8)&\!\! \supset\!\! &{\rm U(4)}  &\!\!\supset\!\!&
{\rm U}(1) &\!\!\times\!\! &{\rm SU}(2)_{\rm S}&\!\!\times\!\!
& {\rm SU}(2)_{\rm T} \\
\pi L\,M 
&& \hf d(\tilde\kappa) &&\! \hf d (\tilde\lambda) && n&& S&&T
\end{array} ,
\label{eq:193} \ee
where   $\hf d(\tilde\kappa)$ and $\hf d(\tilde\lambda)$, with $d=2l+1$, label highest weights for SO(8) and U(4) irreps, respectively, as described below.

As mentioned in Sec.\  \ref{sect:bgl}, the discovery that the two subgroup chains (\ref{eq:193z}) and (\ref{eq:193})
define common basis  states followed the formulation of an SO(8) pairing model by \textcite{FS} who subsequently learned of Bayman's $LST$ coupling scheme.

\subsubsection{The Lie algebras of SO(8) and O$(2l+1)$ and their irreps}

Let $\{ a^\dag_{\sigma m}; \sigma = 1,\dots,4; m =-l,\dots,+l\}$ denote nucleon creation operators, where $\sigma$ indexes the spin-isospin state of a nucleon and $m$ denotes the projection of its orbital angular momentum, $l$, onto the axis of quantization.
The group O$(2l+1)$  is then the  subgroup of U$(2l+1)$ transformations that leave invariant the $L=0$ pair-creation operators
\be \hat {\cal A}_{\sigma\tau}
= \sum_m (-1)^{l+m} a^\dag_{\sigma m} a^\dag_{\tau, -m}
= - \hat{\cal A}_{\tau\sigma} \quad \sigma, \tau = 1, \dots , 4 . \ee

In parallel with all the coupling schemes considered to this point,  
these $L=0$ pair-creation operators and the corresponding pair-annihilation operators $\hat {\cal B}_{\sigma\tau}$, given by their Hermitian adjoints, 
generate the Lie algebra of a group that commutes with the group O$(2l+1)$.
They are in fact the raising and lowering operators of an so(8) Lie algebra, the complex extension of which has a basis given by the $L=0$
angular-momentum-coupled operators
\bea \hat \mathcal{A}_{\sigma\tau} &=&
\sum_{m}a^\dag_{\sigma m} a^\dag_{\tau\bar{m}} , \label{eq:18}\\
\hat \mathcal{B}_{\sigma\tau} &=&  \sum_m
a^{\tau\bar{m}} a^{\sigma m} , \label{eq:19}\\
\hat \mathcal{C}_{\sigma\tau} &=& \sum_m a^\dag_{\sigma m} a^{\tau m}- \hf (2l+1)  \delta_{\sigma,\tau},
\label{eq:U4inO8} \eea
where
\be a^\dag_{\sigma\bar m }=(-1)^{l+m}a^\dag_{\sigma, -m}, \quad
a^{\sigma\bar m }=(-1)^{l+m}a^{\sigma, -m}\ee

The Lie algebra, so$(2l+1)$,  of the group O$(2l+1)$ is the subalgebra of 
u$(2l+1)$ elements that commute with the $\hat \mathcal{A}_{\sigma\tau}$ operators of Eq.\ (\ref{eq:18})
and, hence, with all operators of the so(8) Lie algebra.
In parallel with previous examples, see Eq.\ (\ref{eq:Cartanusp1}), the
Cartan subalgebra of so$(2l+1)$ is spanned by the subset of u$(2l+1)$ operators,
\be \hat C_{mm} = \sum_\sigma
\big( a^\dag_{\sigma m} a^{\sigma m} - a^\dag_{\sigma \bar m} a^{\sigma\bar m}\big), \;\; m=1, \dots , l. \ee
If an so$(2l+1)$ irrep has highest weight 
$\kappa = (\kappa_1, \dots, \kappa_l)$, where $\kappa_m$ is the eigenvalue of the Cartan operator $\hat C_{mm}$ on the highest-weight state for the irrep,  the irrep of so$(2l+1)$ and of its Lie group, SO$(2l+1)$, is denoted by $[\kappa]$.
In addition to elements of its SO$(2l+1)$ subgroup, the group O$(2l+1)$ also
contains reflections and inversions whose matrices have negative 
determinants.  Thus, if we denote by `det' the one-dimensional irrep of 
O$(2l+1)$, in which an element $g\in {\rm O}(2l+1)$ is simply mapped 
to $\det (g)$, then the irreps of O$(2l+1)$ occur in \emph{associated} pairs, $[\kappa]$ and $[\kappa]^*$, that are related by
\be [\kappa]^* = \det \otimes [\kappa] . \ee
In the more convenient notation of \textcite{Littlewood}, if an irrep of O$(N)$ is denoted by $[\kappa]$, the associated irrep $[\kappa]^*$ is denoted by the partition $[\kappa']$, whose conjugate  $\tilde\kappa'$ is defined by
\be \tilde\kappa'_1 = N- \tilde\kappa_1, \quad \tilde\kappa'_i = \tilde\kappa_i,
\quad{\rm for} \; i\not= 1.
\ee
Note, that in replacing $[\kappa]^*$ by $[\kappa']$ in this way, the length 
$l(\kappa') = \tilde\kappa'_1$ of the partition $\kappa'$ will generally exceed 
$N/2$, which is the maximum length of an SO$(N)$ highest weight.
Thus, in the Littlewood convention, the restriction $\tilde\kappa_i \leq N/2$ on an O$(N)$ weight is extended to allow all weights for which
\be \tilde\kappa'_1+\tilde\kappa'_2 \leq N .\ee
The branching rules for the restriction of the irreps of U$(N)$ to O$(N)$,
 compatible with the Littlewood convention,
 have been given, for example, by \textcite{King75}.

A dual pair of SO(8) and O$(2l+1)$ irreps on the Fock space
$\mathbb{F}^{(4(2l+1))}$,  labeled 
 by an O$(2l+1)$  highest weight $\kappa$,
is defined generically by a state
\bea   |\kappa\rangle &=& 
\big( a^\dag_{1,l} a^\dag_{1,l-1} \dots a^\dag_{1, l+1-\tilde\kappa_1}\big)
\label{eq:kappaO2l+1} \\
&& \dots  \big( a^\dag_{4,l} a^\dag_{4,l-1} \dots 
 a^\dag_{4,l+1-\tilde\kappa_4}\big)|0\rangle 
\nonumber  \eea
where $\tilde\kappa$ is the four-row partition conjugate to $\kappa$.
Such a state is of highest weight for an O$(2l+1)$ irrep, denoted in the Littlewood convention by $[\kappa]$, and of lowest weight for an SO(8) irrep of weight given by
\be \langle \kappa | \hat{\cal C}_{\sigma\sigma} |\kappa\rangle = \tilde\kappa_\sigma - \hf(2l+1) . \label{eq:tlleU4wts}
\ee
Thus, if we denote the SO(8) irrep with this lowest weight by 
$[\hf d(\tilde\kappa)]$ with $d=2l+1$, it is seen that each irrep of
${\rm O}(2l+1) \times {\rm SO}(8)$ on a subspace of
$\mathbb{F}^{(4(2l+1))}$ is defined by a unique partition $\kappa$.

\subsubsection{Tabulation of  basis states In $LST$ coupling}

The quantum numbers of basis states defined by the subgroup chain
(\ref{eq:193}) signify irreps of the corresponding groups in the chain as follows.  For a given nucleon number $n$, the labels $\lambda$ and $\tilde\lambda$ of the
${\rm U}(2l+1)\times {\rm U}(4)$ irreps are given by the branching rules
\be \begin{array}{rcl}
\mathrm{U}(4(2l+1)) &\downarrow& {\rm U}(2l+1)\times \mathrm{U}(4) \\
\;:\; \{ 1^n\} &\downarrow& \bigoplus_{\lambda\vdash n}\, \{ \lambda\} \times \{
\tilde\lambda\} ,\end{array} 
\label{eq:brule265}\ee
with $\lambda$ and $\tilde\lambda$ restricted to partitions having
no more than $2l+1$ and  $4$ parts, respectively.

To determine the O$(2l+1)$ irreps $[\kappa]$ in a given U$(2l+1)$ irrep 
$\{\lambda\}$, one needs the coefficients in the known
\be {\rm U}(2l+1)\downarrow{\rm O}(2l+1) \; :\;
 \{ \lambda\} \downarrow \bigoplus_{\kappa} F'_{\kappa\lambda} \,
[\kappa]  \label{eq:UdtoOd}
\ee
branching rules  \cite{King75,RW10}.
Because of the duality relationships, the same coefficients appear in the
\be {\rm SO(8)}\downarrow{\rm U}(4) \; :\;
\langle \hf d(\tilde\kappa)\rangle \downarrow \bigoplus_{\lambda} F'_{\kappa\lambda} \, \{\hf d(\tilde\lambda)\}  \label{eq:SO8toU4}
\ee
branching rules with $d=2l+1$.

The ${\rm U}(4) \downarrow {\rm SU(2)}_S \times {\rm SU(2)}_T$ branching rules are determined sequentially from the known branching rules for
\be {\rm U}(4) \downarrow {\rm O}(4) \downarrow {\rm SO}(4)
\simeq {\rm SU(2)} \times {\rm SU(2)}. \ee
For example,   if an ${\rm SU(2)} \times {\rm SU(2)}$ irrep is denoted by 
$(S,T)$,  the $ {\rm O}(4)\downarrow {\rm SU(2)} \times {\rm SU(2)}$ reduction is given by
\be [\tilde\lambda_1 \tilde\lambda_2] \downarrow 
\Big( \frac{\tilde\lambda_1+\tilde\lambda_2}{2},
\frac{\tilde\lambda_1-\tilde\lambda_2}{2}\Big)
\oplus \Big( \frac{\tilde\lambda_1-\tilde\lambda_2}{2},
\frac{\tilde\lambda_1+\tilde\lambda_2}{2}\Big)
\ee
for $\tilde\lambda_2 \not=0$ and
\be [\tilde\lambda_1, 0] \downarrow
( \hf \tilde\lambda_1,\hf \tilde\lambda_1) ;
\ee
the reduction for a $[\tilde\lambda_1 \tilde\lambda_2]^*$ irrep is identical, e.g.,
\be {\rm U}(4) \downarrow {\rm O}(4)\, ;\, 
\{31^2\} \downarrow [3]^* \oplus [21] \oplus [1]^* \ee
and 
\bea {\rm O}(4) \downarrow {\rm SU(2)} \times {\rm SU(2)} &:& 
[3] \downarrow \textstyle (\frac32,\frac32) \nonumber\\
&:& [21] \downarrow \textstyle (\frac32,\frac12) \oplus (\frac12,\frac32)
\nonumber\\
&:& [1] \downarrow \textstyle (\frac12,\frac12) ,
\eea
give
\bea {\rm U}(4) &\downarrow& {\rm SU(2)} \times {\rm SU(2)} \,:\, \nonumber\\
\{31^2\} &\downarrow&
\textstyle (\frac32,\frac32) \oplus (\frac32,\frac12) \oplus (\frac12,\frac32)
\oplus (\frac12,\frac12) .\eea

As an illustration, the classification of states for $l=1$ by the quantum numbers defined for the subgroup chains (\ref{eq:193z}) and (\ref{eq:193}) is given in Table \ref{tab:su4ul}.
From  this table, the set of U(4) irreps that comprise each SO(8) irrep is  determined by listing all the U(4) irreps $\{\tilde\lambda\}$ that occur in combination with a given 
O$(2l+1)$ irrep (labeled  for O$(2l+1)={\rm O(3)}$  by $L^\pi$) as shown in Table \ref{tab:o8l1}.

\begin{table*}[htb]
\caption{\label{tab:su4ul} Spectrum of nucleon states in the $l=1$ shell classified by nucleon number $n$, for $n=0,\dots 6$, U$(2l+1)= {\rm U}(3)$ symmetry $\{ \lambda\}$,
orbital angular momentum $L$, parity $\pi = (-1)^n$, U(4) symmetry $\{ \tilde\lambda\}$, spin $S$, and isospin $T$. }
\begin{ruledtabular}
\begin{tabular}{ccccccccc}  
\medskip
$n$ &\quad &$\{\lambda\} $  &\quad
& $L$ &$\pi$ &\quad & $\{ \tilde\lambda\}$ & $(S,T)$  \\
\tableline
\smallskip  0&&$\{0\}$ && 0 &$+$ && \{0\} & (0,0)   \\
\smallskip  1 && $\{1\}$ && 1 &$-$  &&$\{1\}$ & $(\frac{1}{2},\frac{1}{2})$ \\
\smallskip  2 && $\{2\}$ && 0, 2 &$+$ &&$\{1^2\}$ & (1,0), (0,1) \\
\smallskip   && $\{1^2\}$ && 1 &$+$ &&$\{2\}$ & (0,0), (1,1) \\
\smallskip  3 && $\{3\}$  && 1, 3  &$-$&&$\{1^3\}$ & $(\frac{1}{2},\frac{1}{2})$ \\
\smallskip   && $\{21\}$  && 1, 2 &$-$ &&$\{21\}$ & $(\frac{1}{2},\frac{1}{2})$,
$(\frac{1}{2},\frac{3}{2})$, $(\frac{3}{2},\frac{1}{2})$\\
\smallskip   && $\{1^3\}$  && 0 &$-$ &&$\{3\}$ & $(\frac{1}{2},\frac{1}{2})$,
$(\frac{3}{2},\frac{3}{2})$ \\
\smallskip  4 && $\{4\}$  && 0, 2, 4 &$+$ &&$\{1^4\}$ & (0,0)\\
\smallskip   && $\{31\}$  && 1, 2, 3 &$+$ &&$\{21^2\}$ & (0,1), (1,0), (1,1) \\
\smallskip   && $\{2^2\}$   && 0, 2 &$+$ &&$\{2^2\}$ & (0,0), (0,2), (2,0), (1,1) \\
\smallskip   && $\{21^2\}$   && 1 &$+$ &&$\{31\}$ & (0,1), (1,0), (1,1), (1,2), (2,1) \\
\smallskip  5 && $\{41\}$  && 1, 2, 3, 4 &$-$ &&$\{21^3\}$ & $(\frac{1}{2},\frac{1}{2})$ \\
\smallskip   && $\{32\}$   && 1, 2, 3 &$-$ &&$\{2^21\}$ & $(\frac{1}{2},\frac{1}{2})$,
$(\frac{1}{2},\frac{3}{2})$, $(\frac{3}{2},\frac{1}{2})$ \\
\smallskip   && $\{31^2\}$ && 0, 2 &$-$ &&$\{31^2\}$ & $(\frac{1}{2},\frac{1}{2})$,
$(\frac{1}{2},\frac{3}{2})$, $(\frac{3}{2},\frac{1}{2})$, $(\frac{3}{2},\frac{3}{2})$
\\
\smallskip   && $\{2^21\}$  && 1 &$-$ &&$\{32\}$ & $(\frac{1}{2},\frac{1}{2})$,
$(\frac{1}{2},\frac{3}{2})$, $(\frac{3}{2},\frac{1}{2})$, $(\frac{1}{2},\frac{5}{2})$,
$(\frac{5}{2},\frac{1}{2})$, $(\frac{3}{2},\frac{3}{2})$ \\
\smallskip  6 && $\{42\}$  && 0, $2^2$, 3, 4 &$+$ &&$\{2^21^2\}$ & (1,0), (0,1) \\
\smallskip   && $\{41^2\}$  && 1, 3 &$+$ &&$\{31^3\}$ & (0,0), (1,1) \\
\smallskip   && $\{33\}$  && 1, 3 &$+$ &&$\{2^3\}$ & (0,0), (1,1) \\
\smallskip   && $\{321\}$  && 1, 2 &$+$ &&$\{321\}$ & (0,1), (1,0),
(0,2), (2,0), 2(1,1), (1,2), (2,1) \\
\smallskip   && $\{2^3\}$  && 0 &$+$ &&$\{3^2\}$ & (0,1), (1,0),
(0,3), (3,0),  (1,2), (2,1)
\smallskip\\
\end{tabular}
\end{ruledtabular}
\end{table*}

\begin{table*}[ht]
\caption{The partitions $\{ \tilde\lambda\}$ defining the
${\rm U(4)}\subset {\rm SO(8)}$ sub-representations  
$\{ \frac32(\tilde\lambda)\}$ 
contained in the ${\rm SO(8)}\times{\rm O}(2l+1)$ irreps for  $l=1$  and even values of $n$.
The ${\rm O}(2l+1)={\rm O}(3)_L$ irreps are labeled by $[\kappa]$ and equivalently by $L^\pi$. 
SO(8) irreps are  labeled  by their highest-weight U(4) subirreps,
$[\frac32(\tilde\lambda)]_{\rm highest}$. \label{tab:o8l1} }
\medskip
\begin{ruledtabular}
\begin{tabular}{cc|c|c|c|c|cc|cc|cc} 
 $n$ &\phantom{\big|}
 & $\{ \tilde\lambda\}$
& $\{ \tilde\lambda\}$
 & $\{ \tilde\lambda\}$
 & $\{ \tilde\lambda\}$
 & $\{ \tilde\lambda\}$\\
\tableline
  0 &\phantom{\big|}&  \{0\}&&&\\
  2 &\phantom{\big|} &$\{1^2\}$& $\{2\}$ & $\{1^2\}$  &\\
  4 &\phantom{\big|} &$\{1^4\},\{2^2\}$& $\{21^2\},\{ 31\}$&
      $\{1^4\},\{ 21^2\},\{2^2\}$&$\{21^2 \}$& $\{1^4 \}$\\
  6 &\phantom{\big|}  &$\{2^21^2\},\{ 3^2\}$&$\{31^3\},\{ 2^3\},\{321\}$
       &2$\{2^21^2\},\{ 321\}$ & $\{ 2^21^2\},\{ 31^3\},\{ 2^3\}$&  $\{
2^21^2\}$  \\
  8 &\phantom{\big|} &$\{2^4\},\{3^21^2\}$&$\{32^21\}, \{ 3^22 \}$&
     $\{2^4\},\{ 32^21\},\{3^21^2\}$& $\{ 32^21\}$&  $\{2^4\}$\\
  10&\phantom{\big|}  &$\{3^22^2\}$& $\{3^31\}$& $\{3^22^2\}$ &\\
  12&\phantom{\big|}  &$\{3^4\}$& && \\ \hline
$[\kappa]$  &\phantom{\big|}&[0] &$[1^2]\equiv [1]^*$ &[2] &$[31]\equiv [3]^*$& [4]     \\
$L^\pi$ &\phantom{\big|}& $0^+$ & $1^+$& $2^+$ &$3^+$ &$4^+$ \\
$[\frac32(\tilde\lambda)]_{\rm highest}$ &\phantom{\big|} & 
$[\frac32,\frac32,\frac32,\frac32]$ & 
$[\frac32,\frac32,\frac12,-\frac12]$ & 
$[\frac32,\frac32,\frac12,\frac12]$ & 
$[\frac32,\frac12,\frac12,-\frac12]$ & 
$[\frac12,\frac12,\frac12,\frac12]$ \\
\end{tabular}
\end{ruledtabular}
\end{table*}

\subsubsection{A simple SO(8) model}

The dynamical subgroup chain (\ref{eq:193}) enables the construction of  simply solvable spin- and isospin-invariant pairing models.
 Consider, for example, the Hamiltonian
\be \hat H  =\varepsilon \hat n -
{\textstyle \frac14} \chi \sum_{\sigma,\tau=1}^2 \hat{\mathcal{A}}_{\sigma\tau} \hat{\mathcal{B}}_{\sigma\tau} , \label{eq:pairingintn}\ee
where
$\hat{\mathcal{A}}_{\sigma\tau}$ and $\hat{\mathcal{B}}_{\sigma\tau}$ are, respectively, so(8) raising and lowering operators.  $\hat H$ could also include terms in the Casimir invariant of the SU(2) subalgebras of U(4).
Such a Hamiltonian can be expressed in terms of Casimir operators of SO(8) and those of its subgroups.  Its spectrum is then immediately determined. As shown in \textcite{RC07}, see also \cite{RW10}, the eigenvalues of the Hamiltonian (\ref{eq:pairingintn})
 are given explicitly in terms of their SO(8) and U(4) labels by
\bea E_{\kappa\lambda}&\!\!=\!\!&\varepsilon n -
{\textstyle \frac14} \chi \sum_\sigma \Big[
(\tilde\kappa_\sigma - \hf d)(\tilde\kappa_\sigma - \hf d +2-2\sigma) \nonumber\\
&& \quad - (\tilde\lambda_\sigma - \hf d)(\tilde\lambda_\sigma - \hf d +2-2\sigma)
\Big] .
\label{eq:5.Ekl}
\eea

A primary value of the SO(8) pair-coupling model is to give  a physical interpretation of the kind of Hamiltonian that is diagonalized by the classification of nuclear shell model states in the $LS$-coupling scheme, defined by the subgroup chain (\ref{eq:193z}). 
In parallel with the SU(2) quasi-spin and SO(5) pairing models, the  SO(8) model has an extension to a multi-shell pairing model together with corresponding shell-model coupling schemes.
Such extensions are important for several reasons:
one reason is that some such models are integrable (see \textcite{LermaEDS07}); another, as we now discuss,  is that they raise the possibility of exploring the competition between pairing and deformation correlations in nuclei.

\subsection{The SU(3) and Sp$(3,\Rb)$ $LST$-coupling models}

In addition to the ${\rm U}(2l+1) \times {\rm U}(4) \supset {\rm O}(2l+1)$ 
coupling scheme, two other $LST$ coupling schemes are of special interest.
The first is based on the subgroup chain
\bea {\rm U}(N) \times {\rm U}(4) &\supset& {\rm SU(3)}\times {\rm O(4)} 
\nonumber\\
&\supset &{\rm SO(3)}\times {\rm SU(2)}_S \times {\rm SU(2)}_T , 
\label{eq:SU3scheme}\eea
 where $N=\sum_l(2l+1)$ is a sum over the $l$ values that occur in a single harmonic oscillator shell, e.g., $l=0$ and 2 for the  $(2s1d$
 shell and $l=1$ and 3 for the $(2p1f)$  shell.
The second is based on the subgroup chain
\be {\rm Sp}(3,\Rb) \times {\rm U}(4) \supset {\rm SU(3)}. \ee
These coupling schemes are important for the microscopic description of nuclear collective states.
The latter was discussed in Sec.\  \ref{sect:Sp(3,R)}.
The former coupling scheme corresponds to Elliott's SU(3) model of rotational states in light nuclei.

It is now apparent that the two coupling schemes, given by Eq.\ 
(\ref{eq:SU3scheme}) and by
\bea {\rm U}(N) \times {\rm U}(4) &\supset& {\rm O}(N) \times {\rm O}(4)
\nonumber\\
&\supset &{\rm SO(3)}\times {\rm SU(2)}_S \times {\rm SU(2)}_T , 
 \label{eq:Onscheme}\eea
define shell model basis states that diagonalize Hamiltonians for nuclei with deformation and pairing correlations, respectively.  
Thus, they provide useful bases for a  study of the competition between these correlations.

A preliminary study of this competition  \cite{RosensteelR07} determined the spectra of  $(2s1d)$-shell nuclei, for which $N=6$, for a Hamiltonian
\be \hat H = -\alpha \,\hat{\Cas} ({\rm su3}) - (1-\alpha)\,\hat{\Cas}({\rm so6}) + 
\beta\, \hat{\Cas}({\rm so3}) ,\ee
where $\hat{\Cas}({\rm su3})$, $\hat{\Cas}({\rm so6})$, and 
$\hat{\Cas}({\rm so3})$ are  Casimir operators for the respective subalgebras of u(6) and $0\leq \alpha \leq 1$.
A remarkable result was found; for $\alpha \lesssim 0.4$ the spectrum of low-energy states and their properties were characteristic of an O(6) phase and for 
$\alpha \gtrsim 0.6$ they became characteristic of an SU(3) phase. Such behavior has been seen in numerous similar studies and has been termed {quasi-dynamical symmetry} (see Sec.~\ref{sect:QDsymmetry}).


\section{Other developments in the application of symmetry methods in physics}
\label{sect:others}

From among the many developments in  symmetry methods of importance in physics, we mention  a few of particular relevance to the topic of this review.

\subsection{Boson mappings}\label{sect:bmaps}

In  low-density situations,  systems of even fermion number often behave   like bosons.  For example, alpha particles are meaningfully approximated as bosons in the interpretation of superfluidity and Bose-Enstein condensation \cite{GriffinSS96,GriffinNZ09}.
Such \emph{quasi-boson approximations} to algebraic models can be obtained  by group contraction methods \cite{InonuW53}
They are widely used in many-body theory \cite{Sawada57};   early
approaches in nuclear physics  were described, for example,  by  \textcite{Rowebook} and 
\textcite{RingS80}.

Thus, it is natural to seek corrections to these approximations in terms of exact boson mappings as given, for example,  by the \textcite{HolsteinP40}   representation of su(2)
\be \begin{array}{c}
\hat J_0 = -j + c^\dag c, \\
\hat J_+ = c^\dag (2j - c^\dag c)^{\frac12} , 
\quad \hat J_- = (2j - c^\dag c)^{\frac12} c ,
\end{array} \label{eq:HPsu2}
\ee
for an arbitrary spin $j$, where $c$ and $c^\dag$ satisfy the boson commutation relations    $[c,c^\dag ]=1$.
A comprehensive review of the many approaches to boson realizations of Lie algebras, initiated by \textcite{BelyaevZ62}, has been given by \textcite{KleinM91}.

Boson realizations are synonymous with coherent state representations.
This is apparent from  the \emph{Bargmann representation} of the Heisenberg-Weyl algebra \cite{Bargmann61}
in which  boson operators are represented in terms of a complex variable $z$ by
\be c^\dag \mapsto z,  \quad c \mapsto d/dz .\ee

Coherent state representations are defined for many Lie algebras.
For example, according to \textcite{Perelomov72, Perelomov86}, SU(2) coherent states are  defined by
$|z\rangle = \exp (z^* \hat J_+) |j,-j\rangle$, where $z$ is a complex variable, and
a coherent-state wave function for a state $|\psi\rangle\in \Hb$ is defined as  the overlap function
\be \psi(z) = \langle z| \psi\rangle = \langle j,-j| e^{z\hat J_-} |\psi\rangle .\ee
A component of SU(2) angular-momentum $J_k$ then has coherent-state representation defined by
\bea \Gamma(J_k) \psi(z) &=& \langle j,-j | e^{z\hat J_-}  \hat J_k |\psi\rangle
\nonumber\\
&=&\langle j,-j | \big( e^{z\hat J_-}  \hat J_k e^{-z\hat J_-} \big) 
e^{z\hat J_-}|\psi\rangle . \quad \eea
Thus, an expansion of  $e^{z\hat J_-}  \hat J_k e^{-z\hat J_-}$ in terms of
$\hat J_0$ and $\hat J_\pm$, and  the identities
\bea \langle j,-j | \hat J_0 e^{z\hat J_-}|\psi\rangle &=& -j \psi(z),\\
\langle j,-j | \hat J_- e^{z\hat J_-}|\psi\rangle &=& \frac{\partial}{\partial z} \psi(z),\\
\langle j,-j | \hat J_+ e^{z\hat J_-}|\psi\rangle &=& 0,
\eea
leads to the su(2) representation
\be \begin{array}{c}
\Gamma(J_0) = -j + z d/d z ,\smallskip \\
\Gamma(J_-) = d/d z, \quad
\Gamma(J_+) = z(2j -d/d z) .
\end{array}
\ee
This representation is now transformed into the Bargmann form of the Holstein-Primakoff representation
\bea
& \Gamma(J_0) = -j + z d/d z ,& \label{eq:HP_Rep}\\
&\Gamma(J_-) = (2j -d/d z)^{\frac12} d/d z, \quad
\Gamma(J_+) = z(2j -d/d z)^{\frac12} , \nonumber
\eea
by a similarity transformation.

More generally, if an irrep of a Lie algebra on a Hilbert space $\Hb$ has a  lowest-weight that is uniquely defined for the vacuum of a commuting set of lowering operators $\{ \hat B_k\}$ and the eigenvalues of a set of weight operator $\{ \hat C_i\}$, i.e., by the formulas
\be \hat B_k|0\rangle =0 , \quad \hat C_i |0\rangle = \lambda_i |0\rangle ,
\ee
then coherent state wave functions are defined in terms of a set
$z=\{ z_k\}$ of complex variables, for each state $|\psi\rangle\in \Hb$,  by
\be  \psi(z)= \langle 0|\exp\Big(\sum_k z_k \hat B_k\Big) |\psi\rangle. \ee
Corresponding coherent-state representations of  the Lie algebra are defined as illustrated above for su(2).
Such coherent state representations were used to determine exact boson mappings for a large variety of semi-simple Lie algebras by 
\textcite{Dobaczewski81a,Dobaczewski81b,Dobaczewski82}.

The standard   theory of coherent states  and coherent-state 
representations has been reviewed by \textcite{KlauderS85} and by \textcite{Perelomov86}. 

\subsection{More general scalar and vector coherent state (VCS) representations}\label{sect:CSreps}

Coherent state methods  are extraordinarily powerful.  In their most general forms  \cite{RoweR91}, they  provide  simple and  versatile constructions of Lie group and Lie algebra representations, which include the methods of \emph{induced representations} of \textcite{Mackey68}.  They also provide an interface between classical and quantum mechanics \cite{BartlettRR02, BartlettRR02b, BartlettR03, Gazeau09, RoweTBP}.

However, the standard construction of  coherent-state representations is  limited  to representations with lowest- and/or highest-weight states that are uniquely defined by sets of commuting raising operators.

Two extensions overcome this limitation.  The first extension is from scalar to vector coherent state (VCS) irreps.
Let $\{ \hat B_k\}$ be a commuting subset of 
lowering operators for an  irrep of a semisimple Lie algebra $\Lfr{g}$ on  Hilbert space $\Hb$ and let $\{ |\nu\rangle\}$ be an orthonormal basis for the subspace 
$\Hb_0\subset \Hb$ of states that are annihilated by these lowering operators.
Subject to certain conditions, a vector-valued coherent state wave function
\be \Psi(z) = 
\sum_\nu |\nu\rangle \langle \nu| \exp\big(\sum_\nu z_k \hat B_k\big) |\psi\rangle ,
\ee
can then be defined for a state $|\psi\rangle\in \Hb$ and a corresponding construction of a coherent-state representation of an element $X\in \Lfr{g}$ is  defined by
\be \Gamma(X) \Psi(z) = \sum_\nu |\nu\rangle \langle \nu |
 \exp\big(\sum_\nu z_k \hat B_k\big) \hat X|\psi\rangle .
\ee
The construction is useful if the  subset of commuting lowering operators is such that the subspace $\Hb_0$ carries a finite-dimensional unitary irrep of a subalgebra  $\Lfr{g}_0 \subset \Lfr{g}$.

Such VCS  irreps were introduced  \cite{Rowe84, RoweRC84, RoweRG85}  for the  purpose of calculating  matrix elements of the non-compact symplectic algebra sp$(3,\Rb)$ in an SU(3) $\supset$ SO(3)  basis as needed in the nuclear symplectic model.
 A   \emph{partial coherent state theory},
which went some way towards solving this problem,
  was also proposed for this purpose \cite{DeenenQ84, DeenenQ85}.
The VCS construction was then applied to calculate the explicit matrices for the irreps of numerous Lie algebras and even some super-algebras, as reviewed by 
\textcite{Hecht87}.
It was also shown  \cite{RoweR91} that VCS irreps are induced representations \cite{Mackey68}
in which an irrep of $\Lfr{g}$ is induced from  an irrep of a subalgebra 
$\Lfr{h} \subset \Lfr{g}$.

A second extension makes use of other kinds of coherent states beside those generated by exponentiating lowering (or raising) operators.
It was introduced  because a standard coherent state irrep of su(3) enables its matrix elements to be computed in an SU(2) basis whereas, in applications with a rotationally-invariant Hamiltonian, one needs  a basis that reduces the SO(3) $\subset$ SU(3) subgroup.
However, as shown by \textcite{Elliott1, Elliott2},  the rotated states 
$\{ |\Omega\rangle = 
\hat R(\Omega) |\lambda\mu\rangle, \Omega\in {\rm SO(3)}\}$, where
$|\lambda\mu\rangle$ is a highest-weight state for a generic su(3) irrep, 
 span the Hilbert space for that irrep.
 Moreover, they are generalized coherent states, as defined by
\textcite{Perelomov86}.
Thus, an arbitrary $|\psi\rangle\in \Hb$ has a scalar coherent state wave function defined by the  overlap function 
\be \psi(\Omega) =  \langle \lambda\mu|\hat R(\Omega) |\psi\rangle . \ee
The corresponding coherent state irrep of the su(3) Lie algebra is then defined as usual  by
\be \Gamma(X) \psi(\Omega) = 
 \langle \lambda\mu|\hat R(\Omega) \hat X|\psi\rangle . \ee

It is interesting to note that, whereas a standard coherent state representation gives a boson realization of a $(\lambda,0)$ irrep of SU(3)  in an SU(2) basis, the new construction leads to a rotor realization of a generic $(\lambda\mu)$ irrep in an SO(3) basis.  Thus quasi-boson and quasi-rotor approximations are obtained, respectively, in large $\lambda$ and/or $\mu$ contraction limits.  
The latter extension can also be applied with other groups besides SO(3) and 
within the framework of the VCS extension so that there is now a wealth of possibilities for handling a large variety of situations.

\subsection{Quasi-dynamical symmetry} \label{sect:QDsymmetry}

Although  dynamical symmetry is  of immense significance, as illustrated by its many applications mentioned in this review,  it is in fact an idealization that is only achieved to  some level of approximation in realistic situtations.
What is remarkable is the extent to which models based on assumed dynamical symmetries are successful.  Thus, as suggested by \textcite{HessAHC02}, they should really be regarded as effective dynamical symmetries.
Quasi-dynamical symmetry was introduced  \cite{RochfordR88} as a mechanism for understanding the nature of these effective symmetries.

Quasi-dynamical symmetry  is an approximate realization of the precise mathematical concept of an \emph{embedded representation} \cite{RoweRR88}   loosely defined as follows.
Let $\Hb$ be a Hilbert space for a representation $T$ that is a direct sum of irreps of a Lie algebra $\Lfr{g}$ and let $\Hb_0\subset \Hb$ be a subspace.
If the matrix elements $\langle \psi |\hat X |\psi'\rangle$ of all 
$\hat X = T(X)$, with $X\in \Lfr{g}$, and all $|\psi\rangle$ and $|\psi'\rangle$ in 
$\Hb_0$, should happen to be equal to those of a representation of $\Lfr{g}$,  we say that this representation is an embedded representation of $\Lfr{g}$.  

Sub-representations and linear combinations of equivalent representations are trivial examples of embedded representations.  
However, there are non-trivial examples for Lie algebras with irreps that are scale related such as rotor-model algebras.
Consider, for example, a set  of irreps of some Lie algebra $\Lfr{g}$ labeled 
 by $\{ \lambda\}$ with basis states $\{ |\lambda LM\rangle\}$ where $L$ and $M$ are SO(3) angular momentum quantum numbers and suppose that elements
$\{ X_i\}$ of $\Lfr{g}$ have matrix elements in this basis that scale in a manner given by an equation
\be \langle \lambda LM | \hat X_i |\lambda' L'M' \rangle =
\delta_{\lambda,\lambda'} f^\lambda_{i\lambda_0}
\langle \lambda_0 LM | \hat X_i |\lambda_0 L'M' \rangle ,\ee
where $f^\lambda_{i\lambda_0}$ is a real proportionality constant.
Matrix elements between states given for each $LM$ by 
\be |LM\rangle = \sum_\lambda C_\lambda |\lambda LM\rangle, \ee
where $C_\lambda$ is an $LM$-independent set of coefficients, are then equal to those of an average irrep $\bar\lambda$ for which
\be f_{i\lambda_0}^{\bar\lambda} = \sum_\lambda |C_\lambda|^2 f_{i\lambda_0}^{\lambda} .\ee

Only a limited number of Lie algebras have irreps that scale precisely in this way. However, most algebras of importance  in physics have  contraction limits with this property and so admit embedded representations approximately.
Thus, quasi-dynamical symmetries as approximate embedded representations  are common and particularly important for the interpretation of symmetry-related phases of physical systems and the transitions between them.
Several examples, were given in a conference report \cite{RoweQDS04}.  Other applications and perspectives have been given by \textcite{HessAHC02, 
Yepez-MCsehH06, MacekDC09,BonatsosMcC10}.  A review of quantum phase transitions and the use of quasi-dynamical symmetry in understanding them has been given by \textcite{CejnarJC10}.

\subsection{Partial dynamical symmetry}

In realistic situations,  approximate  dynamical symmetries   may only be acceptable for a limited number of states of a  system. In fact, mixed symmetry studies have shown that one symmetry may be dominant at low-energies (in a quasi-dynamical symmetry sense) and another at higher energies \cite{CaprioCI08}.
Thus, a theory of partial dynamical symmetry (which also applies to quasi-dynamical symmetry) was introduced by \textcite{AlhassidL92}.  
The occurrence of partial symmetry conservation in nuclear models has subsequently been considered by several authors and 
interesting examples  have been discovered, e.g., by Zamick and colleagues \cite{EscuderosZ06,ZamickI08}.
Such examples and the development of efficient models of partial dynamical symmetry have been reviewed recently by \textcite{Leviatan11}.


\section{Discussion and summary}

Examples have been given in this review of many  results, of importance in physics and mathematics, that  follow from the duality of various group representations.
Examples have been drawn primarily from applications in nuclear and atomic spectroscopy.  However, many more applications of this extraordinarily powerful concept are known and undoubtedly  more remain to be discovered.

We have focused primarily on the subgroup chains of symmetry groups for sequences of Hamiltonians of increasing complexity which define coupling schemes for many particle systems.  Subgroup chains define basis states 
for Hilbert spaces that diagonalize Hamiltonians such as those given by combinations of the Casimir and other invariants of the groups in the chain.  Moreover, they provide basis states for the description of more general Hamiltonians of interest.  Thus, the study of subgroup chains of potential symmetry groups for a system is an important step in understanding the range of possible dynamics that the system can exhibit.  This approach has been emphasized in the many studies of the interacting boson model \cite{IBM}, where such chains of groups are said to define the dynamical symmetries of a model.

A primary motivation for studying dual subgroup chains is that they reveal associations of many phenomenological models of nuclear physics with shell-model coupling schemes.
Thus, we have shown that for many of the groups  in a subgroup chain that define a coupling scheme, there are frequently
other groups  with dual representations on the same or an enlarged  Hilbert space of the system.  
It then follows, as explained and illustrated in this review, that if one group is a dynamical group for a class of Hamiltonians, its dual (if it has one)
is a symmetry group for the same class of Hamiltonians.  
In this situation, the pair of groups with dual representations is of considerably greater value than either group separately.  An even more useful situation arises when each  of the several groups of a subgroup chain that defines a coupling scheme is partnered with a dual group belonging to 
a so-called \emph{dual} subgroup chain.
For then, if one subgroup chain defines a chain of symmetry groups for a sequence of Hamiltonians of decreasing symmetry, the dual subgroup chain consists of dynamical groups for the same sequence of Hamiltonians, albeit in reversed order (i.e., in order of increasing dynamical symmetry).
When such dual chains exist, they augment the tools available for studying the dynamical content of a system considerably.
For example, it means that combinations of the elements of the Lie algebras of both a group and its dual will leave the irreps of both of the groups invariant.  This property was used \cite{RR2003}, for example, to identify subsets of two-body interactions that conserve seniority.

The relationship between algebraic models and shell-model coupling schemes is shown to be invaluable for embedding  models, such as pairing models and collective models, into a more fundamental  microscopic theory.
The examples given were chosen to highlight the 
application of dual pairs of group and subgroup chain representations in the construction of simply solvable algebraic models and for providing useful basis states and coupling schemes for a general theory.
In addition to providing a microscopic interpretation of successful phenomenological models, they also provide the means to identify the appropriate shell model coupling scheme for a microscopic description of phenomena that have a simple model explanation.

We have given many examples of the use of duality relationships to infer the properties of one group from those of another.
For example, Schur-Weyl duality relates the characters of unitary group irreps to irreps of symmetric groups.  
This relationship yields  the extraordinarily valuable result that the characters  of 
 different U$(n)$ groups are given by a 
 common set of Schur functions which are defined by their S$_N$ symmetries 
and take the same form, for each $\lambda\vdash N$, independent of the number $n$ of variables.
Parallel relationships exist between the characters of other dual pairs of group representations and can be used, for example, to determine many branching rules from simpler known rules.
The branching rules for the representations of classical Lie groups have been reviewed by \textcite{King75}, and many of them are related by duality relationships.
For example, it is recalled in Section \ref{sect: impONSpm}
 that the duality of the groups in the chain ${\rm Sp}(N,\Rb) \supset {\rm U}(N)$ with those in the chain ${\rm O}(m) \subset {\rm U}(m)$ on the space of an 
 $Nm$-dimensional harmonic oscillator enables one to calculate the branching rules for the restriction ${\rm Sp}(N,\Rb) \downarrow {\rm U}(N)$ from the known  ${\rm U}(m)\downarrow {\rm O}(m)$ branching rules \cite{RWB}.
This approach was extended by \textcite{KingW85} to include branching rules for U$(p,q)$ and SO$^*(2n)$ to their respective ${\rm U}(p) \times {\rm U}(q)$  and U$(n)$ subgroups.
Such branching rules, which give the restrictions of the  characters of a group to a unitary subgroup, can be more useful than explicit expressions for the characters of  a non-compact group, such as  ${\rm Sp}(N,\Rb)$, whose unitary irreps are infinite-dimensional 
This is because one knows far more about the characters of the unitary and symmetric groups than about those of other groups.
A review of many branching rules for dual reductive group and subgroup pairs has been given by \textcite{HoweTW04} in terms of the Littlewood-Richardson coefficients for the tensor products of symmetric and unitary group characters (cf. Sec.\ \ref{sect:4A}).
Examples are given in this review of how the tensor products of various groups of importance in many-particle spectroscopy can be derived as sums of irreps by methods that result from duality relationships.

It is known that the Clebsch-Gordan coefficients and more general Wigner-Racah algebras for one subgroup chain are related to those for a dual
subgroup chain  \cite{HechtLeBR87,LeBlancH87,LeBlanc87}.
A prototype of such a relationship was given in the papers of \textcite{U5toO6} and \textcite{RoweTh05} and discussed in Sec.\ \ref{sect:6dimHO}.  The example showed   that the
transformation of basis states between a 
${\rm U}(6) \supset {\rm U}(5) \supset {\rm O}(5)$  coupling scheme for a 6-dimensional harmonic oscillator and a
 ${\rm U}(6) \supset {\rm O}(6) \supset {\rm O}(5)$  coupling scheme
 is given simply by SU(1,1) Clebsch-Gordan coefficients which make the dual transformations between
 ${\rm SU(1,1)} \times {\rm SU(1,1)} \supset {\rm SU(1,1)} \supset {\rm U(1)}$ coupled states to 
 ${\rm SU(1,1)} \times {\rm SU(1,1)}  \supset  {\rm U(1)} \times {\rm U(1)} \supset {\rm U(1)}$  coupled states.
As observed, in an analysis of the phase transition from an O(6) to a U(5) coupling scheme with a change of a parameter in the interacting boson model \cite{U5toO6},  the availability of such transformation coefficients enables matrix elements that are diagonal in one coupling scheme to be expressed simply in another.
 The possibility of such relationships has the potential for relating the various couplings schemes in shell-model calculations.   An example, given in Sec.\ 
 \ref{sect: multishellBCS}, is an extension of the single-shell 
 ${\rm U}(2j+1)\supset {\rm USp}(2j+1)$ coupling scheme, relevant for Hamiltonians with strong pairing interactions, to multi-shell coupling schemes.  Such extensions are similarly made for other shell model coupling schemes.

It is mentioned, although it is not discussed in this review, that the Casimir operators of dual pairs of Lie groups are also simply related.
This is to be expected because all multi-linear combinations of the elements of a Lie algebra  are invariants of a dual Lie algebra.  In particular, the Casimir invariant of one Lie algebra is an invariant of the dual Lie algebra.  Moreover, it 
has been seen that the highest weights for dual irreps of a pair of semi-simple or reductive Lie algebras are related.  Thus, the eigenvalues 
of the Casimir and other invariants for dual irreps of such groups, which are expressible in terms of related highest weights, are likewise related.
For example, the Casimir invariant of u$(n)$,
\bea &&\hat{\Cas}(u(n)) = 
 \sum_{i j} \hat C_{ij} \hat C_{ji}  \nonumber \\
&& \qquad = \sum_{i\leq j} (2 \hat C_{ji} \hat C_{ij} +  \hat C_{ii} - \hat C_{jj})
+ \sum_i \hat C_{ii} \hat C_{ii}, \qquad
\eea
is determined to have eigenvalue 
\be \Cas^{(n)}_{\{\lambda\}} = \sum_{i=1}^n \lambda_i (\lambda_i + n+ 1 -2i) ,\ee
for an irrep with highest weight $\lambda$.
Thus, a dual pair of U$(n)$ and U$(m)$ irreps with common highest weights, having the property that $l(\lambda)$ does not exceed either $m$ or $n$, have Casimir invariants with distinct but closely related values.
The Casimir invariants of semisimple Lie algebras  
are described and their values given in terms of highest weights by \cite{Wyb}.

In spite of the simple origin of the relationships between Casimir invariants, the results can nevertheless be useful.  For example, a simply solvable model with a Hamiltonian expressed in terms of the Casimir invariants of chain of subgroups  can also be expressed in terms of the Casimir invariants of a dual subgroup chain.  This can be useful for the same reason that embedding a simple model in a much richer microscopic theory makes it possible to explore many more properties of the system being modeled.

Also not mentioned in this review is the duality between two copies of a single compact  group $G$ 
acting on ${\cal L}^2(G)$ by the left and right regular representations.  The assertion of this duality is essentially the Peter-Weyl Theorem.  
An explicit example,  is given by the
 regular representation of the rotation group, SO(3).
The Hilbert space, ${\cal L}^2$(SO(3)), for this representation
is spanned by the so-called Wigner $\scrD$ functions,
$\{ \scrD^L_{KM};L\!=\!0,1,2, \dots; K\!=\!-L,\dots , +L;  M\!=\!-L, \dots , +L\}$, defined in  terms of bases $\{ |LM\rangle ; M=-L, \dots , +L\}$ for irreps of angular momentum $L$ by
\be \hat T^{(L)}(\Omega)|LM\rangle = \sum_K |LK\rangle \scrD^L_{KM}(\Omega) \ , \quad \Omega \in {\rm SO(3)}. \ee
The left and right regular representations of SO(3) 
are then defined, respectively, for
$\omega\in  {\rm SO}(3)$, by
\bea \hat{L}_{\rm Reg}(\omega)\scrD^L_{KM}(\Omega) &=&
\scrD^L_{KM}(\omega^{-1}\Omega) \nonumber\\
&=& \sum_N \scrD^L_{NM}(\Omega)
\Big( \scrD^{L}_{NK}(\omega) \Big)^* , \quad \\
\hat{R}_{\rm Reg}(\omega)\scrD^{L}_{KM}(\Omega) &=&
\scrD^L_{KM}(\Omega\omega) \nonumber\\
&=& \sum_N \scrD^L_{KN}(\Omega)\scrD^L_{NM}(\omega) , 
\eea
where $\Big( \scrD^{L}_{NK}(\omega)\Big)^*$ 
is the complex conjugate of $\scrD^{L}_{NK}(\omega)$.
It seen that the  Wigner  $\scrD^L$ functions for any given value of $L$
form a basis for an irrep $\scrD^{L*}\times \scrD^{L}$ of the direct product group, 
${\rm SO}(3)\times {\rm SO}(3)$ relative to the left and right actions, respectively, of these SO(3) groups.  Thus, the two copies of SO(3) have dual representations on the Hilbert space of the SO(3) regular representation.
This example has important applications for the representations of the rotor model in nuclear physics, in which the right representations of SO(3) correspond to rotations relative to a space-fixed reference frame and the left representations correspond to intrinsic rotations relative to a frame of reference fixed 
in the body of the rotor.

An aspect of dual group representations that merits further investigation is the related geometry of the systems to which they apply.
It is known, for example, that the dynamics of a central force Hamiltonian for a system with a Euclidean configuration space, $\Rb^n$, can be regarded as a combination of rotations and radial motions.  For such a system the symmetry group O$(n)$ of rotations and inversions and the dynamical group SU$(1,1)$ associated with the radial dynamics are determined to have dual representations on the Hilbert space, $\mathcal{L}^2(\Rb^n)$, of the system.  This dual pair of groups reflects the underlying geometrical structure of the Euclidean space as a product 
manifold of a radial line and a unit $(n-1)$ sphere.
Thus, if $\{ x_i , i=1,\dots,n\}$ is a set of Cartesian coordinates, the group O$(n)$ is the set of all linear transformation of $\Rb^n$ that leave the squared radius of a point, $r^2 = \sum_i x_i^2$, invariant.  Moreover, the set of points generated by all
O$(n)$ transformations  of a point in $\Rb^n$ at distance $r=1$ from the origin is an $(n\!-\!1)$-dimensional unit sphere. Together, the radial coordinate $r$ and a set of coordinates for the unit sphere define spherical polar coordinates for the points of $\Rb^n$.

A generalization of this geometric structure is observed for the collective dynamics of an $N$-particle system with a  Euclidean configuration space
 $\Rb^{Nm}$  \cite{Gelbart}.
For such a system, the symmetry group of the Hamiltonian is the group
O$(N)$ and the dynamical group is Sp$(m,\Rb)$.  These groups have been shown to have dual representations on the Hilbert space
$\mathcal{L}^2(\Rb^{Nm})$.  If $\{ x_{ni}; n=1,\dots,N, i = 1,\dots, m\}$ is a set of Cartesian coordinates for $\Rb^{Nm}$, then O$(N)$ is seen as the set of all linear transformations of  $\Rb^{Nm}$ that leave the quadrupole moments
$Q_{ij} = \sum_{n=1}^N  x_{ni} x_{nj}$ invariant.  The set of points generated by all O$(N)$ transformations of a point in $\Rb^{Nm}$ of unit quadrupole moment is then the generalization of a unit sphere in $\Rb^N$ to a so-called Stieffel manifold
\be S^{N,m} = \{ x\in \Rb^{Nm}\, ; \, \sum_n x_{ni} x_{nj}= \delta_{i,j} \} .\ee
Thus, a set of quadrupole moments (elements of $m\times m$ symmetric matrices) and coordinates for the Stieffel manifold provide a system of collective and intrinsic coordinates for $\Rb^{Nm}$.  The geometrical structure underlying these coordinates introduced by \textcite{Gelbart} in a study of the representations of Sp$(m,\Rb)$ on 
$\mathcal{L}^2(\Rb^{Nm})$, proves to be of considerable significance for the development of the nuclear collective model 
\cite{RosensteelR77,RosensteelR80,Rowe85, RR98}.
Moreover, it transpires that the intertwining of the  representations of the symplectic and orthogonal groups on these spaces accounts for the centrifugal coupling of the dynamics on these two spaces.  
In particular, it leads to an understanding of the role of vorticity degrees of freedom.

The examples chosen in this review to illustrate the range of results that can be obtained from a consideration of dual group representations are primarily from nuclear and atomic physics.
This is the area of physics most familiar to the authors 
and the one in which many of the known  duality relationships 
have been discovered.  However, they are far from complete.
Indeed, we are optimistic that applications in other fields of physics will be brought to light by others.
There are certainly potential applications in quantum optics.  The fascinating concept of dual models in statistical mechanics \cite{Girvin96} also suggests interesting possibilities.
While different systems with common algebraic structures can be expected to exhibit parallel properties, the concept of dual representations of different groups indicates a similarly close parallel between systems with dual algebraic structures.  
We are also aware that many other duality relations are known for which there are undoubtedly applications;
cf., for example, \textcite{KV, Gelbart1, Adams83, Howe85, LeungT94}.
Duality relationships are also known to exist for quantum groups and supersymmetric groups.  For example, \textcite{LuHowe10} have recently explored an application
of a duality relationship between O(3,1) and the orthosymplectic osp(2,2) superalgebra to Maxwell's equations.

\bigskip

\acknowledgments{We are  indebted to Dr.\ Santo \textcite{Dagostino05}, who initiated a study of group duality and its physical applications in his 
Ph.D.\ thesis.
 We are also much appreciative of the careful proof reading of our manuscript by Dr. Trevor Welsh and for  helpful comments and encouragement from Professor Roger Howe.

This work was supported in part by  grants from the
Natural Sciences and Engineering Research Council of Canada.}

\bibliographystyle{apsrmp}
\bibliography{rmpduality}

\end{document}